\documentclass[aps,prapplied,preprint,superscriptaddress]{revtex4-2}

\usepackage{graphicx}
\usepackage{dcolumn}
\usepackage{bm}
\usepackage{amssymb}   
\usepackage{xcolor}
\usepackage{ulem}
\usepackage{soul}
\usepackage{amsmath}
\usepackage{longtable}
\usepackage{float}
\usepackage[colorlinks,linkcolor=green,citecolor=green,urlcolor=blue]{hyperref}
\usepackage[utf8]{inputenc}
\usepackage{physics}

\newcommand{\iitdchem}{Department of Chemistry , Indian Institute of Technology Delhi, Hauz Khas, New Delhi 110016, India}
\newcommand{\iitkphy}{Department of Physics, Indian Institute of Technology Kanpur, Kanpur, 208016, India}
\newcommand{\smpchina}{School of Mathematics and Physics, University of Science and Technology Beijing, Beijing 100083, China}
\newcommand{\iitddmse}{Department of Materials Science and Engineering, Indian Institute of Technology Delhi, Hauz Khas, New Delhi 110016, India}
\newcommand{\nwchem}{Department of Materials Science and Engineering, Northwestern University, Evanston, Illinois 60208, United States}

\usepackage[draft]{changes} 
\usepackage[normalem]{ulem}

\usepackage{soul}
\usepackage{lipsum}

\makeatletter
\def\blfootnote{\xdef\@thefnmark{}\@footnotetext}
\makeatother

\begin{document}
\title{Data-driven Discovery of Novel High-performance Quaternary Chalcogenide Photovoltaics}

\author{Nikhil Singh$^\dagger$}
\affiliation{\iitdchem}
\author{Mohammad Ubaid$^\dagger$}
\blfootnote{$\dagger$ N.S. and M.U. contributed  equally.}
\affiliation{\iitkphy}
\author{Pabitra Kumar Nayak}
\affiliation{\iitdchem}
\author{Jiangang He}
\affiliation{\smpchina}
\author{Dibyajyoti Ghosh}
\affiliation{\iitdchem}
\affiliation{\iitddmse}
\author{Chris Wolverton}
\affiliation{\nwchem}
\author{Koushik Pal}
\email{koushik@iitk.ac.in}
\affiliation{\iitkphy}

\date{\today}

\begin{abstract}
Photovoltaic materials facilitate the conversion of sunlight into electricity by harnessing the interaction between light and matter, offering an eco-friendly and cost-efficient energy solution. Combining data-driven approaches with static and time-dependent density functional theories and nonadiabatic molecular dynamics simulations, we predict 14 high-performance photoabsorber materials from a family of known quaternary semiconductors. Among these, we investigate four  compounds - SrCuGdSe$_3$, SrCuDyTe$_3$, BaCuLaSe$_3$, and BaCuLaTe$_3$ in greater detail. Hybrid density functional theory calculations including spin-orbit coupling reveal that SrCuGdSe$_3$, SrCuDyTe$_3$, BaCuLaSe$_3$  and BaCuLaTe$_3$ possess  direct  band  gaps of 1.65, 1.79, 1.05, and 1.01 eV, respectively.  These band gap values lie close to an optimal range ideal for visible-light absorption. Consequently, the calculated optical absorption coefficient and spectroscopic limited maximum efficiency for these compounds become comparable or larger than crystalline silicon, GaAs, and methylammonium lead iodide. Calculated exciton binding energies for these compounds are relatively small (30-32 meV), signifying easy separation of the electron-hole pairs, and hence enhanced power conversion efficiencies.  Investigations of photoexcited carrier dynamics reveal a relatively long carrier lifetime ($\sim$ 30-40 $ns$), suggesting suppressed nonradiative recombination and enhanced photo-conversion efficiencies. We further determined the defect formation energies in these compounds, which showed that despite the likely formation of cation vacancies and interstitial defects, midgap states remain absent making these defects non-detrimental to carrier recombination. Our theoretical predictions invite experimental verification and encourage further investigations of these and similar compounds in this quaternary semiconductor family.

\end{abstract}

\maketitle

\section{Introduction}

High-performance photovoltaic (PV) materials are in huge demand due to their potential application in solar cell devices. Over the years, remarkable advancements have been made in the discovery, development, and optimization of PV compounds \cite{green2017energy,nayak2019photovoltaic}. For example, the discovery of methylammonium lead iodide (MAPbI$_3$) belonging to the hybrid halide perovskite family has sparked extensive research activity in optoelectronics, that resulted in a dramatic improvement of the power conversion efficiency (PCE) from 3.8\% \cite{kojima2009organometal} in 2009 to 20\% \cite{xing2014low} in 2014.  In general, halide perovskites have gained significant attention due to their remarkable PCE, tunable band gaps, and superior defect tolerance \cite{yin2015halide,yin2014unusual, yin2015origin,kojima2009organometal,Lee2012, stranks2013electron} that originates in part from the ionic bonding and ensuing octahedral coordination environment \cite{yin2014unusual, yin2015origin}. The main challenges associated with halide perovskite are their lack of long-term stability and the toxicity of lead (Pb), which have limited the use of these materials from becoming commercially viable for large-scale applications \cite{yang2023mixed}. On the other hand, despite showing long-term stability, state-of-the-art PV materials like Si, GaAs, and CdTe which possess tetrahedral coordination environment are prone to form point defects and grain boundaries due to the covalent nature of their chemical bonding \cite{deng2013electronic} which degrade their performance.
Therefore, discovery and optimization of novel materials having mixed octahedral and tetrahedral coordination environments could be beneficial for achieving higher efficiencies and improved stability. In this regard, high-throughput (HT) computational screening has emerged as one of the most useful methods, which allows for a systematic exploration of a large number of materials for targeted properties \cite{Singh2025,luo2021high,jain2017high,jiang2021high}.

Recent studies \cite{pal2021accelerated, jiang2021high, ju2017perovskite, huo2018high, deng2024high, zhao2017design} demonstrate that HT screenings of material databases like the Materials Project (MP) \cite{jain2013commentary}, Open Quantum Materials Database (OQMD) \cite{saal2013materials,kirklin2015open}, Automatic Flow for Materials discovery (Aflow) \cite{curtarolo2012aflowlib}, and Inorganic Crystal Structure Database (ICSD) \cite{belsky2002new}, can significantly accelerate the discovery of novel material and property prediction.  Unlike conventional experimental approaches that rely mostly on trial and error-based methods, HT screening  enhances efficiency and clarifies structure-property relationships. Recent HT investigations based on density functional theory (DFT) have successfully contributed to the discovery of materials for various useful properties, leading to progress in chemical synthesis \cite{pal2021accelerated, ju2017perovskite, huo2018high}. For example, Zhao \textit{et al.} \cite{zhao2017design} explored a series of lead-free double perovskites by utilizing the strategy of cation transmutation, highlighting several optimal novel candidates with promising optoelectronic properties for PV applications. Some of these compounds include Cs$_2$SnI$_6$, Cs$_2$AgInCl$_6$, and Cs$_2$AgBiBr$_6$, that have been experimentally verified in recent years \cite{zhou2018synthesis, locardi2018colloidal, mcclure2016cs2agbix6, zhou2018synthesis}. These materials offer the potential for reduced toxicity while maintaining favorable optoelectronic properties. 

Ternary chalcogenides with the formula AMQ$_3$ (A = alkali/alkaline earth metal,  M = transition metal,  Q = chalcogen) represent a family of materials that hold promise for energy applications, fulfilling essential criteria of abundance and stability \cite{Singh2025,shen2023accelerated,pal2019high,jiang2020designing,kuhar2018high}. Many of these materials demonstrate semiconducting behavior and possess useful optoelectronic properties. Sun \textit{et al.} \cite{sun2015chalcogenide} theoretically explored the potential of these ternary chalcogenides for potential PV applications, predicting that some of these materials could match the high performance of MAPbI$_3$. They investigated 18 ternary chalcogenide compounds and identified four promising candidates, namely, BaZrS$_3$, CaTiS$_3$, CaZrSe$_3$, and CaHfSe$_3$, with suitable band gaps. Among these, BaZrS$_3$ has already been synthesized by Perera \textit{et al.} \cite{perera2016chalcogenide}. Ju \textit{et al.} \cite{ju2017perovskite} expanded this study by evaluating another set of 18 ternary chalcogenide compounds with both perfect and distorted structures using DFT calculations. They predicted that SrSnS$_3$ and SrSnSe$_3$ are direct bandgap semiconductors with band gaps ranging from 0.9 to 1.6 eV, proposing that a mixture of SrSnS$_3$ and SrSnSe$_3$ could optimize the band gap for efficient sunlight absorption. Recently, Shen \textit{et al.} \cite{shen2023accelerated} identified 52 AMQ$_3$ candidates as promising materials for PV applications based on another HT study. 
Interestingly, it has been shown that altering the metal-to-chalcogen ratio within these families can provide diverse crystal structures, phases, and band gaps, thereby influencing their optoelectronic properties \cite{mckeever2023functional}.  

Although ternary metal chalcogenides have been investigated heavily in recent years, their quaternary counterparts have remained relatively less explored for optoelectronic applications. Here, we systematically investigate the properties of a family of quaternary chalcogenides, denoted by the formula AMM'Q$_3$ \cite{koscielski2012structural,shahid2023structure,mckeown2021structural,xie2022structure,luo2020high,le2019new} using HT computational screening to discover potential high-performance PV materials. The AMM'Q$_3$ (A = alkali, alkaline earth, or post-transition metals; M/M' = transition metals, lanthanides; Q = S, Se, or Te) family possesses rich chemistry and diverse structure types \cite{koscielski2012structural}. Koscielski \textit{et al.}\cite{koscielski2012structural} explained that the synthesized AMM'Q$_3$ compounds are charge-balanced and therefore many of these AMM'Q$_3$ compounds are expected to be semiconductors. Using HT and machine learning assisted strategies, many thermodynamically stable AMM'Q$_3$ compounds were discovered computationally \cite{pal2019intrinsically,pal2021accelerated}. Some of these predicted compounds are synthesized experimentally \cite{laing2022acuzrq3,berseneva2022transuranium,ishtiyak2021syntheses,grigoriev2022quaternary,eickmeier2022exploring,ruseikina2023synthesis,grigoriev2023single,ruseikina2023synthesis,ruseikina2024syntheses,grigoriev2023single,ruseikina2024elucidating}. Several theoretical and experimental studies have shown that many compounds in this family exhibit ultralow lattice thermal conductivity and promising thermoelectric performance \cite{laing2022acuzrq3,pal2019intrinsically,pal2019unraveling,yu2024substitution}. To our knowledge, 211 compounds have been synthesized in this AMM'Q$_3$ family so far, which are listed in Tables SI, SII, and SIII in the Supplemental Material \cite{supmat}.  These AMM'Q$_3$ compounds have  quasi-3D structures that has mixed ionic and covalent bonding within octahedral and tetrahedral coordination environments, respectively. Whereas ionic bonding environments are expected to give rise to defect tolerance properties, the covalent bonding helps in electrical conduction. Despite these advantages, the potential of AMM'Q$_3$ chalcogenides in PV applications have remained largely unexplored \cite{fabini2019candidate}. This work explores the optoelectronic properties of these 211 experimentally known AMM'Q$_3$ chalcogenides to discover high-performance and stable PV materials.

\section{Computational methods}

\subsection*{Structural optimization}
We performed DFT calculations using the Vienna Ab-initio Simulation Package (VASP)\cite{kresse1996efficiency} utilizing the projector-augmented wave (PAW)\cite{kresse1999ultrasoft} potentials. We represented the exchange-correlation (XC) energies of the electrons employing the  generalized gradient approximation (GGA) \cite{perdew1996generalized} functional utilizing the PBE\cite{perdew1996generalized} parametrization. For structural optimization of the AMM'Q$_3$ compounds including spin-polarization, we adopted the OQMD settings \cite{saal2013materials,kirklin2015open}. These compounds contain either 12 or 24 atoms in their primitive unit cells depending on their crystal symmetry \cite{koscielski2012structural, pal2021accelerated}. The optimized lattice constants agree  well with the corresponding experimental values. In this work, we studied the optoelectronic properties of the four compounds (SrCuGdSe$_3$, SrCuDyTe$_3$, BaCuLaSe$_3$, and BaCuLaTe$_3$) in details, which contain 24 atoms in their primitive unit cells.  We provide the optimized lattice constants for these four compounds in Table SIV in the Supplemental Material \cite{supmat} that agree very well (error $<$ 0.5 \%) with their corresponding experimental values. Spin-polarized calculations helped us identify the magnetic and non-magnetic compounds in the AMM'Q$_3$ family. For exploring optoelectronic properties, we retain only the non-magnetic compounds for further calculations and analysis.

\subsection*{Calculations of electronic structures}
We calculated the electronic density of states (DOS) and  electronic structures of the non-magnetic AMM'Q$_3$ compounds. While calculating the electronic structure, we used the k-path convention of Setyawan and Curtarolo \cite{setyawan2010high}. We used a  kinetic energy cut-off of 520 eV and k-point mesh of 14 $\times$ 5 $\times$ 4 for the static calculations at the GGA level.  To determine the band gap of these compounds with higher accuracy, we performed hybrid density functional calculation using HSE06 \cite{heyd2003hybrid} including spin-orbit coupling (SOC) with a reduced k-point grid (8 $\times$ 2 $\times$ 2). The HSE06 XC energy functional is constructed by mixing 25\% exact exchange and 75\% PBE XC energy with the long-range Coulomb interaction was screened using $\mu$ = 0.207 $\AA^{-1}$.

\subsection*{Calculations of carrier effective masses}
We calculated the carrier effective masses for the electrons and holes for the semiconducting AMM'Q$_3$ compounds using sumo python toolkit \cite{ganose2018sumo}, which employs the parabolic fitting of the band edges. Given the anisotropic nature of these materials,  effective masses are obtained along different high-symmetry directions in the Brillouin zone at the valence band maximum (VBM) and conduction band minimum (CBM), which are then directionally averaged. To provide a few representative values, we provide the average effective masses in Tables SI, SII, and SIII in the Supplemental Material \cite{supmat}.

\subsection*{Calculations of exciton binding energy}
To assess the potential for photon-induced exciton dissociation in the studied compounds, we calculated the exciton binding energy (E$_b$) using the hydrogen-like Wannier-Mott model \cite{fox2010optical}. This model is a widely adopted approach \cite{grunert2024predicting, zhao2017design} that quantitatively relates E$_b$ to the effective masses of holes (m$_h^*$) and electrons (m$_e^*$), and the static dielectric constant ($\epsilon_r$) of the material. The E$_b$ was then calculated by the following expression:
 \begin{equation}
E_b = \frac{\mu_X}{\hbar^2} \left( \frac{e^2}{4\pi \varepsilon_0 \varepsilon_r} \right)^2 
\end{equation}
where $\mu_X$ = ($\frac{1}{m_h^*} + \frac{1}{m_e^*}$) and $\epsilon_0$ are the reduced effective mass and permittivity of free space, respectively. We calculated $\epsilon_r$  using density functional perturbation theory (DFPT) \cite{wu2005systematic,gajdovs2006linear} as implemented in  VASP, within the GGA-PBE exchange-correlation functional.

\subsection*{Calculations of phonon dispersions}
To check the  dynamical stability of the four compounds (SrCuGdSe$_3$, SrCuDyTe$_3$, BaCuLaSe$_3$, and BaCuLaTe$_3$), we have determined their phonon dispersions using  the finite-displacement method as implemented in the Phonopy code \cite{phonopy}. Each of these compounds has a relatively large unit cell containing 24 atoms. To calculate phonon dispersion, we generated nearly isotropic supercells (containing 144 atoms) of each compound using the following supercell transformation matrix (\textbf{a'}, \textbf{b'}, \textbf{c'}) = (\textbf{a}, \textbf{b}, 
\textbf{c}) $ \quad \begin{pmatrix} 0 & 3 & 3\\ 1 & 0  & 1 \\ 1 & 1 & 0 \end{pmatrix}$, where (\textbf{a}, \textbf{b}, \textbf{c}) and (\textbf{a'}, \textbf{b'}, \textbf{c'}) are the lattice vectors for the primitive unit cell and supercell, respectively. To calculate the forces on the atoms in these supercells, we used 2 $\times$ 2 $\times$ 2 k-points. Our calculations show stable phonon dispersions for all compounds, signifying their dynamical stability (Fig. S1 in the Supplemental Material \cite{supmat}).

\subsection*{Calculations of photo-conversion efficiency}
To assess the optoelectronic performance of the AMM'Q$_3$ compounds,  we calculated their frequency dependent complex dielectric function $\epsilon(\omega) = \epsilon_1 (\omega) + i \epsilon_2 (\omega)$, using the PBE functional, which  was then used to calculate the absorption coefficient $\alpha(\omega) = \frac{4\pi}{\lambda}[\frac{(\epsilon_1(\omega)^2 + \epsilon_2(\omega)^2)^{\frac{1}{2}}-\epsilon_1(\omega)}{2}]$ as function of photon absorption frequency ($\omega$) and hence energy with $\lambda$ denoting its wavelength. In our work, we have taken $\omega$ in the unit of energy (eV). Here, $\epsilon_1$ and $\epsilon_2$ are real and imaginary parts of the complex dielectric function, respectively, that depends on the energy of the incident photon ($\omega$). We obtained converged values of $\alpha(\omega)$ with 14 $\times$ 5 $\times$ 4 k-point grid (see Fig. S2 in the Supplemental Material \cite{supmat}). Due to the relatively large primitive unit cell (24 atoms), the calculation of $\epsilon(\omega)$ with HSE06 functional becomes computationally prohibitive. Hence, the calculated absorption spectra were applied a simple scissor correction by shifting them (along the x-axis) with the difference of PBE and the HSE06 calculated band gaps including SOC. Similar approach was adopted in previous studies also \cite{wang2019materials}. We have also calculated $\alpha(\omega)$ for crystalline silicon (c-Si) and MAPbI$_3$ using a similar approach for comparison.  To estimate the photo-conversion efficiency of the quaternary chalcogenides, we calculated their spectroscopic limited maximum efficiency (SLME)\cite{yu2012identification} as a function of layer thickness utilizing VASPKIT \cite{VASPKIT}. SLME provides a realistic estimation of the maximum efficiency of an absorber layer by accounting for its actual absorption and interaction with the solar spectrum, assuming a single-junction solar cell construction. In the SLME approach, the absorption probability $a(\omega)$ is approximated by
\begin{equation}
    a(\omega) = 1 - \exp^{-2\alpha(\omega)L}
\end{equation}
where $L$ is the thickness of the solar absorber material.

\subsection*{Calculations of defect formation energy}
We modelled point defects by constructing  3 $\times$ 1 $\times$ 1 supercell of each compound that contains 72 atoms, making them nearly isotropic in the spatial dimensions and used  4 $\times$ 4 $\times$ 3 $\Gamma$-centered k-point grid for the static calculations. Due to the complex crystal structures of these compounds, we calculated the defect formation energy of neutral (q = 0) point defect (X) using the following equation \cite{zhang1998defect,freysoldt2014first} at the PBE level:
\begin{equation}
    E^{f}[X^q] = E_{tot}[X^q] - E_{tot}[bulk] - \Sigma_im_i\mu_i +qE_F,
\end{equation}
where $E_{tot}[X^q]$ and $E_{tot}[bulk]$ are the total energies of the  AMM'Q$_3$ supercell with and without defect along with a charge $q$ = 0. The integer $m_i$ indicates that $m$ atoms of species $i$ are being removed ($m_i < $ 0) or added ($m_i > $ 0), and $\mu_i$ is the corresponding chemical potential of the atom. Standard values of $\mu_i$'s are taken from the reference \cite{kirklin2015open} which are also obtained from calculations at the PBE level. $E_F$ is the position of the Fermi level relative to VBM in the bulk.

\subsection*{Ab-initio molecular dynamics simulations}
We evaluated the structural stability of the screened materials at ambient condition (300K) by performing \textit{ab initio} molecular dynamics (AIMD) simulations using VASP \cite{kresse1996efficiency}. We considered a relatively large 3 $\times$ 1 $\times$ 1 supercell containing 72 atoms, used $\Gamma$-point for the calculations with a time step of 1 fs, and a plane-wave energy cutoff of 420 eV. The PBE exchange-correlation functional was utilized for these calculations due to the prohibitive cost of employing the HSE06 functional. For AIMD simulations, we begin the calculations with the DFT-optimized structures, then heated them to 300K through repeated velocity rescaling over a duration of 3 ps. To ensure thermal equilibrium, an additional 4 ps of trajectories were generated using the canonical ensemble.  Finally, we conducted 15 ps trajectories in the microcanonical ensemble and used 5 ps of these trajectories (with a 1fs timestep) for calculations involving nonadiabatic coupling at the $\Gamma$-point. 

\subsection*{Nonadiabatic molecular dynamics and time-dependent density functional theory calculations}
We used nonadiabatic molecular dynamics (NAMD) simulations and time-dependent density functional theory (TDDFT) calculations to understand the carrier dynamics of the photo-excited charge carriers.
The combination of classical and quantum approaches in NAMD simulations, specifically using the decoherence-induced surface hopping (DISH) method \cite{Akimov2013}, has been applied to explore the behavior of excited state charge carriers in these quaternary chalcogenides. In this method, electrons are treated quantum mechanically, while the behavior of nuclei is approached classically. Electron–hole recombination dynamics were analyzed using the LIBRA code. \cite{Akimov2016} This analysis involved 5,000 geometries extracted from AIMD trajectories, each undergoing 500 stochastic realizations of the DISH processes. To explore longer timescale recombination dynamics, the nonadiabatic Hamiltonian derived from the 5 ps trajectory was iteratively reused. These simulations quantified the real-time evolution of nonradiatively recombined carrier populations driven by nonadiabatic interactions between multiple potential energy surfaces. We emphasize on the electron-hole recombination happening across the band gap, highlighting the dynamic structural characteristics that facilitate electron-phonon interactions and limit carrier lifetime. The fitting function $f(t) = 1 - exp(-t/\tau)$ has been used to calculate the recombination lifetime ($\tau$). The pure dephasing function is the destruction of quantum coherence due to elastic electron-phonon scattering, which can be evaluated by the 2nd-order approximation of optical response formalism \cite{hamm2005principles} 
\begin{equation}
    D_{ij}(t) = \exp\left\{ -\frac{1}{\hbar}\int_0^t dt' \int_0^{t'} dt'' \, C_{ij}(t'') \right\}
    \label{eq:placeholder}
\end{equation}
$C_{ij}(t)$ is the unnormalized auto-correlation function of the fundamental energy gap; $C_{ij}(t) = \langle \delta E_{ij}(t) \delta E_{ij}(0) \rangle$, where $\delta E_{ij}(t) = E_{ij}(t) - \langle E_{ij} \rangle$ is the fluctuation of energy gap between $i^{\text{th}}$ and $j^{\text{th}}$ states from the canonical ensemble average value. We estimated the decoherence time by applying the concept of pure-dephasing time, as defined in optical response theory.  \cite{hamm2005principles}

\section{Results and discussion}

\subsection{High-throughput screening}

We conducted  HT computational screening of 211 experimentally reported AMM'Q$_3$ chalcogenides to identify promising candidates for PV applications. These AMM'Q$_3$ compounds possess layered quasi-three dimensional crystal structures belonging to four different space groups (Cmcm, Pnma, C2/m, P2$_1$/m) and are categorized into three classes, namely Type-I (A$^{1+}$M$^{1+}$M'$^{4+}$Q$_3^{2-}$), Type-II (A$^{2+}$M$^{1+}$M'$^{3+}$Q$_3^{2-}$), and Type-III (A$^{1+}$M$^{2+}$M'$^{3+}$Q$_3^{2-}$) depending on the formal valence charges of the constituent cations \cite{koscielski2012structural, pal2021accelerated}. 
The primitive unit cells of the AMM'Q$_3$ compounds contain either 12 or 24 atoms, and most of the compounds in this family possess  Cmcm space group followed by  Pnma space group.
Before performing any calculations and analysis for AMM'Q$_3$ compounds to explore their optoelectronic properties, we reviewed  literature for well-known and high-performance PV materials. We examined their PCE as a function of band gaps (calculated and experimentally measured) and effective carrier masses (electrons and holes). A list of 33 such compounds is curated, and their properties are given in Table SV in the Supplemental Material \cite{supmat}. First, we find that all of these compounds are non-magnetic. This is our first criterion while screening the AMM'Q$_3$ compounds. Therefore, we retained only non-magnetic AMM'Q$_3$ compounds to further explore their PV properties.

\begin{figure*}[t]
\centering
\includegraphics[clip, trim=5cm 0cm 9cm 0cm, width=1\textwidth]{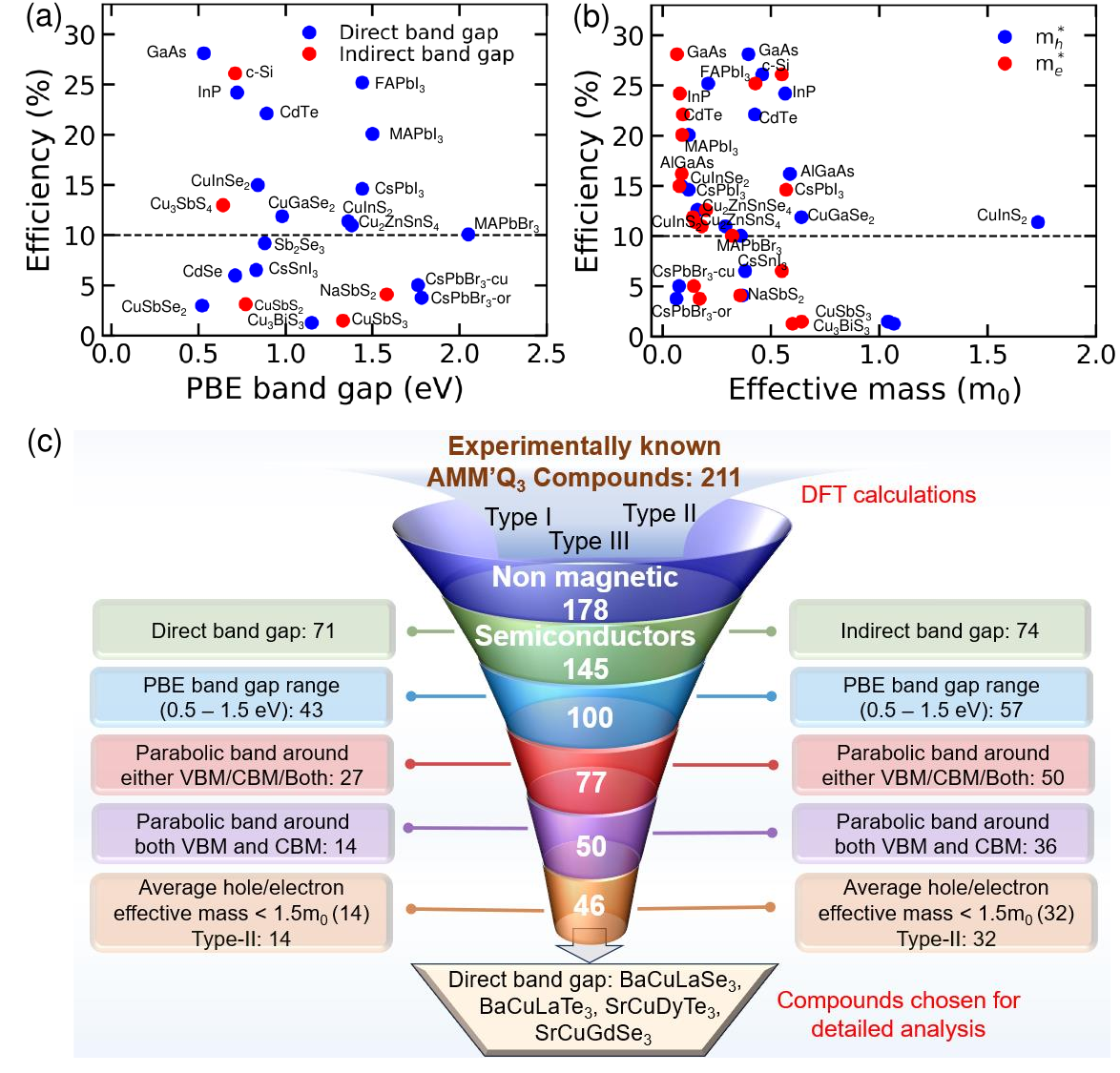}
\caption{Photo-conversion efficiencies of experimentally realized well-known photo-absorber materials plotted with respect to their (a) PBE functional calculated band gap and (b) average effective masses for the holes ($m_h^*$) and electrons ($m_e^*$). (c) The funnel diagram showing step-wise computational screening of experimentally known AMM'Q$_3$ chalcogenides to find potential high-performance photo-absorber materials.}\label{figfunnel}
\end{figure*}

Since HT-DFT calculations are typically done at the PBE level, we first collected and conducted calculations (wherever DFT-PBE data were missing) of the PBE band gaps of these 33 compounds and plotted their PCE (Fig.~\ref{figfunnel}(a)). Taking a cut-off of PCE $\ge$ 10 \% as a criterion for a high-performance PV material, we identify that most well-known PV materials have DFT-PBE band gaps between 0.5 to 1.5 eV. In addition, most high-efficiency PV materials have direct band gaps. 
Therefore, this would be our second and very important screening criterion based on PBE calculated electronic structures and band gaps of the AMM'Q$_3$ chalcogenides.
We will retain those AMM'Q$_3$ compounds as our first choice which have direct PBE band gaps between 0.5 to 1.5 eV. Since band gap is one of the most critical factors for high-performance solar absorber materials, we analyzed the band gaps of the curated compounds in greater detail. 
A plot of PCE as a function of their experimental band gaps (see Fig. S3(a) in the Supplemental Material \cite{supmat}) shows that the high-performance compounds have experimental band gaps between 0.8 eV to 2.3 eV. This range is higher than the DFT PBE band gaps because of the well known underestimations of  band gaps in DFT PBE calculations. 
Hybrid calculations based on HSE06 functional,  which improve upon the DFT PBE-based results show an increased range of band gap ($\sim$ 1-2.6 eV) for these high-performance PV materials (see Fig. S3(b) in the Supplemental Material \cite{supmat}). The HSE06 band gaps for our predicted compounds also fall within this range (see Table \ref{table1}). Plot of DFT PBE band gaps for the curated compounds also reveal the expected trend that these band gaps (see Fig. S3(c) in the Supplemental Material \cite{supmat}) are underestimated with respect to experimental values. However, their HSE06 band gaps {\textcolor{blue}(see Fig. S3(d) in the Supplemental Material \cite{supmat})} are in much better agreement. The comparison plot between HSE06 and PBE band gaps for the curated and some of the AMM'Q$_3$ compounds further depicts that PBE gaps are generally underestimated with respect to the HSE06 gaps \textcolor{blue}{}{(see Fig. S3(e) in the Supplemental Material \cite{supmat})}, as expected.

Further, we notice that all these high-performance PV compounds in the curated list have parabolic band dispersions around their band extrema in the Brillouin zone \cite{bouarissa1999effective,yang2016more,shabaev2015energy,sa2022unveiling,pandech2020effects,ghaithan2020density,wang2020structural}. Therefore, this would be our third screening criterion.  A critical factor in determining the transport properties of a semiconductor is the effective mass of charge carriers (electrons and holes). Materials with low effective masses are preferred because they facilitate higher carrier mobility, which is essential for efficient device performance. Therefore, we considered the average effective mass of the charge carriers as another critical screening criterion. We plotted the PCE of those 33 curated compounds as a function of their average electron (m$_e^*$) and hole effective masses (m$_h^*$) (Fig.~\ref{figfunnel}(b)), which reveal that all of the high-performance PV compounds have m$_e^*$ or m$_h^*$ $\le$ 1.5 m$_0$, where m$_0$ is the rest mass of an electron. This would be our final screening criterion for the AMM'Q$_3$ compounds, where we keep compounds with average effective masses less than 1.5 m$_0$ for further explorations in potential PV applications.

The OQMD contains many of the experimentally known 211 AMM'Q$_3$ compounds and their DFT calculated data. Therefore, their optimized structures, band gaps, and magnetic information are already present at the DFT-PBE level. We have utilized these data in our screening steps. We performed similar calculations for the rest of the known AMM'Q$_3$ compounds to use them in the screening procedure (Fig.~\ref{figfunnel}(c)). We found 33 magnetic AMM'Q$_3$ compounds from the above data, hence removed them from further screening. Afterward, we analyzed DOS for the remaining 178 AMM'Q$_3$ compounds at the DFT-PBE level, revealing 145 compounds with finite band gaps. We performed electronic structure calculations on these 145 compounds. Next, we applied the band gap criterion on these 145 compounds, which allowed us to retain 100 (43 and 57 AMM'Q$_3$ compounds with direct and indirect band gaps, respectively) with band gap values between 0.5 and 1.5 eV. From these, we keep only those compounds that have parabolic bands around either valence band maximum (VBM), conduction band minimum (CBM), or both, which reduces the number of compounds to 77, among which 27 and 50 AMM'Q$_3$ compounds possess direct and indirect band gaps, respectively. Further, we found that there are 50 compounds that exhibit parabolic bands around both VBM and CBM simultaneously, among which 14 and 36 AMM'Q$_3$ compounds possess direct and indirect band gaps, respectively. We retain only these 50 compounds for further screening. 

\begin{figure}[h!]
\centering
\includegraphics[clip, trim=0cm 4cm 15cm 1.5cm, width=0.5\textwidth]{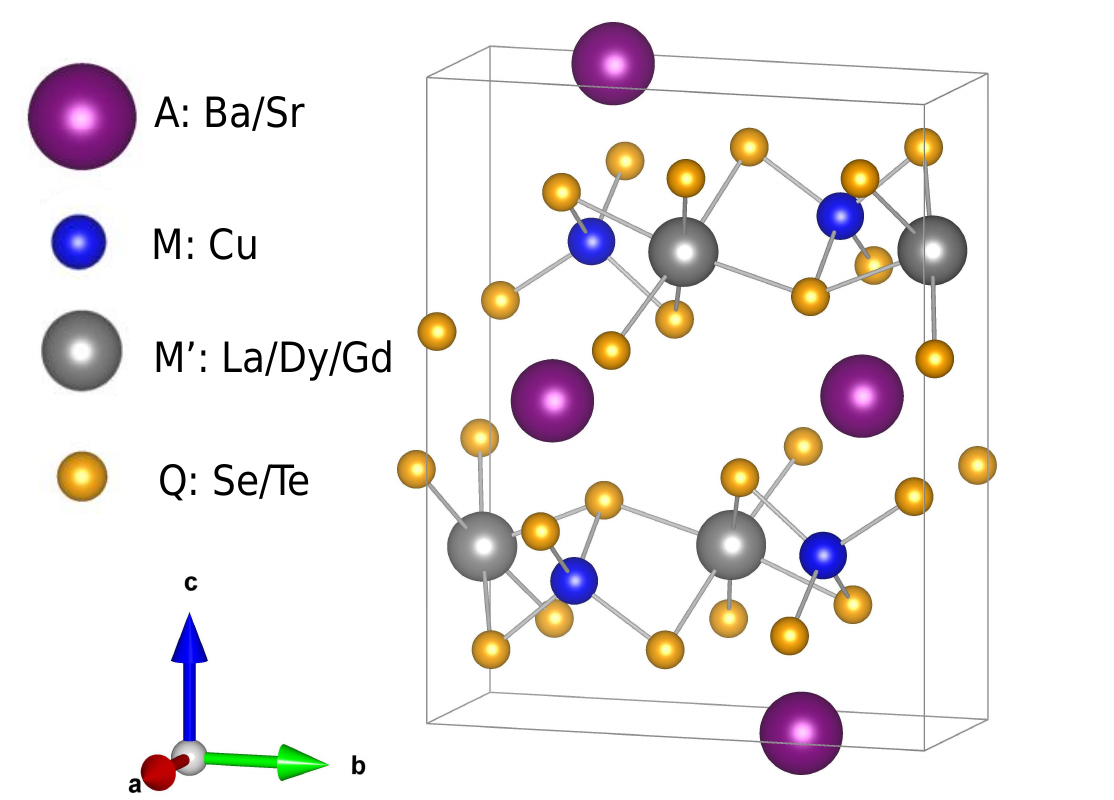}
\caption{Crystal structure of AMM'Q$_3$ quaternary compounds, namely, SrCuGdSe$_3$, SrCuDyTe$_3$, BaCuLaSe$_3$, and BaCuLaTe$_3$. M and M' in this structure possess tetrahedral and octahedral coordination environments, respectively.}\label{figcstr}
\end{figure}

In the next step, we applied the effective carrier mass criterion to the 50 compounds obtained in the previous steps, which removed only 4 compounds having indirect band gaps and reduces the number of potential candidates to 46. Therefore, after the final step of the screening we got 14 AMM'Q$_3$ compounds with direct band gaps and 32 compounds having indirect band gaps. Since most of the known high-performance PV materials have direct band gaps, the direct band gap AMM'Q$_3$ compounds would be our preferred choice for further exploration for potential high-performance PV applications. All these screening steps are summarized in (Fig.~\ref{figfunnel}(c)) in a funnel diagram. A list and electronic structures of these 14 direct band gap compounds are given in Table \ref{table1} and Fig. S4 in the Supplemental Material \cite{supmat}, respectively. The list and electronic structures of the  32 indirect band gap compounds are given in the Supplemental Material (Table SVI and Fig. S5)\cite{supmat}, which may also be interesting for PV applications but have not been explored further in this study. Interestingly, all of these 46 AMM'Q$_3$ compounds belong to Type-II category and crystallize in the orthorhombic Pnma (\# 62) space group in which M and M' atoms form tetrahedral  (MQ$_4$) and octahedral (M'Q$_6$) coordination environments (as shown in Fig. ~\ref{figcstr}), respectively.   Finally, from the list of 14 direct band gap AMM'Q$_3$ compounds, we choose four compounds SrCuGdSe$_3$, SrCuDyTe$_3$, BaCuLaSe$_3$, BaCuLaTe$_3$ for detailed study and analysis. We evaluate their properties using state-of-the-art first-principles calculations that have been employed to study stability, defect,  electronic and optoelectronic properties of several PV materials \cite{wallace2017candidate, frost2014atomistic, chen2009crystal, yin2015halide, eames2015ionic, xiao2017searching, zhu2017i2}.

\begin{table}
\caption{List of 14 direct band gap AMM'Q$_3$ compounds obtained in the final step of our HT screening and their PBE calculated band gap (E$_g^{PBE}$ in eV), , average effective mass of the holes (m$_h^*$ in terms of free electron mass, $m_0$), electrons (m$_e^*$ in terms of $m_0$), exciton binding energy (E$_b$ in meV), hybrid functional calculated band gap with spin-orbit coupling (E$_g^{HSE+SOC}$ in eV), and the calculated spectroscopic limited maximum efficiency (SLME in $\%$). These compounds are predicted to be potential high-performance PVs.}\label{table1}
	\begin{center}
		\begin{tabular}{l@{\hskip 0.05in} c@{\hskip 0.05in} c@{\hskip 0.1in}  c@{\hskip 0.1in} c@{\hskip 0.05in} c@{\hskip 0.05in} c@{\hskip 0.05in}}
		\hline 
		\hline
		Compounds  & E$_g^{PBE}$ &  m$_h^*$ & m$_e^*$ & E$_b$&E$_g^{HSE+SOC}$  & SLME \\
		\hline
		\hline
		BaCuErSe$_3$ & 0.97 &  0.896 & 0.729 & 39 & 1.71 & 28 \\
		SrCuDyTe$_3$ & 0.63 &  0.823 & 0.848 & 25 & 1.01 & 31\\
		SrCuErTe$_3$ & 0.62 &  0.824 & 0.840 & 34 & 1.04 & 31\\ 
		SrCuHoTe$_3$ & 0.60 &  0.821 & 0.854 & 33 & 1.02& 31\\
		SrCuSmTe$_3$ & 0.58 &  0.936 & 0.900 & 35 & 0.98 & 31\\
		SrCuTbTe$_3$ & 0.59 &  0.868 & 0.852 & 34 & 1.00 & 31\\
		SrCuTmS$_3$  & 1.32 &  1.334 & 0.782 & 59 & 2.27 & 17\\
		SrCuTmTe$_3$ & 0.63 &  0.804 & 0.830 & 33 & 1.02 & 31\\
		BaCuGdTe$_3$ & 0.52 &  0.654 & 0.711 & 27 & 0.93& 30\\
		BaCuLaSe$_3$ & 1.20 &  1.277 & 0.797 & 56 & 1.79 & 27\\
		BaCuLaTe$_3$ & 0.81 &  0.878 & 0.788 & 9 & 1.05 & 32\\
		SrCuGdS$_3$  & 1.26 &  1.389 & 0.828 & 58 & 2.18& 19\\
		SrCuGdSe$_3$ & 1.05 &  1.160 & 0.865 & 80 & 1.65 & 30\\
		SrCuGdTe$_3$ & 0.59 &  0.874 & 0.870 & 34 & 0.99& 31\\
		\hline
		\hline
		\end{tabular} 
		\end{center}
\end{table}

\begin{figure*}
	\centering
	\includegraphics[clip, trim=0cm 0cm 0.5cm 0cm, width=1\textwidth]{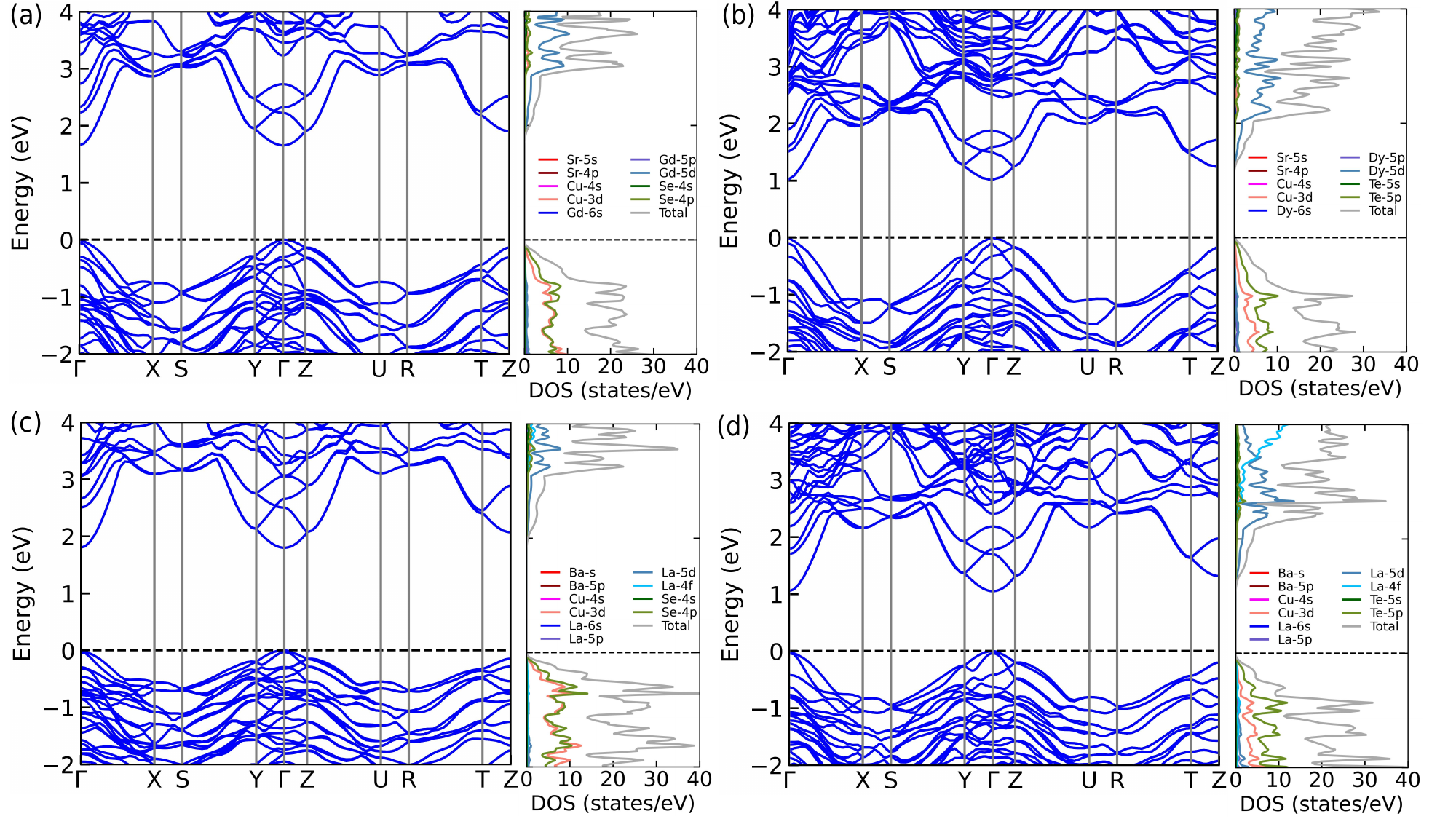}
	\caption{Electronic structures and orbital projected density of states of SrCuGdSe$_3$ (a), SrCuDyTe$_3$ (b), BaCuLaSe$_3$ (c) and BaCuLaTe$_3$ (d)  calculated using the HSE06 functional including spin-orbit coupling.}\label{figest}
\end{figure*}

\subsection{Electronic structure and optical properties}

Electronic structures of the selected four compounds calculated using HSE06 functional including spin-orbit coupling (SOC) are shown in Fig. \ref{figest}, which reveal direct band gaps at $\Gamma$ point and parabolic band dispersions along $\Gamma$-X, $\Gamma$-Y, and $\Gamma$-Z directions in the Brillouin zone. 
The calculated band gaps (HSE06+SOC) of SrCuGdSe$_3$, SrCuDyTe$_3$, BaCuLaSe$_3$, and BaCuLaTe$_3$ are 1.65 eV, 1.01 eV, 1.79 eV and 1.05 eV, respectively, which lie in the ideal band gap range ($\sim$ 1-1.7 eV) required for strong photo-absorber materials in the visible range of the solar spectrum \cite{wallace2017candidate}. 
Since these four compounds possess a centro-symmetric space group \textit{Pnma}, the allowed optical transition is determined by the parity of the VBM and CBM. According to Fermi's golden rule, the optical absorption from the valence band ($|v\rangle$) to the conduction band ($|c\rangle$) is allowed if the transition matrix $\langle v|\hat{p}|c\rangle$ becomes non-zero, where $\hat{p}$ is the momentum operator. This will be true  for material with inversion symmetry if $|v\rangle$ and $|c\rangle$ have opposite parity. 
To analyze the parity of the VBM and CBM for the four compounds, we have analyzed s, p, and d-orbital projected electronic structures (see Fig. S6 in the Supplemental Material\cite{supmat}), and projected density of states (PDOS). We found that for SrCuDyTe$_3$ (Fig. \ref{figest}(a)) and BaCuLaTe$_3$ (Fig. \ref{figest}(d)), Te-5p orbitals mainly contribute to the VBM. The CBM of SrCuDyTe$_3$ and BaCuLaTe$_3$ are primarily contributed by the Dy-4d and La-5d orbitals, respectively. For SrCuGdSe$_3$ (Fig. \ref{figest}(b)) and BaCuLaSe$_3$ (Fig. \ref{figest}(c)), the VBM predominantly consists of Se-4p orbitals and to some extent Cu-3d orbitals. The CBM is largely made up of Gd-5d orbitals in SrCuGdSe$_3$ and La-5d orbitals in BaCuLaSe$_3$. As VBM and CBM are mainly constituted of orbitals with opposite parity (i.e., p and d orbitals, respectively), the optical absorption coefficient of BaCuLaTe$_3$ and SrCuDyTe$_3$ are expected to be higher than that of SrCuGdSe$_3$ and BaCuLaSe$_3$ for which the VBM have mixtures of p and d-orbitals.

The effective mass of the charge carriers is a good indicator of their mobility, which is an important factor in determining device performance. A lower effective mass leads to higher mobility, allowing charge carriers to travel longer distances without dissipation. In line with our screening criteria, charge carriers with effective masses below 1.5m$_0$ are desirable, as they facilitate efficient transport of photo-generated carriers. 
We calculated the average electron and hole effective masses at the CBM and VBM from their HSE06+SOC electronic structures.
Table \ref{table1} provides the average effective masses of the final 14 direct band gap AMM'Q$_3$ compounds that include these four compounds.  
The average effective masses of the electrons for the four compounds are SrCuGdSe$_3$ (0.865m$_o$), SrCuDyTe$_3$ (0.848m$_o$), BaCuLaSe$_3$ (0.797m$_o$), BaCuLaTe$_3$ (0.788m$_o$) that are less than 1m$_o$ at the CBM. On the other hand, the hole effective masses become larger than 1m$_o$ for SrCuGdSe$_3$ (1.160m$_o$) and BaCuLaSe$_3$ (1.277m$_o$), while  it remains less than 1m$_o$ at the VBM for SrCuDyTe$_3$ (0.823m$_o$) and BaCuLaTe$_3$ (0.878m$_o$). 

\begin{figure*}
	\centering
	\includegraphics[clip, trim=0cm 9cm 4.5cm 0cm, width=0.9\textwidth]{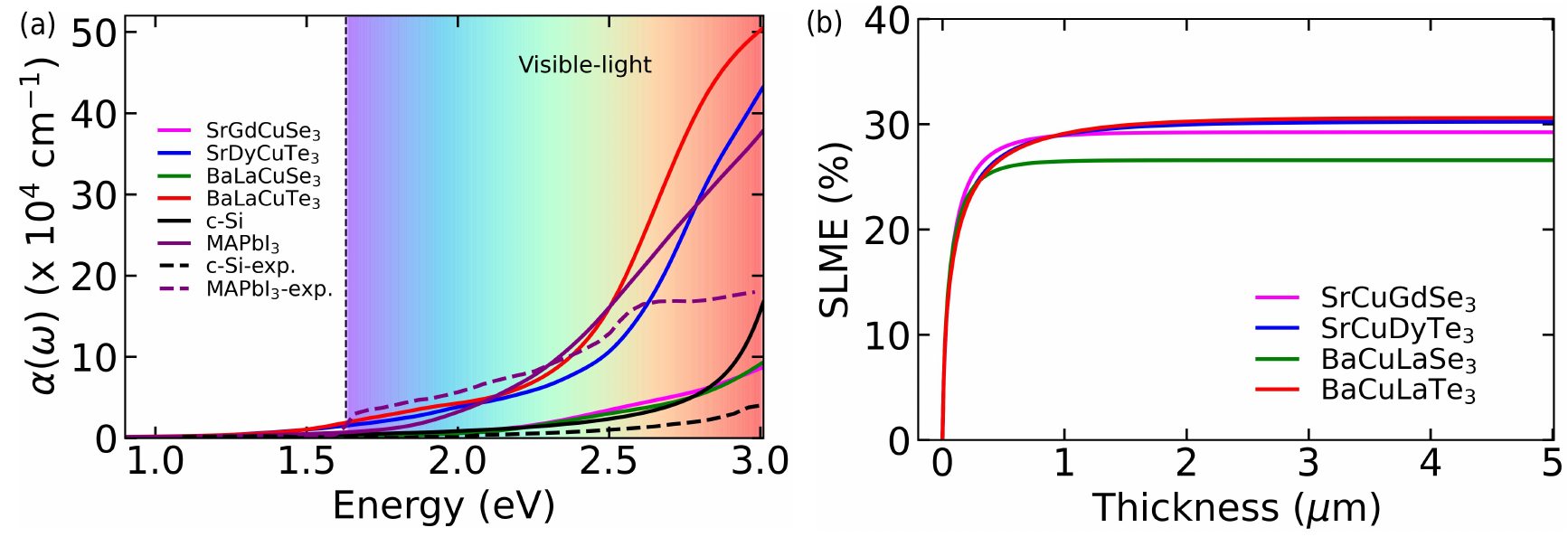}
	\caption{Calculated optical absorption coefficient ($\alpha(\omega)$) for SrCuGdSe$_3$, SrCuDyTe$_3$, BaCuLaSe$_3$, and BaCuLaTe$_3$ compared with $\alpha(\omega)$ of crystalline-Si and  hybrid perovskite MAPbI$_3$ calculated at the same level of theory over the energy range $\sim$ 1–3 eV. Experimentally measured $\alpha(\omega)$ of crystalline-Si and MAPbI$_3$ are also plotted, which are taken from Ref. \cite{de2014organometallic}.}
	\label{figabs}
\end{figure*}

Larger dielectric constants imply greater defect tolerance of a crystal \cite{brandt2015identifying}.  The calculated optical dielectric constants ($\epsilon_\infty$) of SrCuGdSe$_3$ (7.55), SrCuDyTe$_3$ (9.26), BaCuLaSe$_3$ (7.18), and BaCuLaTe$_3$ (8.60) are slightly higher than that of MAPbI$_3$ (5.20) \cite{brivio2013structural} calculated within the same level of theory \cite{wallace2017candidate}. This signifies the enhanced defect tolerance capability due to the presence of the mixed tetrahedral and octahedral coordination environments in their crystal structures, consistent with the theoretical analysis of Wang et al. \cite{wang2019materials}. 
To assess the efficiency of these compounds as potential photo absorbers in the visible range of the solar spectrum, we calculated the absorption coefficient ($\alpha(\omega)$) as a function of the incident photon energy and compared their absorption strength with the calculated as well as experimentally measured $\alpha(\omega)$ of c-Si and MAPbI$_3$ (Fig. \ref{figabs}(a)). 
The calculated $\alpha(\omega)$ for SrCuGdSe$_3$ and BaCuLaSe$_3$ are similar. Although the absorption coefficients of these two compounds ($\approx$ 3.50 $\times$ 10$^4$) at 2.50 eV are much smaller than that of  MAPbI$_3$ (16 $\times$ 10$^4$), they are still higher than crystalline silicon (c-Si). 
On the other hand, $\alpha(\omega)$ of BaCuLaTe$_3$ (16 $\times$ 10$^4$) is much higher than the other three AMM'Q$_3$ compounds, that surpasses the absorption strength of  MAPbI$_3$ above 2.50 eV. For SrCuDyTe$_3$, the absorption coefficient $\alpha(\omega)$ is slightly smaller than that of MAPbI$_3$ for photon energies at 2.50 eV but is larger than that of SrCuGdSe$_3$, BaCuLaSe$_3$, and c-Si. SrDyCuTe$_3$ exhibits a stronger absorption coefficient than that of MAPbI$_3$ above 2.75 eV.

Highly efficient PV materials generally possess low exciton binding energy ($<$ 100 meV), which enables efficient dissociation of photon-induced electron–hole pairs and enhances charge carrier collection efficiency \cite{zhao2017design, d2014excitons}. As shown in Table~\ref{table1}, the low values of E$_b$ (9–80 meV) for all compounds  indicate their potential as PV absorber materials. Among the studied quaternary compounds, SrCuDyTe$_3$ and BaCuLaTe$_3$ exhibit the lowest E$_b$, making them particularly promising candidates for PV applications.

Next, we calculated the SLME \cite{fabini2019candidate} for SrCuGdSe$_3$, SrCuDyTe$_3$, BaCuLaSe$_3$, and BaCuLaTe$_3$ as a function of the thickness of the photo-absorber materials. SLME defines the maximum achievable solar cell efficiency, which depends on the band gap and optical absorption coefficient $\alpha(\omega)$. The thickness-dependent SLME values are shown in Fig. \ref{figabs}(b), which shows that SLME increases sharply with film thickness up to 0.80 $\mu$m, then slowly increases to 1 $\mu$m before becoming constant. 
For thicknesses exceeding 1 $\mu$m, BaCuLaTe$_3$, SrCuDyTe$_3$, and SrCuGdSe$_3$ achieve  efficiencies of 32 \%, 31 \%, and 30 \% respectively. In contrast, BaCuLaSe$_3$ exhibits a relatively lower SLME of 27 \%. The SLME values for BaCuLaTe$_3$, SrCuDyTe$_3$, and SrCuGdSe$_3$ are notably higher than those of other well-known solar cell materials, such as GaAs (27.8 \%) \cite{yu2012identification}, Sb$_2$Se$_3$ (29.9 \%) \cite{yu2012identification}, and Cs$_2$AgBiBr$_6$ (25.88 \%) \cite{kaur2022tuning}. We calculated the SLME values for all 14 direct band gap AMM'Q$_3$ compounds obtained in the final step of our screening, which are provided in Table \ref{table1}. Therefore, high absorption coefficients on the order of 10$^5$ cm$^{-1}$, coupled with suitable band gaps and significantly higher SLME values, confirm that these quaternary chalcogenides are promising candidates for PV applications.

\subsection{Temperature-dependent dynamic structural and electronic properties}

\begin{figure*}
	\centering
	\includegraphics[width=1\textwidth]{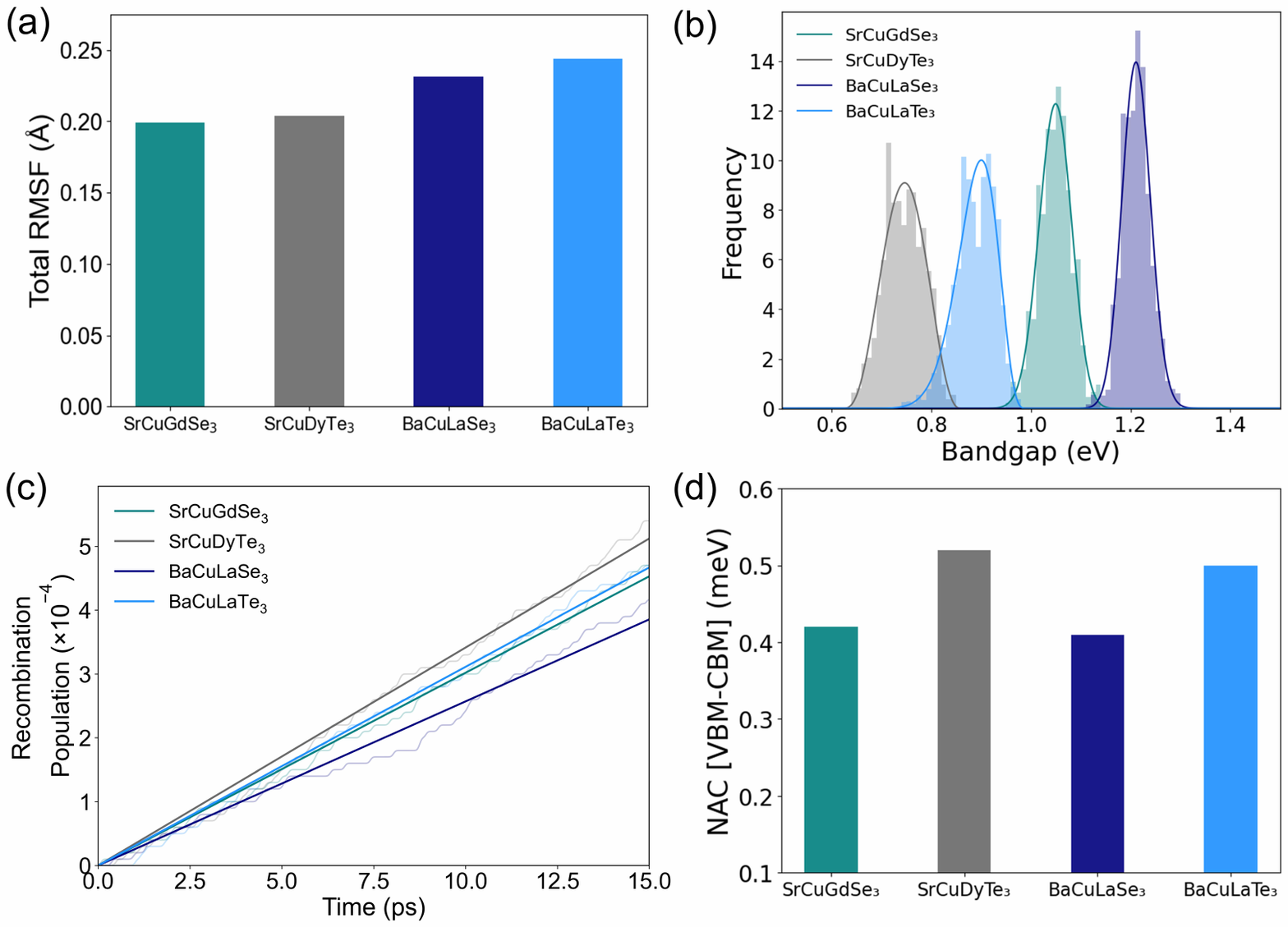}
	\caption{(a) Bar chart for the root mean square fluctuation (RMSF) of the structures of the four quaternary chalcogenides. (b) Histogram plot showing variation of the band gaps along the AIMD trajectories of these compounds for 5$ps$ (5000 data points). (c) The population of nonradiatively recombined electron-hole pairs over time. The function $f(t) = 1 - exp(-t/\tau)$ is used to fit the population rise with time, where $\tau$ is the electron-hole recombination time. (d) Time-averaged nonadiabatic coupling (NAC) constant between the VBM and CBM for all four chalcogenides.}
	\label{fig5u}
\end{figure*}

We  performed AIMD simulations to investigate the structural fluctuations of SrCuGdSe$_3$, SrCuDyTe$_3$,  BaCuLaSe$_3$, and BaCuLaTe$_3$ at 300 K. 
The root-mean-square fluctuation (RMSF), which computes the real-time deviation of a particular atom from its mean position over the trajectories, quantifies the extent of thermal fluctuations. The calculated overall RMSF values, as plotted in Fig. \ref{fig5u}(a), depict that Te-based compounds have higher RMSF than Se-based compounds. The RMSFs for the individual atom types are given in the Supplemental Material (Fig. S7(a))\cite{supmat}, further demonstrating that most of the constituent elements of the tellurides exhibit larger deviation from their mean positions than the atoms in the selenides. Notably, the total RMSF values of these compounds are comparable to those of other well-known optoelectronic materials \cite{Ghosh2022}, indicating their structural stability within the expected thermal fluctuation limits. These lattice fluctuations impact the dynamic electronic properties, such as the energies of the VBM and CBM (see Fig. S8) in semiconductors and excited charge carrier lifetimes to different extents. 
The variation in band gaps due to lattice fluctuations at finite temperatures manifests the strength of electron-phonon interactions in these chalcogenide semiconductors. The distributions of band gap over 5 \textit{ps} (5000 data points) for the four systems are shown in Fig. \ref{fig5u}(b). The fitted distributions reveal that SrCuDyTe$_3$ and BaCuLaTe$_3$ exhibit higher band gap fluctuations than their Se-based counterparts. 
The narrowest distribution of the band gap is observed in BaCuLaSe$_3$ with a low standard deviation (SD = 28.58 meV) under ambient condition, indicating minimal fluctuation in its dynamic band gap, compared to BaCuLaTe$_3$ (SD = 32.30 meV), SrCuGdSe$_3$ (SD = 40.03 meV), and SrCuDyTe$_3$ (SD = 40.21 meV).  Such low fluctuations are smaller than those observed in conventional systems such as hybrid halide perovskites \cite{Nayak2025,Nayak2024}, highlighting the overall electronic stability of these quaternary chalcogenides. This enhanced  stability is particularly advantageous for PV applications, as minimal band gap fluctuations contribute to efficient charge carrier transport and reduced recombination losses. 
We also note that the ensemble-averaged band gaps are 0.75, 1.05, 1.21, and 0.89 eV, respectively, for SrCuDyTe$_3$, SrCuGdSe$_3$, BaCuLaSe$_3$, and, BaCuLaTe$_3$, which follow a trend similar to what has been found from static electronic structure calculations, i.e., Te-based systems have narrower band gaps than Se-based ones.\\

\subsection{Charge carrier dynamics}

Understanding the excited-state charge carrier dynamics in the AMM'Q$_3$ quaternary chalcogenides is a complex task due to the structural complexities of these compounds and the significant impact of local structural fluctuations. 
Computational exploration of electron-phonon interactions and their subsequent effects on charge transport poses a challenge. 
To better understand the complexities of nonradiative carrier recombination processes, we used a combination of NAMD and TDDFT simulations. 
This approach allows for a dynamic investigation of electrons and lattice vibrations interactions, shedding light on critical factors such as carrier mobility, recombination rates, and overall device performance in these unique quaternary chalcogenide compounds. 
The nonradiative recombination of the charge carriers across the band gap is tracked over time by computing the increased carrier population at the ground state (Fig. \ref{fig5u}(c)). 
The faster the increase in population, the faster is the nonradiative carrier relaxation. 
Fig. \ref{fig5u}(c) reveals that SrCuDyTe$_3$ exhibits the fastest nonradiative recombination rate, whereas BaCuLaSe$_3$ has the slowest. By utilizing the short-time linear approximation method to analyze the exponentially growing function\cite{Zhang2018,Ghosh2020}, the carrier lifetimes for  SrCuGdSe$_3$, SrCuDyTe$_3$, BaCuLaSe$_3$, and BaCuLaTe$_3$ are determined to be 33.12 $ns$, 29.28 $ns$,  39.90 $ns$, and 32.13 $ns$, respectively. 
These carrier lifetimes are larger than those of conventional inorganic halide perovskites (e.g., CsPbBr$_3$) and hybrid organic-inorganic layered halide perovskites, which generally lie in a range of a few $ns$ (1-20 $ns$). \cite{li2017pbcl2,he2018halide}  The extended carrier lifetime indicates the potential of these quaternary chalcogenides for optoelectronic applications.
\\
\indent
Fermi's golden rule asserts that the electron-hole recombination rate, which involves two energy states, is inversely proportional to the magnitude of the energy gap between those states. 
The nonradiative recombination rate is also influenced by the square of the nonadiabatic coupling (NAC) strength, which indicates the nonadiabatic transition probability. 
In general, high time-averaged NAC values between the VBM and CBM  indicate significant electron-phonon interactions and, as a result, a rapid non-radiative charge recombination process occurring within the electronic band gap. Moreover, a higher quantity of closely situated energy eigenstates (within an energy range of $\sim$ 50 meV) near the band edges can actively engage in the real-time processes of charge carrier relaxation, hence expediting the nonradiative recombination. We closely investigate these two dominant factors to understand the trend of carrier lifetimes in these quaternary chalcogenides. 
As shown in Fig. \ref{fig5u}(b), the time-averaged band gaps primarily influence the carrier lifetimes of these chalcogenides: the smaller the band gap, the faster the nonradiative carrier recombination process that agrees with the calculated ensemble averaged band gaps of these compounds. We further explore the NAC in these chalcogenides to quantify the coupling strength between electronic and phononic subsystems. 
The photoexcited electrons at the CBM couple with phonon modes to dissipate excess energy and subsequently relax to the ground state, signifying stronger electron-phonon coupling (i.e., higher NAC values) boosts the energy dissipation of photoexcited electrons that accelerates the nonradiative carrier relaxation. 
The time-averaged NACs between the VBM and CBM in Fig. \ref{fig5u}(d) illustrate that tellurides have higher electron-phonon coupling strength than the selenides. 
Fundamentally, the higher extent of structural fluctuations and relatively smaller energy gaps for BaCuLaTe$_3$ and SrCuDyTe$_3$ give rise to higher NAC values and, hence, higher nonradiative recombination rates of the charge carriers. These detailed analyses provide an intricate relationship between the structural dynamic and charge carrier recombination processes in these AMM'Q$_3$ chalcogenides.

\subsection{Influence spectra}  

We assess the influence spectra for the four quaternary chalcogenides to understand the effect of the electron-phonon interactions better. These spectra are obtained by performing Fourier transformation on the autocorrelation functions of the fluctuations in the electronic energy gaps generated by phonons. The influence spectra are used to describe the phonon modes that are coupled to the electronic subsystem. The identification of phonon modes that exhibit active coupling with the electronic structures and contribute to nonradiative electron-hole recombination can be achieved by analyzing the peak positions observed in these spectra. 
For the studied quaternary chalcogenides, all major peaks in the influence spectra emerge below 300 cm$^{-1}$ (Fig. S7(b)), signifying the active phonon modes that participate in the nonradiative carrier relaxation.  From this analysis, it is evident that tellurides have sharp and higher peaks up to 200 cm$^{-1}$ in comparison to selenides, signifying the presence of selective vibrational modes with stronger electron-phonon coupling that strongly contribute to nonradiative charge carrier relaxation processes. 
Thus, enhanced electron-phonon coupling in SrCuDyTe$_3$ and BaCuLaTe$_3$ results in higher NAC values, leading to faster nonradiative recombination compared to SrCuGdSe$_3$ and BaCuLaSe$_3$ systems.

\subsection{Dephasing process}

The nonadiabatic coupling between the two energy states in electronic structures in real-time facilitates the occurrence of quantum mechanical transitions through the building up of wavefunction superposition. Elastic electron-phonon interactions hinder quantum coherence and disrupt the superposition of electronic states. Consequently, the wavefunction collapses to a single state, limiting the corresponding transition probability. Therefore, an accelerated decoherence process between the band edges would increase electron-hole recombination time. Quantum decoherence in solids generally occurs within a timeframe of tens of femtoseconds ($fs$), significantly faster than the recombination times of the electron-hole pairs. This suggests that it is imperative to consider this phenomenon while conducting simulations of carrier relaxation, as depicted in Fig. \ref{fig5u}(c). The decoherence time can be determined using linear response formalism, treating it as the pure dephasing time \cite{hamm2005principles}. The calculated decoherence times for four quaternary chalcogenides are SrCuGdSe$_3$ (12.35 $fs$), SrCuDyTe$_3$ (10.93 $fs$),  BaCuLaSe$_3$ (11.01 $fs$), and BaCuLaTe$_3$ (12.47 $fs$).  Notably, the decoherence lifetimes are similar across all these materials, indicating the limited influence of elastic electron-phonon interactions on the trend of the charge carrier relaxation process occurring between the VBM and CBM.

Upon examining the various influencing factors, we conclude that the nonradiative carrier recombination rate has a complex relationship with electronic and vibrational properties in the AMM'Q$_3$ chalcogenides. 
The primary contributors to the recombination process are the differences in time-averaged band gaps (as shown in Fig. \ref{fig5u}(b)), enhanced electron-phonon coupling and the NAC values between the band edge states (illustrated in Fig. \ref{fig5u}(d)).


\subsection{Defect tolerance}  

Defects play a very important role in the performance of PV devices. Therefore, it is imperative to assess the formation of defects and examine their effects on material properties. When defects induce extra states that reside deep inside the band gap, they are called deep-level defects. These deep defect states often act as traps for the photo-generated charge carriers, which accelerate the nonradiative charge carrier recombination, reducing carrier lifetime and charge separation efficiency, and hence a reduced photoconversion efficiency. 
On the other hand, when defect-induced states lie close to the band edges, they form shallow-level states, which are often non-detrimental to the performance and give rise to defect tolerance in a material. For example, MAPbI$_3$  which is a well-known high-performance PV material, has been shown to host iodine vacancy (V$_I$), lead vacancy (V$_{Pb}$), iodine interstitial (I$_i$) and MA interstitial (MA$_i$) with low formation energies \cite{chu2020low, shan2017segregation}. These point defects form shallow defect levels that do not accelerate the nonradiative charge-carrier recombination processes \cite{chu2020low}.  To this end, we assess the defect tolerance of the four candidates (SrCuGdSe$_3$, SrCuDyTe$_3$,  BaCuLaSe$_3$, and BaCuLaTe$_3$) against various point defects.  To uncover the key trend of point-defect induced extra states, we considered atomic vacancies at the A, M, M', and  Q sites (i.e., $V_A$, $V_M$, $V_M$, $V_M'$, $V_{Q'}$) and interstitials ($A_i$, $M_i$, $M'_i$, $Q_i$). Due to the relatively large and complex crystal structures of these quaternary compounds and to keep the defect calculations and analysis computationally tractable, we considered all defects in neutral charge states.  We calculated the defect formation energies for all of the above point defects (Table \ref{table2}) using  the PBE exchange-correlation functional. 
Our analysis reveals that vacancy formation at the M site (i.e., $V_{Cu}$) has the lowest energy in SrCuGdSe$_3$ and SrCuDyTe$_3$. In contrast, the defect formation energies for the Q and A interstitials (i.e., $Q_i$ and $A_i$) are the lowest in BaCuLaSe$_3$ and BaCuLaTe$_3$, respectively. We calculated  electronic structures and DOS in these compounds containing Cu-vacancy (see Fig. S9 in the Supplemental Material \cite{supmat}) and interstitial defects (see Fig. S10 in the Supplemental Material \cite{supmat}), which reveal the absence of any mid-gap states, signifying the defect tolerance of these AMM'Q$_3$ chalcogenides.

\begin{table}
\caption{Vacancy ($V$) and interstitial ($i$) defect formation energies (in eV/defect) at the $A$, $M$, $M'$ and $Q$ sites of SrCuGdSe$_3$, SrCuDyTe$_3$, BaCuLaSe$_3$, and BaCuLaTe$_3$.} \label{table2}
	\begin{center}
		\begin{tabular}{l @{\hskip 0.1in}c @{\hskip 0.1in}c@{\hskip 0.1in}  c@{\hskip 0.1in} c@{\hskip 0.1in}}
		\hline 
		\hline
		  & SrCuGdSe$_3$ & SrCuDyTe$_3$  & BaCuLaSe$_3$ & BaCuLaTe$_3$ \\
		\hline
		\hline
		V$_A$ & 4.842  & 3.852 & 5.413 & 4.733  \\
		V$_M$ & 0.841  & 0.525 & 1.027 & 0.799 \\
		V$_M'$ & 6.535 & 4.994 & 7.285 & 5.971 \\ 
		V$_Q$ & 3.242  & 2.880 & 3.761 & 3.242 \\
        A$_i$ & 1.240  & 1.008 & 1.173 & 0.614  \\
		M$_i$ & 1.403  & 1.051 & 1.434 & 1.030 \\
		M'$_i$ & 2.152 & 1.957 & 1.856 & 1.405 \\ 
		Q$_i$ & 1.158  & 1.467 & 0.513 & 0.792 \\
		\hline
		\hline
		\end{tabular} 
		\end{center}
\end{table} 

Furthermore, it has been shown that semiconductors that possess an antibonding valence state below the Fermi level exhibit defect-tolerant properties \cite{zakutayev2014defect}. 
To understand the chemical bonding in these quaternary chalcogenides, we performed crystal orbital Hamilton population analysis (COHP), which reveals the presence of large anti-bonding states below the Fermi level for all four compounds (see Fig. S11 in the Supplemental Material \cite{supmat}). 
These analyses further affirm the defect tolerance of those quaternary chalcogenides. 

\section{Conclusions}

Data-driven computational methods offer a powerful and efficient strategy to systematically screen a large number of materials for desired properties and applications. In this work, we utilized HT calculations based on DFT to systematically screen potential high-performance PV materials in a family of experimentally known quaternary chalcogenides denoted by AMM'Q$_3$. Guided by rational design criteria obtained from studying the already known high-performance PV materials, we predicted 14 direct band gap and 32 indirect band gap AMM'Q$_3$ compounds to be potentially high-performance PVs.  Among the 14 direct band gap candidates, we chose four compounds,  SrCuGdSe$_3$, SrCuDyTe$_3$, BaCuLaSe$_3$ and BaCuLaTe$_3$ for detailed investigations of their electronic structures, absorptivity, defect tolerance and photo-conversion efficiencies.  Furthermore, we employed AIMD, TDDFT, and  NAMD simulations to unravel the charge carrier dynamics and the effect of temperature on the structural stability and electronic band energies. 
These four compounds exhibit excellent PCE's exceeding 20\%.
Interestingly, our detailed electronic structure analysis of the AMM'Q$_3$ compounds reveals that all 46 compounds obtained at the final step of our screening belong to the Type-II (A$^{2+}$M$^{1+}$M'$^{3+}$Q$_3^{2-}$) category having Pnma space group and contain 24 atoms in their primitive unit cells. 
Type-I (A$^{1+}$M$^{1+}$M'$^{4+}$Q$_3^{2-}$) and Type-III (A$^{1+}$M$^{2+}$M'$^{3+}$Q$_3^{2-}$) compounds generally do not have parabolic bands around either their VBM or CBM, and hence, do not qualify as potential high-performance PV materials.

The crystal structures of the 14 direct band gap AMM'Q$_3$ compounds that include SrCuGdSe$_3$, SrCuDyTe$_3$, BaCuLaSe$_3$ and BaCuLaTe$_3$ possess mixed tetrahedral and  octahedral coordination environments, which  are expected to exhibit  properties complementary to that of tetrahedrally 
(e.g., c-Si, CdTe) and octahedrally (e.g., MAPbI$_3$) bonded conventional high-performance PV materials. 
While the calculated $\alpha(\omega)$ of SrCuDyTe$_3$ and BaCuLaTe$_3$ are stronger than that of MAPbI$_3$ at high photon energy, $\alpha(\omega)$ of BaCuLaTe$_3$ is much higher than those of SrCuGdSe$_3$, SrCuDyTe$_3$, and BaCuLaSe$_3$. For these four quaternary compounds, $\alpha(\omega)$ surpasses the absorptivity of c-Si. Calculated exciton binding energies (E$_b$) for BaCuLaTe$_3$ and SrCuDyTe$_3$ exhibit the lowest values, making them particularly promising candidates for high-performance PV applications.
Analysis of the chemical bonding reveals that the valence band of these AMM'Q$_3$ compounds have antibonding states just below the Fermi level, which indicates defect tolerance. 
To assess the formation of defects and examine their effects on the properties, we calculated various point defects, namely vacancies and interstitials. Our calculations reveal that the formation of Cu vacancy and Q$_i$/A$_i$ interstitial defects are likely to occur more frequently in these compounds. 
We separately calculated the electronic structures and DOS of those four compounds containing Cu vacancy and various interstitial defects, which do not show the presence of any extra midgap states. 
These calculations signify that the formation of these defects would not deteriorate the radiative recombination rates of the photo-generated carriers, and hence, high photo-conversion efficiency would be maintained. 

AIMD calculations reveal that these compounds possess relatively low root mean squared fluctuations of the structural moieties, which signify the stability of these compounds. 
Our analysis reveals that structural fluctuation for the selenides is lower than that of the tellurides. NAMD simulations reveal low nonadiabatic coupling between the electronic and ionic degrees of freedom, consistent with the low RMSF values. Smaller values of NAC  lead to weak electron-phonon coupling, reducing the pathways for nonradiative carrier recombination in these compounds. We further determined the nonradiative recombination rates of photo-generated carriers between the VBM and CBM, revealing relatively large nonradiative lifetimes ($\sim$ 30-40 ns). 
These large lifetimes allow efficient separation of the electron and holes and hence, higher photo-conversion efficiency. 
Therefore, our work predicts a new family of high-performance photo-absorber materials that would encourage further experimental and theoretical investigations to pursue high-efficiency PV technology.

\section{Acknowledgments}
K.P. acknowledges financial support from an  Initiation Grant from Indian Institute of Technology Kanpur.   M. U. thanks Indian Institute of Technology Kanpur for an Institute Postdoctoral Fellowship. We acknowledge computational resources provided by the (a) HPC2013 and Param Sanganak computing facilities provided by IIT Kanpur. 
D.G. acknowledges the IIT Delhi SEED Grant (PLN12/04MS), the Science and Engineering Research Board (SERB), Department of Science and Technology (DST), India for Start-up Research Grant SRG/2022/00l234, CSIR-Human Resource Development Group (HRDG) for ExtraMural Research-II Grant 01/3136/23/EMR-II and the IIT Delhi HPC facility for computational resources. 
This work was performed, in part, at the Center for Integrated Nanotechnologies, an Office of Science User Facility operated for the U.S. Department of Energy (DOE) Office of Science by Los Alamos National Laboratory (Contract 89233218CNA000001) and Sandia National Laboratories (Contract DE-NA-0003525).

\bibliography{prx-ref}

\begin{thebibliography}{102}%
\makeatletter
\providecommand \@ifxundefined [1]{%
 \@ifx{#1\undefined}
}%
\providecommand \@ifnum [1]{%
 \ifnum #1\expandafter \@firstoftwo
 \else \expandafter \@secondoftwo
 \fi
}%
\providecommand \@ifx [1]{%
 \ifx #1\expandafter \@firstoftwo
 \else \expandafter \@secondoftwo
 \fi
}%
\providecommand \natexlab [1]{#1}%
\providecommand \enquote  [1]{``#1''}%
\providecommand \bibnamefont  [1]{#1}%
\providecommand \bibfnamefont [1]{#1}%
\providecommand \citenamefont [1]{#1}%
\providecommand \href@noop [0]{\@secondoftwo}%
\providecommand \href [0]{\begingroup \@sanitize@url \@href}%
\providecommand \@href[1]{\@@startlink{#1}\@@href}%
\providecommand \@@href[1]{\endgroup#1\@@endlink}%
\providecommand \@sanitize@url [0]{\catcode `\\12\catcode `\$12\catcode
  `\&12\catcode `\#12\catcode `\^12\catcode `\_12\catcode `\%12\relax}%
\providecommand \@@startlink[1]{}%
\providecommand \@@endlink[0]{}%
\providecommand \url  [0]{\begingroup\@sanitize@url \@url }%
\providecommand \@url [1]{\endgroup\@href {#1}{\urlprefix }}%
\providecommand \urlprefix  [0]{URL }%
\providecommand \Eprint [0]{\href }%
\providecommand \doibase [0]{https://doi.org/}%
\providecommand \selectlanguage [0]{\@gobble}%
\providecommand \bibinfo  [0]{\@secondoftwo}%
\providecommand \bibfield  [0]{\@secondoftwo}%
\providecommand \translation [1]{[#1]}%
\providecommand \BibitemOpen [0]{}%
\providecommand \bibitemStop [0]{}%
\providecommand \bibitemNoStop [0]{.\EOS\space}%
\providecommand \EOS [0]{\spacefactor3000\relax}%
\providecommand \BibitemShut  [1]{\csname bibitem#1\endcsname}%
\let\auto@bib@innerbib\@empty
\bibitem [{\citenamefont {Green}\ and\ \citenamefont
  {Bremner}(2017)}]{green2017energy}%
  \BibitemOpen
  \bibfield  {author} {\bibinfo {author} {\bibfnamefont {M.~A.}\ \bibnamefont
  {Green}}\ and\ \bibinfo {author} {\bibfnamefont {S.~P.}\ \bibnamefont
  {Bremner}},\ }\bibfield  {title} {\bibinfo {title} {Energy conversion
  approaches and materials for high-efficiency photovoltaics},\ }\href
  {https://doi.org/10.1038/nmat4676} {\bibfield  {journal} {\bibinfo  {journal}
  {Nature materials}\ }\textbf {\bibinfo {volume} {16}},\ \bibinfo {pages} {23}
  (\bibinfo {year} {2017})}\BibitemShut {NoStop}%
\bibitem [{\citenamefont {Nayak}\ \emph {et~al.}(2019)\citenamefont {Nayak},
  \citenamefont {Mahesh}, \citenamefont {Snaith},\ and\ \citenamefont
  {Cahen}}]{nayak2019photovoltaic}%
  \BibitemOpen
  \bibfield  {author} {\bibinfo {author} {\bibfnamefont {P.~K.}\ \bibnamefont
  {Nayak}}, \bibinfo {author} {\bibfnamefont {S.}~\bibnamefont {Mahesh}},
  \bibinfo {author} {\bibfnamefont {H.~J.}\ \bibnamefont {Snaith}},\ and\
  \bibinfo {author} {\bibfnamefont {D.}~\bibnamefont {Cahen}},\ }\bibfield
  {title} {\bibinfo {title} {Photovoltaic solar cell technologies: analysing
  the state of the art},\ }\href {https://doi.org/10.1038/s41578-019-0097-0}
  {\bibfield  {journal} {\bibinfo  {journal} {Nature Reviews Materials}\
  }\textbf {\bibinfo {volume} {4}},\ \bibinfo {pages} {269} (\bibinfo {year}
  {2019})}\BibitemShut {NoStop}%
\bibitem [{\citenamefont {Kojima}\ \emph {et~al.}(2009)\citenamefont {Kojima},
  \citenamefont {Teshima}, \citenamefont {Shirai},\ and\ \citenamefont
  {Miyasaka}}]{kojima2009organometal}%
  \BibitemOpen
  \bibfield  {author} {\bibinfo {author} {\bibfnamefont {A.}~\bibnamefont
  {Kojima}}, \bibinfo {author} {\bibfnamefont {K.}~\bibnamefont {Teshima}},
  \bibinfo {author} {\bibfnamefont {Y.}~\bibnamefont {Shirai}},\ and\ \bibinfo
  {author} {\bibfnamefont {T.}~\bibnamefont {Miyasaka}},\ }\bibfield  {title}
  {\bibinfo {title} {Organometal halide perovskites as visible-light
  sensitizers for photovoltaic cells},\ }\href
  {https://pubs.acs.org/doi/10.1021/ja809598r} {\bibfield  {journal} {\bibinfo
  {journal} {Journal of the American Chemical Society}\ }\textbf {\bibinfo
  {volume} {131}},\ \bibinfo {pages} {6050} (\bibinfo {year}
  {2009})}\BibitemShut {NoStop}%
\bibitem [{\citenamefont {Xing}\ \emph {et~al.}(2014)\citenamefont {Xing},
  \citenamefont {Mathews}, \citenamefont {Lim}, \citenamefont {Yantara},
  \citenamefont {Liu}, \citenamefont {Sabba}, \citenamefont {Gr{\"a}tzel},
  \citenamefont {Mhaisalkar},\ and\ \citenamefont {Sum}}]{xing2014low}%
  \BibitemOpen
  \bibfield  {author} {\bibinfo {author} {\bibfnamefont {G.}~\bibnamefont
  {Xing}}, \bibinfo {author} {\bibfnamefont {N.}~\bibnamefont {Mathews}},
  \bibinfo {author} {\bibfnamefont {S.~S.}\ \bibnamefont {Lim}}, \bibinfo
  {author} {\bibfnamefont {N.}~\bibnamefont {Yantara}}, \bibinfo {author}
  {\bibfnamefont {X.}~\bibnamefont {Liu}}, \bibinfo {author} {\bibfnamefont
  {D.}~\bibnamefont {Sabba}}, \bibinfo {author} {\bibfnamefont
  {M.}~\bibnamefont {Gr{\"a}tzel}}, \bibinfo {author} {\bibfnamefont
  {S.}~\bibnamefont {Mhaisalkar}},\ and\ \bibinfo {author} {\bibfnamefont
  {T.~C.}\ \bibnamefont {Sum}},\ }\bibfield  {title} {\bibinfo {title}
  {Low-temperature solution-processed wavelength-tunable perovskites for
  lasing},\ }\href {https://doi.org/10.1038/nmat3911} {\bibfield  {journal}
  {\bibinfo  {journal} {Nature materials}\ }\textbf {\bibinfo {volume} {13}},\
  \bibinfo {pages} {476} (\bibinfo {year} {2014})}\BibitemShut {NoStop}%
\bibitem [{\citenamefont {Yin}\ \emph {et~al.}(2015{\natexlab{a}})\citenamefont
  {Yin}, \citenamefont {Yang}, \citenamefont {Kang}, \citenamefont {Yan},\ and\
  \citenamefont {Wei}}]{yin2015halide}%
  \BibitemOpen
  \bibfield  {author} {\bibinfo {author} {\bibfnamefont {W.-J.}\ \bibnamefont
  {Yin}}, \bibinfo {author} {\bibfnamefont {J.-H.}\ \bibnamefont {Yang}},
  \bibinfo {author} {\bibfnamefont {J.}~\bibnamefont {Kang}}, \bibinfo {author}
  {\bibfnamefont {Y.}~\bibnamefont {Yan}},\ and\ \bibinfo {author}
  {\bibfnamefont {S.-H.}\ \bibnamefont {Wei}},\ }\bibfield  {title} {\bibinfo
  {title} {Halide perovskite materials for solar cells: a theoretical review},\
  }\href {https://doi.org/10.1039/C4TA05033A} {\bibfield  {journal} {\bibinfo
  {journal} {Journal of Materials Chemistry A}\ }\textbf {\bibinfo {volume}
  {3}},\ \bibinfo {pages} {8926} (\bibinfo {year}
  {2015}{\natexlab{a}})}\BibitemShut {NoStop}%
\bibitem [{\citenamefont {Yin}\ \emph {et~al.}(2014)\citenamefont {Yin},
  \citenamefont {Shi},\ and\ \citenamefont {Yan}}]{yin2014unusual}%
  \BibitemOpen
  \bibfield  {author} {\bibinfo {author} {\bibfnamefont {W.-J.}\ \bibnamefont
  {Yin}}, \bibinfo {author} {\bibfnamefont {T.}~\bibnamefont {Shi}},\ and\
  \bibinfo {author} {\bibfnamefont {Y.}~\bibnamefont {Yan}},\ }\bibfield
  {title} {\bibinfo {title} {Unusual defect physics in
  {CH}$_3${NH}$_3${P}b{I}$_3$ perovskite solar cell absorber},\ }\href
  {https://pubs.aip.org/aip/apl/article/104/6/063903/237158/Unusual-defect-physics-in-CH3NH3PbI3-perovskite}
  {\bibfield  {journal} {\bibinfo  {journal} {Applied Physics Letters}\
  }\textbf {\bibinfo {volume} {104}},\ \bibinfo {pages} {063903} (\bibinfo
  {year} {2014})}\BibitemShut {NoStop}%
\bibitem [{\citenamefont {Yin}\ \emph {et~al.}(2015{\natexlab{b}})\citenamefont
  {Yin}, \citenamefont {Chen}, \citenamefont {Shi}, \citenamefont {Wei},\ and\
  \citenamefont {Yan}}]{yin2015origin}%
  \BibitemOpen
  \bibfield  {author} {\bibinfo {author} {\bibfnamefont {W.-J.}\ \bibnamefont
  {Yin}}, \bibinfo {author} {\bibfnamefont {H.}~\bibnamefont {Chen}}, \bibinfo
  {author} {\bibfnamefont {T.}~\bibnamefont {Shi}}, \bibinfo {author}
  {\bibfnamefont {S.-H.}\ \bibnamefont {Wei}},\ and\ \bibinfo {author}
  {\bibfnamefont {Y.}~\bibnamefont {Yan}},\ }\bibfield  {title} {\bibinfo
  {title} {Origin of high electronic quality in structurally disordered
  {CH}$_3${NH}$_3${P}b{I}$_3$ and the passivation effect of {C}l and {O} at
  grain boundaries},\ }\href
  {https://advanced.onlinelibrary.wiley.com/doi/full/10.1002/aelm.201500044}
  {\bibfield  {journal} {\bibinfo  {journal} {Advanced Electronic Materials}\
  }\textbf {\bibinfo {volume} {1}},\ \bibinfo {pages} {1500044} (\bibinfo
  {year} {2015}{\natexlab{b}})}\BibitemShut {NoStop}%
\bibitem [{\citenamefont {Lee}\ \emph {et~al.}(2012)\citenamefont {Lee},
  \citenamefont {Teuscher}, \citenamefont {Miyasaka}, \citenamefont
  {Murakami},\ and\ \citenamefont {Snaith}}]{Lee2012}%
  \BibitemOpen
  \bibfield  {author} {\bibinfo {author} {\bibfnamefont {M.~M.}\ \bibnamefont
  {Lee}}, \bibinfo {author} {\bibfnamefont {J.}~\bibnamefont {Teuscher}},
  \bibinfo {author} {\bibfnamefont {T.}~\bibnamefont {Miyasaka}}, \bibinfo
  {author} {\bibfnamefont {T.~N.}\ \bibnamefont {Murakami}},\ and\ \bibinfo
  {author} {\bibfnamefont {H.~J.}\ \bibnamefont {Snaith}},\ }\bibfield  {title}
  {\bibinfo {title} {Efficient hybrid solar cells based on meso-superstructured
  organometal halide perovskites},\ }\href
  {https://doi.org/10.1126/science.1228604} {\bibfield  {journal} {\bibinfo
  {journal} {Science}\ }\textbf {\bibinfo {volume} {338}},\ \bibinfo {pages}
  {643} (\bibinfo {year} {2012})}\BibitemShut {NoStop}%
\bibitem [{\citenamefont {Stranks}\ \emph {et~al.}(2013)\citenamefont
  {Stranks}, \citenamefont {Eperon}, \citenamefont {Grancini}, \citenamefont
  {Menelaou}, \citenamefont {Alcocer}, \citenamefont {Leijtens}, \citenamefont
  {Herz}, \citenamefont {Petrozza},\ and\ \citenamefont
  {Snaith}}]{stranks2013electron}%
  \BibitemOpen
  \bibfield  {author} {\bibinfo {author} {\bibfnamefont {S.~D.}\ \bibnamefont
  {Stranks}}, \bibinfo {author} {\bibfnamefont {G.~E.}\ \bibnamefont {Eperon}},
  \bibinfo {author} {\bibfnamefont {G.}~\bibnamefont {Grancini}}, \bibinfo
  {author} {\bibfnamefont {C.}~\bibnamefont {Menelaou}}, \bibinfo {author}
  {\bibfnamefont {M.~J.}\ \bibnamefont {Alcocer}}, \bibinfo {author}
  {\bibfnamefont {T.}~\bibnamefont {Leijtens}}, \bibinfo {author}
  {\bibfnamefont {L.~M.}\ \bibnamefont {Herz}}, \bibinfo {author}
  {\bibfnamefont {A.}~\bibnamefont {Petrozza}},\ and\ \bibinfo {author}
  {\bibfnamefont {H.~J.}\ \bibnamefont {Snaith}},\ }\bibfield  {title}
  {\bibinfo {title} {Electron-hole diffusion lengths exceeding 1 micrometer in
  an organometal trihalide perovskite absorber},\ }\href
  {https://www.science.org/doi/10.1126/science.1243982} {\bibfield  {journal}
  {\bibinfo  {journal} {Science}\ }\textbf {\bibinfo {volume} {342}},\ \bibinfo
  {pages} {341} (\bibinfo {year} {2013})}\BibitemShut {NoStop}%
\bibitem [{\citenamefont {Yang}\ \emph {et~al.}(2023)\citenamefont {Yang},
  \citenamefont {Wang}, \citenamefont {Cai},\ and\ \citenamefont
  {Zang}}]{yang2023mixed}%
  \BibitemOpen
  \bibfield  {author} {\bibinfo {author} {\bibfnamefont {M.}~\bibnamefont
  {Yang}}, \bibinfo {author} {\bibfnamefont {H.}~\bibnamefont {Wang}}, \bibinfo
  {author} {\bibfnamefont {W.}~\bibnamefont {Cai}},\ and\ \bibinfo {author}
  {\bibfnamefont {Z.}~\bibnamefont {Zang}},\ }\bibfield  {title} {\bibinfo
  {title} {Mixed-halide inorganic perovskite solar cells: Opportunities and
  challenges},\ }\href {https://doi.org/10.1002/adom.202301052} {\bibfield
  {journal} {\bibinfo  {journal} {Advanced Optical Materials}\ }\textbf
  {\bibinfo {volume} {11}},\ \bibinfo {pages} {2301052} (\bibinfo {year}
  {2023})}\BibitemShut {NoStop}%
\bibitem [{\citenamefont {Deng}\ \emph {et~al.}(2013)\citenamefont {Deng},
  \citenamefont {Wei}, \citenamefont {Li}, \citenamefont {Li},\ and\
  \citenamefont {Walsh}}]{deng2013electronic}%
  \BibitemOpen
  \bibfield  {author} {\bibinfo {author} {\bibfnamefont {H.-X.}\ \bibnamefont
  {Deng}}, \bibinfo {author} {\bibfnamefont {S.-H.}\ \bibnamefont {Wei}},
  \bibinfo {author} {\bibfnamefont {S.-S.}\ \bibnamefont {Li}}, \bibinfo
  {author} {\bibfnamefont {J.}~\bibnamefont {Li}},\ and\ \bibinfo {author}
  {\bibfnamefont {A.}~\bibnamefont {Walsh}},\ }\bibfield  {title} {\bibinfo
  {title} {Electronic origin of the conductivity imbalance between covalent and
  ionic amorphous semiconductors},\ }\href
  {https://journals.aps.org/prb/abstract/10.1103/PhysRevB.87.125203} {\bibfield
   {journal} {\bibinfo  {journal} {Physical Review B}\ }\textbf {\bibinfo
  {volume} {87}},\ \bibinfo {pages} {125203} (\bibinfo {year}
  {2013})}\BibitemShut {NoStop}%
\bibitem [{\citenamefont {Singh}\ \emph {et~al.}(2025)\citenamefont {Singh},
  \citenamefont {Samanta}, \citenamefont {Maharana}, \citenamefont {Pal},
  \citenamefont {Tretiak}, \citenamefont {Talapatra},\ and\ \citenamefont
  {Ghosh}}]{Singh2025}%
  \BibitemOpen
  \bibfield  {author} {\bibinfo {author} {\bibfnamefont {N.}~\bibnamefont
  {Singh}}, \bibinfo {author} {\bibfnamefont {K.}~\bibnamefont {Samanta}},
  \bibinfo {author} {\bibfnamefont {S.~K.}\ \bibnamefont {Maharana}}, \bibinfo
  {author} {\bibfnamefont {K.}~\bibnamefont {Pal}}, \bibinfo {author}
  {\bibfnamefont {S.}~\bibnamefont {Tretiak}}, \bibinfo {author} {\bibfnamefont
  {A.}~\bibnamefont {Talapatra}},\ and\ \bibinfo {author} {\bibfnamefont
  {D.}~\bibnamefont {Ghosh}},\ }\bibfield  {title} {\bibinfo {title}
  {High-throughput and data-driven search for stable optoelectronic amse$_3$
  materials},\ }\href {https://doi.org/10.1039/D4TA08867K} {\bibfield
  {journal} {\bibinfo  {journal} {J. Mater. Chem. A}\ }\textbf {\bibinfo
  {volume} {13}},\ \bibinfo {pages} {9192} (\bibinfo {year}
  {2025})}\BibitemShut {NoStop}%
\bibitem [{\citenamefont {Luo}\ \emph {et~al.}(2021)\citenamefont {Luo},
  \citenamefont {Li}, \citenamefont {Wang}, \citenamefont {Faizan},\ and\
  \citenamefont {Zhang}}]{luo2021high}%
  \BibitemOpen
  \bibfield  {author} {\bibinfo {author} {\bibfnamefont {S.}~\bibnamefont
  {Luo}}, \bibinfo {author} {\bibfnamefont {T.}~\bibnamefont {Li}}, \bibinfo
  {author} {\bibfnamefont {X.}~\bibnamefont {Wang}}, \bibinfo {author}
  {\bibfnamefont {M.}~\bibnamefont {Faizan}},\ and\ \bibinfo {author}
  {\bibfnamefont {L.}~\bibnamefont {Zhang}},\ }\bibfield  {title} {\bibinfo
  {title} {High-throughput computational materials screening and discovery of
  optoelectronic semiconductors},\ }\href {https://doi.org/10.1002/wcms.1489}
  {\bibfield  {journal} {\bibinfo  {journal} {Wiley Interdisciplinary Reviews:
  Computational Molecular Science}\ }\textbf {\bibinfo {volume} {11}},\
  \bibinfo {pages} {e1489} (\bibinfo {year} {2021})}\BibitemShut {NoStop}%
\bibitem [{\citenamefont {Jain}\ \emph {et~al.}(2017)\citenamefont {Jain},
  \citenamefont {Voznyy},\ and\ \citenamefont {Sargent}}]{jain2017high}%
  \BibitemOpen
  \bibfield  {author} {\bibinfo {author} {\bibfnamefont {A.}~\bibnamefont
  {Jain}}, \bibinfo {author} {\bibfnamefont {O.}~\bibnamefont {Voznyy}},\ and\
  \bibinfo {author} {\bibfnamefont {E.~H.}\ \bibnamefont {Sargent}},\
  }\bibfield  {title} {\bibinfo {title} {High-throughput screening of lead-free
  perovskite-like materials for optoelectronic applications},\ }\href
  {https://doi.org/10.1021/acs.jpcc.7b02221} {\bibfield  {journal} {\bibinfo
  {journal} {The Journal of Physical Chemistry C}\ }\textbf {\bibinfo {volume}
  {121}},\ \bibinfo {pages} {7183} (\bibinfo {year} {2017})}\BibitemShut
  {NoStop}%
\bibitem [{\citenamefont {Jiang}\ and\ \citenamefont
  {Yin}(2021)}]{jiang2021high}%
  \BibitemOpen
  \bibfield  {author} {\bibinfo {author} {\bibfnamefont {X.}~\bibnamefont
  {Jiang}}\ and\ \bibinfo {author} {\bibfnamefont {W.-J.}\ \bibnamefont
  {Yin}},\ }\bibfield  {title} {\bibinfo {title} {High-throughput computational
  screening of oxide double perovskites for optoelectronic and photocatalysis
  applications},\ }\href {https://doi.org/10.1016/j.jechem.2020.08.046}
  {\bibfield  {journal} {\bibinfo  {journal} {Journal of Energy Chemistry}\
  }\textbf {\bibinfo {volume} {57}},\ \bibinfo {pages} {351} (\bibinfo {year}
  {2021})}\BibitemShut {NoStop}%
\bibitem [{\citenamefont {Pal}\ \emph {et~al.}(2021)\citenamefont {Pal},
  \citenamefont {Xia}, \citenamefont {Shen}, \citenamefont {He}, \citenamefont
  {Luo}, \citenamefont {Kanatzidis},\ and\ \citenamefont
  {Wolverton}}]{pal2021accelerated}%
  \BibitemOpen
  \bibfield  {author} {\bibinfo {author} {\bibfnamefont {K.}~\bibnamefont
  {Pal}}, \bibinfo {author} {\bibfnamefont {Y.}~\bibnamefont {Xia}}, \bibinfo
  {author} {\bibfnamefont {J.}~\bibnamefont {Shen}}, \bibinfo {author}
  {\bibfnamefont {J.}~\bibnamefont {He}}, \bibinfo {author} {\bibfnamefont
  {Y.}~\bibnamefont {Luo}}, \bibinfo {author} {\bibfnamefont {M.~G.}\
  \bibnamefont {Kanatzidis}},\ and\ \bibinfo {author} {\bibfnamefont
  {C.}~\bibnamefont {Wolverton}},\ }\bibfield  {title} {\bibinfo {title}
  {Accelerated discovery of a large family of quaternary chalcogenides with
  very low lattice thermal conductivity},\ }\href
  {https://doi.org/10.1038/s41524-021-00549-x} {\bibfield  {journal} {\bibinfo
  {journal} {npj Computational Materials}\ }\textbf {\bibinfo {volume} {7}},\
  \bibinfo {pages} {82} (\bibinfo {year} {2021})}\BibitemShut {NoStop}%
\bibitem [{\citenamefont {Ju}\ \emph {et~al.}(2017)\citenamefont {Ju},
  \citenamefont {Dai}, \citenamefont {Ma},\ and\ \citenamefont
  {Zeng}}]{ju2017perovskite}%
  \BibitemOpen
  \bibfield  {author} {\bibinfo {author} {\bibfnamefont {M.-G.}\ \bibnamefont
  {Ju}}, \bibinfo {author} {\bibfnamefont {J.}~\bibnamefont {Dai}}, \bibinfo
  {author} {\bibfnamefont {L.}~\bibnamefont {Ma}},\ and\ \bibinfo {author}
  {\bibfnamefont {X.~C.}\ \bibnamefont {Zeng}},\ }\bibfield  {title} {\bibinfo
  {title} {Perovskite chalcogenides with optimal bandgap and desired optical
  absorption for photovoltaic devices},\ }\href
  {https://doi.org/10.1002/aenm.201700216} {\bibfield  {journal} {\bibinfo
  {journal} {Advanced Energy Materials}\ }\textbf {\bibinfo {volume} {7}},\
  \bibinfo {pages} {1700216} (\bibinfo {year} {2017})}\BibitemShut {NoStop}%
\bibitem [{\citenamefont {Huo}\ \emph {et~al.}(2018)\citenamefont {Huo},
  \citenamefont {Wei},\ and\ \citenamefont {Yin}}]{huo2018high}%
  \BibitemOpen
  \bibfield  {author} {\bibinfo {author} {\bibfnamefont {Z.}~\bibnamefont
  {Huo}}, \bibinfo {author} {\bibfnamefont {S.-H.}\ \bibnamefont {Wei}},\ and\
  \bibinfo {author} {\bibfnamefont {W.-J.}\ \bibnamefont {Yin}},\ }\bibfield
  {title} {\bibinfo {title} {High-throughput screening of chalcogenide single
  perovskites by first-principles calculations for photovoltaics},\ }\href
  {https://iopscience.iop.org/article/10.1088/1361-6463/aae1ee/meta#Acknowledgments}
  {\bibfield  {journal} {\bibinfo  {journal} {Journal Of Physics D: Applied
  Physics}\ }\textbf {\bibinfo {volume} {51}},\ \bibinfo {pages} {474003}
  (\bibinfo {year} {2018})}\BibitemShut {NoStop}%
\bibitem [{\citenamefont {Deng}\ \emph {et~al.}(2024)\citenamefont {Deng},
  \citenamefont {Qiu}, \citenamefont {Yin}, \citenamefont {Li}, \citenamefont
  {Yang}, \citenamefont {Wei},\ and\ \citenamefont {Shi}}]{deng2024high}%
  \BibitemOpen
  \bibfield  {author} {\bibinfo {author} {\bibfnamefont {T.}~\bibnamefont
  {Deng}}, \bibinfo {author} {\bibfnamefont {P.}~\bibnamefont {Qiu}}, \bibinfo
  {author} {\bibfnamefont {T.}~\bibnamefont {Yin}}, \bibinfo {author}
  {\bibfnamefont {Z.}~\bibnamefont {Li}}, \bibinfo {author} {\bibfnamefont
  {J.}~\bibnamefont {Yang}}, \bibinfo {author} {\bibfnamefont {T.}~\bibnamefont
  {Wei}},\ and\ \bibinfo {author} {\bibfnamefont {X.}~\bibnamefont {Shi}},\
  }\bibfield  {title} {\bibinfo {title} {High-throughput strategies in the
  discovery of thermoelectric materials},\ }\href
  {https://doi.org/10.1002/adma.202311278} {\bibfield  {journal} {\bibinfo
  {journal} {Advanced Materials}\ }\textbf {\bibinfo {volume} {36}},\ \bibinfo
  {pages} {2311278} (\bibinfo {year} {2024})}\BibitemShut {NoStop}%
\bibitem [{\citenamefont {Zhao}\ \emph {et~al.}(2017)\citenamefont {Zhao},
  \citenamefont {Yang}, \citenamefont {Fu}, \citenamefont {Yang}, \citenamefont
  {Xu}, \citenamefont {Yu}, \citenamefont {Wei},\ and\ \citenamefont
  {Zhang}}]{zhao2017design}%
  \BibitemOpen
  \bibfield  {author} {\bibinfo {author} {\bibfnamefont {X.-G.}\ \bibnamefont
  {Zhao}}, \bibinfo {author} {\bibfnamefont {J.-H.}\ \bibnamefont {Yang}},
  \bibinfo {author} {\bibfnamefont {Y.}~\bibnamefont {Fu}}, \bibinfo {author}
  {\bibfnamefont {D.}~\bibnamefont {Yang}}, \bibinfo {author} {\bibfnamefont
  {Q.}~\bibnamefont {Xu}}, \bibinfo {author} {\bibfnamefont {L.}~\bibnamefont
  {Yu}}, \bibinfo {author} {\bibfnamefont {S.-H.}\ \bibnamefont {Wei}},\ and\
  \bibinfo {author} {\bibfnamefont {L.}~\bibnamefont {Zhang}},\ }\bibfield
  {title} {\bibinfo {title} {Design of lead-free inorganic halide perovskites
  for solar cells via cation-transmutation},\ }\href
  {https://doi.org/10.1021/jacs.6b09645} {\bibfield  {journal} {\bibinfo
  {journal} {Journal of the American Chemical Society}\ }\textbf {\bibinfo
  {volume} {139}},\ \bibinfo {pages} {2630} (\bibinfo {year}
  {2017})}\BibitemShut {NoStop}%
\bibitem [{\citenamefont {Jain}\ \emph {et~al.}(2013)\citenamefont {Jain},
  \citenamefont {Ong}, \citenamefont {Hautier}, \citenamefont {Chen},
  \citenamefont {Richards}, \citenamefont {Dacek}, \citenamefont {Cholia},
  \citenamefont {Gunter}, \citenamefont {Skinner}, \citenamefont {Ceder},\ and\
  \citenamefont {Persson}}]{jain2013commentary}%
  \BibitemOpen
  \bibfield  {author} {\bibinfo {author} {\bibfnamefont {A.}~\bibnamefont
  {Jain}}, \bibinfo {author} {\bibfnamefont {S.~P.}\ \bibnamefont {Ong}},
  \bibinfo {author} {\bibfnamefont {G.}~\bibnamefont {Hautier}}, \bibinfo
  {author} {\bibfnamefont {W.}~\bibnamefont {Chen}}, \bibinfo {author}
  {\bibfnamefont {W.~D.}\ \bibnamefont {Richards}}, \bibinfo {author}
  {\bibfnamefont {S.}~\bibnamefont {Dacek}}, \bibinfo {author} {\bibfnamefont
  {S.}~\bibnamefont {Cholia}}, \bibinfo {author} {\bibfnamefont
  {D.}~\bibnamefont {Gunter}}, \bibinfo {author} {\bibfnamefont
  {D.}~\bibnamefont {Skinner}}, \bibinfo {author} {\bibfnamefont
  {G.}~\bibnamefont {Ceder}},\ and\ \bibinfo {author} {\bibfnamefont {K.~A.}\
  \bibnamefont {Persson}},\ }\bibfield  {title} {\bibinfo {title} {Commentary:
  The materials project: A materials genome approach to accelerating materials
  innovation},\ }\href {https://doi.org/10.1063/1.4812323} {\bibfield
  {journal} {\bibinfo  {journal} {APL materials}\ }\textbf {\bibinfo {volume}
  {1}} (\bibinfo {year} {2013})}\BibitemShut {NoStop}%
\bibitem [{\citenamefont {Saal}\ \emph {et~al.}(2013)\citenamefont {Saal},
  \citenamefont {Kirklin}, \citenamefont {Aykol}, \citenamefont {Meredig},\
  and\ \citenamefont {Wolverton}}]{saal2013materials}%
  \BibitemOpen
  \bibfield  {author} {\bibinfo {author} {\bibfnamefont {J.~E.}\ \bibnamefont
  {Saal}}, \bibinfo {author} {\bibfnamefont {S.}~\bibnamefont {Kirklin}},
  \bibinfo {author} {\bibfnamefont {M.}~\bibnamefont {Aykol}}, \bibinfo
  {author} {\bibfnamefont {B.}~\bibnamefont {Meredig}},\ and\ \bibinfo {author}
  {\bibfnamefont {C.}~\bibnamefont {Wolverton}},\ }\bibfield  {title} {\bibinfo
  {title} {Materials design and discovery with high-throughput density
  functional theory: the open quantum materials database ({OQMD})},\ }\href
  {https://doi.org/10.1007/s11837-013-0755-4} {\bibfield  {journal} {\bibinfo
  {journal} {Jom}\ }\textbf {\bibinfo {volume} {65}},\ \bibinfo {pages} {1501}
  (\bibinfo {year} {2013})}\BibitemShut {NoStop}%
\bibitem [{\citenamefont {Kirklin}\ \emph {et~al.}(2015)\citenamefont
  {Kirklin}, \citenamefont {Saal}, \citenamefont {Meredig}, \citenamefont
  {Thompson}, \citenamefont {Doak}, \citenamefont {Aykol}, \citenamefont
  {R{\"u}hl},\ and\ \citenamefont {Wolverton}}]{kirklin2015open}%
  \BibitemOpen
  \bibfield  {author} {\bibinfo {author} {\bibfnamefont {S.}~\bibnamefont
  {Kirklin}}, \bibinfo {author} {\bibfnamefont {J.~E.}\ \bibnamefont {Saal}},
  \bibinfo {author} {\bibfnamefont {B.}~\bibnamefont {Meredig}}, \bibinfo
  {author} {\bibfnamefont {A.}~\bibnamefont {Thompson}}, \bibinfo {author}
  {\bibfnamefont {J.~W.}\ \bibnamefont {Doak}}, \bibinfo {author}
  {\bibfnamefont {M.}~\bibnamefont {Aykol}}, \bibinfo {author} {\bibfnamefont
  {S.}~\bibnamefont {R{\"u}hl}},\ and\ \bibinfo {author} {\bibfnamefont
  {C.}~\bibnamefont {Wolverton}},\ }\bibfield  {title} {\bibinfo {title} {The
  open quantum materials database ({OQMD}): assessing the accuracy of dft
  formation energies},\ }\href {https://doi.org/10.1038/npjcompumats.2015.10}
  {\bibfield  {journal} {\bibinfo  {journal} {npj Computational Materials}\
  }\textbf {\bibinfo {volume} {1}},\ \bibinfo {pages} {1} (\bibinfo {year}
  {2015})}\BibitemShut {NoStop}%
\bibitem [{\citenamefont {Curtarolo}\ \emph {et~al.}(2012)\citenamefont
  {Curtarolo}, \citenamefont {Setyawan}, \citenamefont {Wang}, \citenamefont
  {Xue}, \citenamefont {Yang}, \citenamefont {Taylor}, \citenamefont {Nelson},
  \citenamefont {Hart}, \citenamefont {Sanvito}, \citenamefont
  {Buongiorno-Nardelli}, \citenamefont {Mingo},\ and\ \citenamefont
  {Levy}}]{curtarolo2012aflowlib}%
  \BibitemOpen
  \bibfield  {author} {\bibinfo {author} {\bibfnamefont {S.}~\bibnamefont
  {Curtarolo}}, \bibinfo {author} {\bibfnamefont {W.}~\bibnamefont {Setyawan}},
  \bibinfo {author} {\bibfnamefont {S.}~\bibnamefont {Wang}}, \bibinfo {author}
  {\bibfnamefont {J.}~\bibnamefont {Xue}}, \bibinfo {author} {\bibfnamefont
  {K.}~\bibnamefont {Yang}}, \bibinfo {author} {\bibfnamefont {R.~H.}\
  \bibnamefont {Taylor}}, \bibinfo {author} {\bibfnamefont {L.~J.}\
  \bibnamefont {Nelson}}, \bibinfo {author} {\bibfnamefont {G.~L.}\
  \bibnamefont {Hart}}, \bibinfo {author} {\bibfnamefont {S.}~\bibnamefont
  {Sanvito}}, \bibinfo {author} {\bibfnamefont {M.}~\bibnamefont
  {Buongiorno-Nardelli}}, \bibinfo {author} {\bibfnamefont {N.}~\bibnamefont
  {Mingo}},\ and\ \bibinfo {author} {\bibfnamefont {O.}~\bibnamefont {Levy}},\
  }\bibfield  {title} {\bibinfo {title} {{AFLOWLIB}. {ORG}: A distributed
  materials properties repository from high-throughput ab initio
  calculations},\ }\href {https://doi.org/10.1016/j.commatsci.2012.02.002}
  {\bibfield  {journal} {\bibinfo  {journal} {Computational Materials Science}\
  }\textbf {\bibinfo {volume} {58}},\ \bibinfo {pages} {227} (\bibinfo {year}
  {2012})}\BibitemShut {NoStop}%
\bibitem [{\citenamefont {Belsky}\ \emph {et~al.}(2002)\citenamefont {Belsky},
  \citenamefont {Hellenbrandt}, \citenamefont {Karen},\ and\ \citenamefont
  {Luksch}}]{belsky2002new}%
  \BibitemOpen
  \bibfield  {author} {\bibinfo {author} {\bibfnamefont {A.}~\bibnamefont
  {Belsky}}, \bibinfo {author} {\bibfnamefont {M.}~\bibnamefont
  {Hellenbrandt}}, \bibinfo {author} {\bibfnamefont {V.~L.}\ \bibnamefont
  {Karen}},\ and\ \bibinfo {author} {\bibfnamefont {P.}~\bibnamefont
  {Luksch}},\ }\bibfield  {title} {\bibinfo {title} {New developments in the
  inorganic crystal structure database ({ICSD}): accessibility in support of
  materials research and design},\ }\href
  {https://doi.org/10.1107/S0108768102006948} {\bibfield  {journal} {\bibinfo
  {journal} {Acta Crystallographica Section B: Structural Science}\ }\textbf
  {\bibinfo {volume} {58}},\ \bibinfo {pages} {364} (\bibinfo {year}
  {2002})}\BibitemShut {NoStop}%
\bibitem [{\citenamefont {Zhou}\ \emph {et~al.}(2018)\citenamefont {Zhou},
  \citenamefont {Xu}, \citenamefont {Chen}, \citenamefont {Kuang},\ and\
  \citenamefont {Su}}]{zhou2018synthesis}%
  \BibitemOpen
  \bibfield  {author} {\bibinfo {author} {\bibfnamefont {L.}~\bibnamefont
  {Zhou}}, \bibinfo {author} {\bibfnamefont {Y.-F.}\ \bibnamefont {Xu}},
  \bibinfo {author} {\bibfnamefont {B.-X.}\ \bibnamefont {Chen}}, \bibinfo
  {author} {\bibfnamefont {D.-B.}\ \bibnamefont {Kuang}},\ and\ \bibinfo
  {author} {\bibfnamefont {C.-Y.}\ \bibnamefont {Su}},\ }\bibfield  {title}
  {\bibinfo {title} {Synthesis and photocatalytic application of stable
  lead-free {C}s$_2${A}g{B}i{B}r$_6$ perovskite nanocrystals},\ }\href
  {https://doi.org/10.1002/smll.201703762} {\bibfield  {journal} {\bibinfo
  {journal} {Small}\ }\textbf {\bibinfo {volume} {14}},\ \bibinfo {pages}
  {1703762} (\bibinfo {year} {2018})}\BibitemShut {NoStop}%
\bibitem [{\citenamefont {Locardi}\ \emph {et~al.}(2018)\citenamefont
  {Locardi}, \citenamefont {Cirignano}, \citenamefont {Baranov}, \citenamefont
  {Dang}, \citenamefont {Prato}, \citenamefont {Drago}, \citenamefont
  {Ferretti}, \citenamefont {Pinchetti}, \citenamefont {Fanciulli},
  \citenamefont {Brovelli},\ and\ \citenamefont
  {Trizio}}]{locardi2018colloidal}%
  \BibitemOpen
  \bibfield  {author} {\bibinfo {author} {\bibfnamefont {F.}~\bibnamefont
  {Locardi}}, \bibinfo {author} {\bibfnamefont {M.}~\bibnamefont {Cirignano}},
  \bibinfo {author} {\bibfnamefont {D.}~\bibnamefont {Baranov}}, \bibinfo
  {author} {\bibfnamefont {Z.}~\bibnamefont {Dang}}, \bibinfo {author}
  {\bibfnamefont {M.}~\bibnamefont {Prato}}, \bibinfo {author} {\bibfnamefont
  {F.}~\bibnamefont {Drago}}, \bibinfo {author} {\bibfnamefont
  {M.}~\bibnamefont {Ferretti}}, \bibinfo {author} {\bibfnamefont
  {V.}~\bibnamefont {Pinchetti}}, \bibinfo {author} {\bibfnamefont
  {M.}~\bibnamefont {Fanciulli}}, \bibinfo {author} {\bibfnamefont
  {S.}~\bibnamefont {Brovelli}},\ and\ \bibinfo {author} {\bibfnamefont
  {M.~L.}\ \bibnamefont {Trizio}, \bibfnamefont {Luca~De}},\ }\bibfield
  {title} {\bibinfo {title} {Colloidal synthesis of double perovskite
  {C}s$_2${A}g{I}n{C}l$_6$ and {M}n-doped {C}s$_2${A}g{I}n{C}l$_6$
  nanocrystals},\ }\href {https://doi.org/10.1021/jacs.8b07983} {\bibfield
  {journal} {\bibinfo  {journal} {Journal of the American Chemical Society}\
  }\textbf {\bibinfo {volume} {140}},\ \bibinfo {pages} {12989} (\bibinfo
  {year} {2018})}\BibitemShut {NoStop}%
\bibitem [{\citenamefont {McClure}\ \emph {et~al.}(2016)\citenamefont
  {McClure}, \citenamefont {Ball}, \citenamefont {Windl},\ and\ \citenamefont
  {Woodward}}]{mcclure2016cs2agbix6}%
  \BibitemOpen
  \bibfield  {author} {\bibinfo {author} {\bibfnamefont {E.~T.}\ \bibnamefont
  {McClure}}, \bibinfo {author} {\bibfnamefont {M.~R.}\ \bibnamefont {Ball}},
  \bibinfo {author} {\bibfnamefont {W.}~\bibnamefont {Windl}},\ and\ \bibinfo
  {author} {\bibfnamefont {P.~M.}\ \bibnamefont {Woodward}},\ }\bibfield
  {title} {\bibinfo {title} {{C}s$_2${A}g{B}i{X}$_6$ ({X} = {B}r, {C}l): new
  visible light absorbing, lead-free halide perovskite semiconductors},\ }\href
  {https://doi.org/10.1021/acs.chemmater.5b04231} {\bibfield  {journal}
  {\bibinfo  {journal} {Chemistry of Materials}\ }\textbf {\bibinfo {volume}
  {28}},\ \bibinfo {pages} {1348} (\bibinfo {year} {2016})}\BibitemShut
  {NoStop}%
\bibitem [{\citenamefont {Shen}\ \emph {et~al.}(2023)\citenamefont {Shen},
  \citenamefont {Li}, \citenamefont {Zhang}, \citenamefont {Xie}, \citenamefont
  {Long}, \citenamefont {Fortunato}, \citenamefont {Liang}, \citenamefont
  {Dai}, \citenamefont {Shen}, \citenamefont {Wolverton},\ and\ \citenamefont
  {Zhang}}]{shen2023accelerated}%
  \BibitemOpen
  \bibfield  {author} {\bibinfo {author} {\bibfnamefont {C.}~\bibnamefont
  {Shen}}, \bibinfo {author} {\bibfnamefont {T.}~\bibnamefont {Li}}, \bibinfo
  {author} {\bibfnamefont {Y.}~\bibnamefont {Zhang}}, \bibinfo {author}
  {\bibfnamefont {R.}~\bibnamefont {Xie}}, \bibinfo {author} {\bibfnamefont
  {T.}~\bibnamefont {Long}}, \bibinfo {author} {\bibfnamefont {N.~M.}\
  \bibnamefont {Fortunato}}, \bibinfo {author} {\bibfnamefont {F.}~\bibnamefont
  {Liang}}, \bibinfo {author} {\bibfnamefont {M.}~\bibnamefont {Dai}}, \bibinfo
  {author} {\bibfnamefont {J.}~\bibnamefont {Shen}}, \bibinfo {author}
  {\bibfnamefont {C.~M.}\ \bibnamefont {Wolverton}},\ and\ \bibinfo {author}
  {\bibfnamefont {H.}~\bibnamefont {Zhang}},\ }\bibfield  {title} {\bibinfo
  {title} {Accelerated screening of ternary chalcogenides for potential
  photovoltaic applications},\ }\href {https://doi.org/10.1021/jacs.3c06207}
  {\bibfield  {journal} {\bibinfo  {journal} {Journal of the American Chemical
  Society}\ }\textbf {\bibinfo {volume} {145}},\ \bibinfo {pages} {21925}
  (\bibinfo {year} {2023})}\BibitemShut {NoStop}%
\bibitem [{\citenamefont {Pal}\ \emph {et~al.}(2019{\natexlab{a}})\citenamefont
  {Pal}, \citenamefont {Xia}, \citenamefont {He},\ and\ \citenamefont
  {Wolverton}}]{pal2019high}%
  \BibitemOpen
  \bibfield  {author} {\bibinfo {author} {\bibfnamefont {K.}~\bibnamefont
  {Pal}}, \bibinfo {author} {\bibfnamefont {Y.}~\bibnamefont {Xia}}, \bibinfo
  {author} {\bibfnamefont {J.}~\bibnamefont {He}},\ and\ \bibinfo {author}
  {\bibfnamefont {C.}~\bibnamefont {Wolverton}},\ }\bibfield  {title} {\bibinfo
  {title} {High thermoelectric performance in {B}a{A}g{Y}{T}e$_3$ via low
  lattice thermal conductivity induced by bonding heterogeneity},\ }\href
  {https://doi.org/10.1103/PhysRevMaterials.3.085402} {\bibfield  {journal}
  {\bibinfo  {journal} {Physical Review Materials}\ }\textbf {\bibinfo {volume}
  {3}},\ \bibinfo {pages} {085402} (\bibinfo {year}
  {2019}{\natexlab{a}})}\BibitemShut {NoStop}%
\bibitem [{\citenamefont {Jiang}\ and\ \citenamefont
  {Yin}(2020)}]{jiang2020designing}%
  \BibitemOpen
  \bibfield  {author} {\bibinfo {author} {\bibfnamefont {X.}~\bibnamefont
  {Jiang}}\ and\ \bibinfo {author} {\bibfnamefont {W.-J.}\ \bibnamefont
  {Yin}},\ }\bibfield  {title} {\bibinfo {title} {Designing solar-cell absorber
  materials through computational high-throughput screening},\ }\href
  {http://dx.doi.org/10.1088/1674-1056/ab6655} {\bibfield  {journal} {\bibinfo
  {journal} {Chinese Physics B}\ }\textbf {\bibinfo {volume} {29}},\ \bibinfo
  {pages} {028803} (\bibinfo {year} {2020})}\BibitemShut {NoStop}%
\bibitem [{\citenamefont {Kuhar}\ \emph {et~al.}(2018)\citenamefont {Kuhar},
  \citenamefont {Pandey}, \citenamefont {Thygesen},\ and\ \citenamefont
  {Jacobsen}}]{kuhar2018high}%
  \BibitemOpen
  \bibfield  {author} {\bibinfo {author} {\bibfnamefont {K.}~\bibnamefont
  {Kuhar}}, \bibinfo {author} {\bibfnamefont {M.}~\bibnamefont {Pandey}},
  \bibinfo {author} {\bibfnamefont {K.~S.}\ \bibnamefont {Thygesen}},\ and\
  \bibinfo {author} {\bibfnamefont {K.~W.}\ \bibnamefont {Jacobsen}},\
  }\bibfield  {title} {\bibinfo {title} {High-throughput computational
  assessment of previously synthesized semiconductors for photovoltaic and
  photoelectrochemical devices},\ }\href
  {https://doi.org/10.1021/acsenergylett.7b01312} {\bibfield  {journal}
  {\bibinfo  {journal} {ACS Energy Letters}\ }\textbf {\bibinfo {volume} {3}},\
  \bibinfo {pages} {436} (\bibinfo {year} {2018})}\BibitemShut {NoStop}%
\bibitem [{\citenamefont {Sun}\ \emph {et~al.}(2015)\citenamefont {Sun},
  \citenamefont {Agiorgousis}, \citenamefont {Zhang},\ and\ \citenamefont
  {Zhang}}]{sun2015chalcogenide}%
  \BibitemOpen
  \bibfield  {author} {\bibinfo {author} {\bibfnamefont {Y.-Y.}\ \bibnamefont
  {Sun}}, \bibinfo {author} {\bibfnamefont {M.~L.}\ \bibnamefont
  {Agiorgousis}}, \bibinfo {author} {\bibfnamefont {P.}~\bibnamefont {Zhang}},\
  and\ \bibinfo {author} {\bibfnamefont {S.}~\bibnamefont {Zhang}},\ }\bibfield
   {title} {\bibinfo {title} {Chalcogenide perovskites for photovoltaics},\
  }\href {https://doi.org/10.1021/nl504046x} {\bibfield  {journal} {\bibinfo
  {journal} {Nano letters}\ }\textbf {\bibinfo {volume} {15}},\ \bibinfo
  {pages} {581} (\bibinfo {year} {2015})}\BibitemShut {NoStop}%
\bibitem [{\citenamefont {Perera}\ \emph {et~al.}(2016)\citenamefont {Perera},
  \citenamefont {Hui}, \citenamefont {Zhao}, \citenamefont {Xue}, \citenamefont
  {Sun}, \citenamefont {Deng}, \citenamefont {Gross}, \citenamefont
  {Milleville}, \citenamefont {Xu}, \citenamefont {Watson}, \citenamefont
  {Weinstein}, \citenamefont {Sun}, \citenamefont {Zhang},\ and\ \citenamefont
  {Zeng}}]{perera2016chalcogenide}%
  \BibitemOpen
  \bibfield  {author} {\bibinfo {author} {\bibfnamefont {S.}~\bibnamefont
  {Perera}}, \bibinfo {author} {\bibfnamefont {H.}~\bibnamefont {Hui}},
  \bibinfo {author} {\bibfnamefont {C.}~\bibnamefont {Zhao}}, \bibinfo {author}
  {\bibfnamefont {H.}~\bibnamefont {Xue}}, \bibinfo {author} {\bibfnamefont
  {F.}~\bibnamefont {Sun}}, \bibinfo {author} {\bibfnamefont {C.}~\bibnamefont
  {Deng}}, \bibinfo {author} {\bibfnamefont {N.}~\bibnamefont {Gross}},
  \bibinfo {author} {\bibfnamefont {C.}~\bibnamefont {Milleville}}, \bibinfo
  {author} {\bibfnamefont {X.}~\bibnamefont {Xu}}, \bibinfo {author}
  {\bibfnamefont {D.~F.}\ \bibnamefont {Watson}}, \bibinfo {author}
  {\bibfnamefont {B.}~\bibnamefont {Weinstein}}, \bibinfo {author}
  {\bibfnamefont {Y.-Y.}\ \bibnamefont {Sun}}, \bibinfo {author} {\bibfnamefont
  {S.}~\bibnamefont {Zhang}},\ and\ \bibinfo {author} {\bibfnamefont
  {H.}~\bibnamefont {Zeng}},\ }\bibfield  {title} {\bibinfo {title}
  {Chalcogenide perovskites--an emerging class of ionic semiconductors},\
  }\href {https://doi.org/10.1016/j.nanoen.2016.02.020} {\bibfield  {journal}
  {\bibinfo  {journal} {Nano Energy}\ }\textbf {\bibinfo {volume} {22}},\
  \bibinfo {pages} {129} (\bibinfo {year} {2016})}\BibitemShut {NoStop}%
\bibitem [{\citenamefont {McKeever}\ \emph {et~al.}(2023)\citenamefont
  {McKeever}, \citenamefont {Patil}, \citenamefont {Palabathuni},\ and\
  \citenamefont {Singh}}]{mckeever2023functional}%
  \BibitemOpen
  \bibfield  {author} {\bibinfo {author} {\bibfnamefont {H.}~\bibnamefont
  {McKeever}}, \bibinfo {author} {\bibfnamefont {N.~N.}\ \bibnamefont {Patil}},
  \bibinfo {author} {\bibfnamefont {M.}~\bibnamefont {Palabathuni}},\ and\
  \bibinfo {author} {\bibfnamefont {S.}~\bibnamefont {Singh}},\ }\bibfield
  {title} {\bibinfo {title} {Functional alkali metal-based ternary
  chalcogenides: design, properties, and opportunities},\ }\href
  {https://doi.org/10.1021/acs.chemmater.3c01652} {\bibfield  {journal}
  {\bibinfo  {journal} {Chemistry of Materials}\ }\textbf {\bibinfo {volume}
  {35}},\ \bibinfo {pages} {9833} (\bibinfo {year} {2023})}\BibitemShut
  {NoStop}%
\bibitem [{\citenamefont {Koscielski}\ and\ \citenamefont
  {Ibers}(2012)}]{koscielski2012structural}%
  \BibitemOpen
  \bibfield  {author} {\bibinfo {author} {\bibfnamefont {L.~A.}\ \bibnamefont
  {Koscielski}}\ and\ \bibinfo {author} {\bibfnamefont {J.~A.}\ \bibnamefont
  {Ibers}},\ }\bibfield  {title} {\bibinfo {title} {The structural chemistry of
  quaternary chalcogenides of the type {AMM}'{Q}$_3$},\ }\href
  {https://onlinelibrary.wiley.com/doi/full/10.1002/zaac.201200301} {\bibfield
  {journal} {\bibinfo  {journal} {Zeitschrift f{\"u}r anorganische und
  allgemeine Chemie}\ }\textbf {\bibinfo {volume} {638}},\ \bibinfo {pages}
  {2585} (\bibinfo {year} {2012})}\BibitemShut {NoStop}%
\bibitem [{\citenamefont {Shahid}\ \emph {et~al.}(2023)\citenamefont {Shahid},
  \citenamefont {Ray}, \citenamefont {Yadav}, \citenamefont {Deepa},
  \citenamefont {Niranjan},\ and\ \citenamefont
  {Prakash}}]{shahid2023structure}%
  \BibitemOpen
  \bibfield  {author} {\bibinfo {author} {\bibfnamefont {O.}~\bibnamefont
  {Shahid}}, \bibinfo {author} {\bibfnamefont {A.~K.}\ \bibnamefont {Ray}},
  \bibinfo {author} {\bibfnamefont {S.}~\bibnamefont {Yadav}}, \bibinfo
  {author} {\bibfnamefont {M.}~\bibnamefont {Deepa}}, \bibinfo {author}
  {\bibfnamefont {M.~K.}\ \bibnamefont {Niranjan}},\ and\ \bibinfo {author}
  {\bibfnamefont {J.}~\bibnamefont {Prakash}},\ }\bibfield  {title} {\bibinfo
  {title} {Structure-property relationships and dft studies of three quaternary
  chalcogenides: {B}a{C}e{C}u{S}e$_3$, {B}a{C}e{A}g{S}$_3$, and
  {B}a{C}e{A}g{S}e$_3$},\ }\href
  {https://doi.org/10.1016/j.materresbull.2023.112469} {\bibfield  {journal}
  {\bibinfo  {journal} {Materials Research Bulletin}\ }\textbf {\bibinfo
  {volume} {168}},\ \bibinfo {pages} {112469} (\bibinfo {year}
  {2023})}\BibitemShut {NoStop}%
\bibitem [{\citenamefont {McKeown~Wessler}\ \emph {et~al.}(2021)\citenamefont
  {McKeown~Wessler}, \citenamefont {Wang}, \citenamefont {Sun}, \citenamefont
  {Liao}, \citenamefont {Fischer}, \citenamefont {Blum},\ and\ \citenamefont
  {Mitzi}}]{mckeown2021structural}%
  \BibitemOpen
  \bibfield  {author} {\bibinfo {author} {\bibfnamefont {G.~C.}\ \bibnamefont
  {McKeown~Wessler}}, \bibinfo {author} {\bibfnamefont {T.}~\bibnamefont
  {Wang}}, \bibinfo {author} {\bibfnamefont {J.-P.}\ \bibnamefont {Sun}},
  \bibinfo {author} {\bibfnamefont {Y.}~\bibnamefont {Liao}}, \bibinfo {author}
  {\bibfnamefont {M.~C.}\ \bibnamefont {Fischer}}, \bibinfo {author}
  {\bibfnamefont {V.}~\bibnamefont {Blum}},\ and\ \bibinfo {author}
  {\bibfnamefont {D.~B.}\ \bibnamefont {Mitzi}},\ }\bibfield  {title} {\bibinfo
  {title} {Structural, optical, and electronic properties of two quaternary
  chalcogenide semiconductors: {A}g$_2${S}r{S}i{S}$_4$ and
  {A}g$_2${S}r{G}e{S}$_4$},\ }\href
  {https://doi.org/10.1021/acs.inorgchem.1c01416} {\bibfield  {journal}
  {\bibinfo  {journal} {Inorganic Chemistry}\ }\textbf {\bibinfo {volume}
  {60}},\ \bibinfo {pages} {12206} (\bibinfo {year} {2021})}\BibitemShut
  {NoStop}%
\bibitem [{\citenamefont {Xie}\ \emph {et~al.}(2022)\citenamefont {Xie},
  \citenamefont {Husremovic}, \citenamefont {Gonzalez}, \citenamefont {Craig},\
  and\ \citenamefont {Bediako}}]{xie2022structure}%
  \BibitemOpen
  \bibfield  {author} {\bibinfo {author} {\bibfnamefont {L.~S.}\ \bibnamefont
  {Xie}}, \bibinfo {author} {\bibfnamefont {S.}~\bibnamefont {Husremovic}},
  \bibinfo {author} {\bibfnamefont {O.}~\bibnamefont {Gonzalez}}, \bibinfo
  {author} {\bibfnamefont {I.~M.}\ \bibnamefont {Craig}},\ and\ \bibinfo
  {author} {\bibfnamefont {D.~K.}\ \bibnamefont {Bediako}},\ }\bibfield
  {title} {\bibinfo {title} {Structure and magnetism of iron-and
  chromium-intercalated niobium and tantalum disulfides},\ }\href
  {https://doi.org/10.1021/jacs.1c12975} {\bibfield  {journal} {\bibinfo
  {journal} {Journal of the American Chemical Society}\ }\textbf {\bibinfo
  {volume} {144}},\ \bibinfo {pages} {9525} (\bibinfo {year}
  {2022})}\BibitemShut {NoStop}%
\bibitem [{\citenamefont {Luo}\ \emph {et~al.}(2020)\citenamefont {Luo},
  \citenamefont {Hao}, \citenamefont {Cai}, \citenamefont {Slade},
  \citenamefont {Luo}, \citenamefont {Dravid}, \citenamefont {Wolverton},
  \citenamefont {Yan},\ and\ \citenamefont {Kanatzidis}}]{luo2020high}%
  \BibitemOpen
  \bibfield  {author} {\bibinfo {author} {\bibfnamefont {Y.}~\bibnamefont
  {Luo}}, \bibinfo {author} {\bibfnamefont {S.}~\bibnamefont {Hao}}, \bibinfo
  {author} {\bibfnamefont {S.}~\bibnamefont {Cai}}, \bibinfo {author}
  {\bibfnamefont {T.~J.}\ \bibnamefont {Slade}}, \bibinfo {author}
  {\bibfnamefont {Z.~Z.}\ \bibnamefont {Luo}}, \bibinfo {author} {\bibfnamefont
  {V.~P.}\ \bibnamefont {Dravid}}, \bibinfo {author} {\bibfnamefont
  {C.}~\bibnamefont {Wolverton}}, \bibinfo {author} {\bibfnamefont
  {Q.}~\bibnamefont {Yan}},\ and\ \bibinfo {author} {\bibfnamefont {M.~G.}\
  \bibnamefont {Kanatzidis}},\ }\bibfield  {title} {\bibinfo {title} {High
  thermoelectric performance in the new cubic semiconductor
  {A}g{S}n{S}b{S}e$_3$ by high-entropy engineering},\ }\href
  {https://doi.org/10.1021/jacs.0c07803} {\bibfield  {journal} {\bibinfo
  {journal} {Journal of the American Chemical Society}\ }\textbf {\bibinfo
  {volume} {142}},\ \bibinfo {pages} {15187} (\bibinfo {year}
  {2020})}\BibitemShut {NoStop}%
\bibitem [{\citenamefont {Le~Donne}\ \emph {et~al.}(2019)\citenamefont
  {Le~Donne}, \citenamefont {Trifiletti},\ and\ \citenamefont
  {Binetti}}]{le2019new}%
  \BibitemOpen
  \bibfield  {author} {\bibinfo {author} {\bibfnamefont {A.}~\bibnamefont
  {Le~Donne}}, \bibinfo {author} {\bibfnamefont {V.}~\bibnamefont
  {Trifiletti}},\ and\ \bibinfo {author} {\bibfnamefont {S.}~\bibnamefont
  {Binetti}},\ }\bibfield  {title} {\bibinfo {title} {New earth-abundant thin
  film solar cells based on chalcogenides},\ }\href
  {https://doi.org/10.3389/fchem.2019.00297} {\bibfield  {journal} {\bibinfo
  {journal} {Frontiers in chemistry}\ }\textbf {\bibinfo {volume} {7}},\
  \bibinfo {pages} {297} (\bibinfo {year} {2019})}\BibitemShut {NoStop}%
\bibitem [{\citenamefont {Pal}\ \emph {et~al.}(2019{\natexlab{b}})\citenamefont
  {Pal}, \citenamefont {Xia}, \citenamefont {He},\ and\ \citenamefont
  {Wolverton}}]{pal2019intrinsically}%
  \BibitemOpen
  \bibfield  {author} {\bibinfo {author} {\bibfnamefont {K.}~\bibnamefont
  {Pal}}, \bibinfo {author} {\bibfnamefont {Y.}~\bibnamefont {Xia}}, \bibinfo
  {author} {\bibfnamefont {J.}~\bibnamefont {He}},\ and\ \bibinfo {author}
  {\bibfnamefont {C.}~\bibnamefont {Wolverton}},\ }\bibfield  {title} {\bibinfo
  {title} {Intrinsically low lattice thermal conductivity derived from rattler
  cations in an {AMM'}{Q}$_3$ family of chalcogenides},\ }\href
  {https://doi.org/10.1021/acs.chemmater.9b02484} {\bibfield  {journal}
  {\bibinfo  {journal} {Chemistry of Materials}\ }\textbf {\bibinfo {volume}
  {31}},\ \bibinfo {pages} {8734} (\bibinfo {year}
  {2019}{\natexlab{b}})}\BibitemShut {NoStop}%
\bibitem [{\citenamefont {Laing}\ \emph {et~al.}(2022)\citenamefont {Laing},
  \citenamefont {Weiss}, \citenamefont {Pal}, \citenamefont {Quintero},
  \citenamefont {Xie}, \citenamefont {Zhou}, \citenamefont {Shen},
  \citenamefont {Chung}, \citenamefont {Wolverton},\ and\ \citenamefont
  {Kanatzidis}}]{laing2022acuzrq3}%
  \BibitemOpen
  \bibfield  {author} {\bibinfo {author} {\bibfnamefont {C.~C.}\ \bibnamefont
  {Laing}}, \bibinfo {author} {\bibfnamefont {B.~E.}\ \bibnamefont {Weiss}},
  \bibinfo {author} {\bibfnamefont {K.}~\bibnamefont {Pal}}, \bibinfo {author}
  {\bibfnamefont {M.~A.}\ \bibnamefont {Quintero}}, \bibinfo {author}
  {\bibfnamefont {H.}~\bibnamefont {Xie}}, \bibinfo {author} {\bibfnamefont
  {X.}~\bibnamefont {Zhou}}, \bibinfo {author} {\bibfnamefont {J.}~\bibnamefont
  {Shen}}, \bibinfo {author} {\bibfnamefont {D.~Y.}\ \bibnamefont {Chung}},
  \bibinfo {author} {\bibfnamefont {C.}~\bibnamefont {Wolverton}},\ and\
  \bibinfo {author} {\bibfnamefont {M.~G.}\ \bibnamefont {Kanatzidis}},\
  }\bibfield  {title} {\bibinfo {title} {{A}{C}u{Z}r{Q}$_3$ ({A} = {R}b, {C}s;
  {Q} = {S}, {S}e, {T}e): direct bandgap semiconductors and metals with
  ultralow thermal conductivity},\ }\href
  {https://doi.org/10.1021/acs.chemmater.2c02104} {\bibfield  {journal}
  {\bibinfo  {journal} {Chemistry of Materials}\ }\textbf {\bibinfo {volume}
  {34}},\ \bibinfo {pages} {8389} (\bibinfo {year} {2022})}\BibitemShut
  {NoStop}%
\bibitem [{\citenamefont {Berseneva}\ \emph {et~al.}(2022)\citenamefont
  {Berseneva}, \citenamefont {Klepov}, \citenamefont {Pal}, \citenamefont
  {Seeley}, \citenamefont {Koury}, \citenamefont {Schaeperkoetter},
  \citenamefont {Wright}, \citenamefont {Misture}, \citenamefont {Kanatzidis},
  \citenamefont {Wolverton},\ and\ \citenamefont
  {Gelis}}]{berseneva2022transuranium}%
  \BibitemOpen
  \bibfield  {author} {\bibinfo {author} {\bibfnamefont {A.~A.}\ \bibnamefont
  {Berseneva}}, \bibinfo {author} {\bibfnamefont {V.~V.}\ \bibnamefont
  {Klepov}}, \bibinfo {author} {\bibfnamefont {K.}~\bibnamefont {Pal}},
  \bibinfo {author} {\bibfnamefont {K.}~\bibnamefont {Seeley}}, \bibinfo
  {author} {\bibfnamefont {D.}~\bibnamefont {Koury}}, \bibinfo {author}
  {\bibfnamefont {J.}~\bibnamefont {Schaeperkoetter}}, \bibinfo {author}
  {\bibfnamefont {J.~T.}\ \bibnamefont {Wright}}, \bibinfo {author}
  {\bibfnamefont {S.~T.}\ \bibnamefont {Misture}}, \bibinfo {author}
  {\bibfnamefont {M.~G.}\ \bibnamefont {Kanatzidis}}, \bibinfo {author}
  {\bibfnamefont {C.}~\bibnamefont {Wolverton}},\ and\ \bibinfo {author}
  {\bibfnamefont {H.-C.~z.}\ \bibnamefont {Gelis}, \bibfnamefont {Artem
  V.and~Loye}},\ }\bibfield  {title} {\bibinfo {title} {Transuranium sulfide
  via the boron chalcogen mixture method and reversible water uptake in the
  {N}a{C}u{T}{S}$_3$ family},\ }\href
  {https://pubs.acs.org/doi/full/10.1021/jacs.2c04783} {\bibfield  {journal}
  {\bibinfo  {journal} {Journal of the American Chemical Society}\ }\textbf
  {\bibinfo {volume} {144}},\ \bibinfo {pages} {13773} (\bibinfo {year}
  {2022})}\BibitemShut {NoStop}%
\bibitem [{\citenamefont {Ishtiyak}\ \emph {et~al.}(2021)\citenamefont
  {Ishtiyak}, \citenamefont {Jana}, \citenamefont {Karthikeyan}, \citenamefont
  {Ramesh}, \citenamefont {Tripathy}, \citenamefont {Malladi}, \citenamefont
  {Niranjan},\ and\ \citenamefont {Prakash}}]{ishtiyak2021syntheses}%
  \BibitemOpen
  \bibfield  {author} {\bibinfo {author} {\bibfnamefont {M.}~\bibnamefont
  {Ishtiyak}}, \bibinfo {author} {\bibfnamefont {S.}~\bibnamefont {Jana}},
  \bibinfo {author} {\bibfnamefont {R.}~\bibnamefont {Karthikeyan}}, \bibinfo
  {author} {\bibfnamefont {M.}~\bibnamefont {Ramesh}}, \bibinfo {author}
  {\bibfnamefont {B.}~\bibnamefont {Tripathy}}, \bibinfo {author}
  {\bibfnamefont {S.~K.}\ \bibnamefont {Malladi}}, \bibinfo {author}
  {\bibfnamefont {M.~K.}\ \bibnamefont {Niranjan}},\ and\ \bibinfo {author}
  {\bibfnamefont {J.}~\bibnamefont {Prakash}},\ }\bibfield  {title} {\bibinfo
  {title} {Syntheses of five new layered quaternary chalcogenides
  {S}r{S}c{C}u{S}e$_3$, {S}r{S}c{C}u{T}e$_3$, {B}a{S}c{C}u{S}e$_3$,
  {B}a{S}c{C}u{T}e$_3$, and {B}a{S}c{A}g{T}e$_3$: Crystal structures,
  thermoelectric properties, and electronic structures},\ }\href
  {https://pubs.rsc.org/en/content/articlehtml/2021/qi/d1qi00717c} {\bibfield
  {journal} {\bibinfo  {journal} {Inorganic Chemistry Frontiers}\ }\textbf
  {\bibinfo {volume} {8}},\ \bibinfo {pages} {4086} (\bibinfo {year}
  {2021})}\BibitemShut {NoStop}%
\bibitem [{\citenamefont {Grigoriev}\ \emph {et~al.}(2022)\citenamefont
  {Grigoriev}, \citenamefont {Solovyov}, \citenamefont {Ruseikina},
  \citenamefont {Aleksandrovsky}, \citenamefont {Chernyshev}, \citenamefont
  {Velikanov}, \citenamefont {Garmonov}, \citenamefont {Molokeev},
  \citenamefont {Oreshonkov}, \citenamefont {Shestakov}, \citenamefont
  {Matigorov}, \citenamefont {Volkova}, \citenamefont {Ostapchuk},
  \citenamefont {Kertman}, \citenamefont {Schleid},\ and\ \citenamefont
  {Safin}}]{grigoriev2022quaternary}%
  \BibitemOpen
  \bibfield  {author} {\bibinfo {author} {\bibfnamefont {M.~V.}\ \bibnamefont
  {Grigoriev}}, \bibinfo {author} {\bibfnamefont {L.~A.}\ \bibnamefont
  {Solovyov}}, \bibinfo {author} {\bibfnamefont {A.~V.}\ \bibnamefont
  {Ruseikina}}, \bibinfo {author} {\bibfnamefont {A.~S.}\ \bibnamefont
  {Aleksandrovsky}}, \bibinfo {author} {\bibfnamefont {V.~A.}\ \bibnamefont
  {Chernyshev}}, \bibinfo {author} {\bibfnamefont {D.~A.}\ \bibnamefont
  {Velikanov}}, \bibinfo {author} {\bibfnamefont {A.~A.}\ \bibnamefont
  {Garmonov}}, \bibinfo {author} {\bibfnamefont {M.~S.}\ \bibnamefont
  {Molokeev}}, \bibinfo {author} {\bibfnamefont {A.~S.}\ \bibnamefont
  {Oreshonkov}}, \bibinfo {author} {\bibfnamefont {N.~P.}\ \bibnamefont
  {Shestakov}}, \bibinfo {author} {\bibfnamefont {A.~V.}\ \bibnamefont
  {Matigorov}}, \bibinfo {author} {\bibfnamefont {S.~S.}\ \bibnamefont
  {Volkova}}, \bibinfo {author} {\bibfnamefont {E.~A.}\ \bibnamefont
  {Ostapchuk}}, \bibinfo {author} {\bibfnamefont {A.~V.}\ \bibnamefont
  {Kertman}}, \bibinfo {author} {\bibfnamefont {T.}~\bibnamefont {Schleid}},\
  and\ \bibinfo {author} {\bibfnamefont {D.~A.}\ \bibnamefont {Safin}},\
  }\bibfield  {title} {\bibinfo {title} {Quaternary selenides
  {E}u{L}n{C}u{S}e$_3$: synthesis, structures, properties and in silico
  studies},\ }\href {https://www.mdpi.com/1422-0067/23/3/1503} {\bibfield
  {journal} {\bibinfo  {journal} {International Journal of Molecular Sciences}\
  }\textbf {\bibinfo {volume} {23}},\ \bibinfo {pages} {1503} (\bibinfo {year}
  {2022})}\BibitemShut {NoStop}%
\bibitem [{\citenamefont {Eickmeier}\ \emph {et~al.}(2022)\citenamefont
  {Eickmeier}, \citenamefont {Poschkamp}, \citenamefont {Dronskowski},\ and\
  \citenamefont {Steinberg}}]{eickmeier2022exploring}%
  \BibitemOpen
  \bibfield  {author} {\bibinfo {author} {\bibfnamefont {K.}~\bibnamefont
  {Eickmeier}}, \bibinfo {author} {\bibfnamefont {R.}~\bibnamefont
  {Poschkamp}}, \bibinfo {author} {\bibfnamefont {R.}~\bibnamefont
  {Dronskowski}},\ and\ \bibinfo {author} {\bibfnamefont {S.}~\bibnamefont
  {Steinberg}},\ }\bibfield  {title} {\bibinfo {title} {Exploring the impact of
  lone pairs on the structural features of alkaline-earth ({A})
  transition-metal ({M, M’}) chalcogenides ({Q}) {AMM'}{Q}$_3$},\ }\href
  {https://chemistry-europe.onlinelibrary.wiley.com/doi/full/10.1002/ejic.202200360}
  {\bibfield  {journal} {\bibinfo  {journal} {European Journal of Inorganic
  Chemistry}\ }\textbf {\bibinfo {volume} {2022}},\ \bibinfo {pages}
  {e202200360} (\bibinfo {year} {2022})}\BibitemShut {NoStop}%
\bibitem [{\citenamefont {Ruseikina}\ \emph {et~al.}(2023)\citenamefont
  {Ruseikina}, \citenamefont {Grigoriev}, \citenamefont {Molokeev},
  \citenamefont {Garmonov}, \citenamefont {Elyshev}, \citenamefont {Locke},\
  and\ \citenamefont {Schleid}}]{ruseikina2023synthesis}%
  \BibitemOpen
  \bibfield  {author} {\bibinfo {author} {\bibfnamefont {A.~V.}\ \bibnamefont
  {Ruseikina}}, \bibinfo {author} {\bibfnamefont {M.~V.}\ \bibnamefont
  {Grigoriev}}, \bibinfo {author} {\bibfnamefont {M.~S.}\ \bibnamefont
  {Molokeev}}, \bibinfo {author} {\bibfnamefont {A.~A.}\ \bibnamefont
  {Garmonov}}, \bibinfo {author} {\bibfnamefont {A.~V.}\ \bibnamefont
  {Elyshev}}, \bibinfo {author} {\bibfnamefont {R.~J.}\ \bibnamefont {Locke}},\
  and\ \bibinfo {author} {\bibfnamefont {T.}~\bibnamefont {Schleid}},\
  }\bibfield  {title} {\bibinfo {title} {Synthesis, crystal structure and
  properties of the new laminar quaternary tellurides sr ln {C}u{T}e$_3$ ({L}n
  = {S}m, {G}d--{T}m and {L}u)},\ }\href
  {https://www.mdpi.com/2073-4352/13/2/291} {\bibfield  {journal} {\bibinfo
  {journal} {Crystals}\ }\textbf {\bibinfo {volume} {13}},\ \bibinfo {pages}
  {291} (\bibinfo {year} {2023})}\BibitemShut {NoStop}%
\bibitem [{\citenamefont {Grigoriev}\ \emph {et~al.}(2023)\citenamefont
  {Grigoriev}, \citenamefont {Ruseikina}, \citenamefont {Chernyshev},
  \citenamefont {Oreshonkov}, \citenamefont {Garmonov}, \citenamefont
  {Molokeev}, \citenamefont {Locke}, \citenamefont {Elyshev},\ and\
  \citenamefont {Schleid}}]{grigoriev2023single}%
  \BibitemOpen
  \bibfield  {author} {\bibinfo {author} {\bibfnamefont {M.~V.}\ \bibnamefont
  {Grigoriev}}, \bibinfo {author} {\bibfnamefont {A.~V.}\ \bibnamefont
  {Ruseikina}}, \bibinfo {author} {\bibfnamefont {V.~A.}\ \bibnamefont
  {Chernyshev}}, \bibinfo {author} {\bibfnamefont {A.~S.}\ \bibnamefont
  {Oreshonkov}}, \bibinfo {author} {\bibfnamefont {A.~A.}\ \bibnamefont
  {Garmonov}}, \bibinfo {author} {\bibfnamefont {M.~S.}\ \bibnamefont
  {Molokeev}}, \bibinfo {author} {\bibfnamefont {R.~J.}\ \bibnamefont {Locke}},
  \bibinfo {author} {\bibfnamefont {A.~V.}\ \bibnamefont {Elyshev}},\ and\
  \bibinfo {author} {\bibfnamefont {T.}~\bibnamefont {Schleid}},\ }\bibfield
  {title} {\bibinfo {title} {Single crystals of {E}u{S}c{C}u{S}e$_3$:
  synthesis, experimental and dft investigations},\ }\href
  {https://www.mdpi.com/1996-1944/16/4/1555} {\bibfield  {journal} {\bibinfo
  {journal} {Materials}\ }\textbf {\bibinfo {volume} {16}},\ \bibinfo {pages}
  {1555} (\bibinfo {year} {2023})}\BibitemShut {NoStop}%
\bibitem [{\citenamefont {Ruseikina}\ \emph
  {et~al.}(2024{\natexlab{a}})\citenamefont {Ruseikina}, \citenamefont
  {Grigoriev}, \citenamefont {Locke}, \citenamefont {Chernyshev},\ and\
  \citenamefont {Schleid}}]{ruseikina2024syntheses}%
  \BibitemOpen
  \bibfield  {author} {\bibinfo {author} {\bibfnamefont {A.~V.}\ \bibnamefont
  {Ruseikina}}, \bibinfo {author} {\bibfnamefont {M.~V.}\ \bibnamefont
  {Grigoriev}}, \bibinfo {author} {\bibfnamefont {R.~J.}\ \bibnamefont
  {Locke}}, \bibinfo {author} {\bibfnamefont {V.~A.}\ \bibnamefont
  {Chernyshev}},\ and\ \bibinfo {author} {\bibfnamefont {T.}~\bibnamefont
  {Schleid}},\ }\bibfield  {title} {\bibinfo {title} {Syntheses and patterns of
  changes in structural parameters of the new quaternary tellurides
  {E}u{RE}{C}u{T}e$_3$ ({RE} = {H}o, {T}m, and {S}c): Experiment and theory},\
  }\href {https://www.mdpi.com/1996-1944/17/14/3378} {\bibfield  {journal}
  {\bibinfo  {journal} {Materials}\ }\textbf {\bibinfo {volume} {17}},\
  \bibinfo {pages} {3378} (\bibinfo {year} {2024}{\natexlab{a}})}\BibitemShut
  {NoStop}%
\bibitem [{\citenamefont {Ruseikina}\ \emph
  {et~al.}(2024{\natexlab{b}})\citenamefont {Ruseikina}, \citenamefont
  {Pinigina}, \citenamefont {Grigoriev},\ and\ \citenamefont
  {Safin}}]{ruseikina2024elucidating}%
  \BibitemOpen
  \bibfield  {author} {\bibinfo {author} {\bibfnamefont {A.~V.}\ \bibnamefont
  {Ruseikina}}, \bibinfo {author} {\bibfnamefont {A.~N.}\ \bibnamefont
  {Pinigina}}, \bibinfo {author} {\bibfnamefont {M.~V.}\ \bibnamefont
  {Grigoriev}},\ and\ \bibinfo {author} {\bibfnamefont {D.~A.}\ \bibnamefont
  {Safin}},\ }\bibfield  {title} {\bibinfo {title} {Elucidating a series of the
  quaternary selenides {B}a{RE}{C}u{S}e$_3$ with a rich library of optical
  properties},\ }\href {https://pubs.acs.org/doi/full/10.1021/acs.cgd.3c01479}
  {\bibfield  {journal} {\bibinfo  {journal} {Crystal Growth \& Design}\
  }\textbf {\bibinfo {volume} {24}},\ \bibinfo {pages} {2485} (\bibinfo {year}
  {2024}{\natexlab{b}})}\BibitemShut {NoStop}%
\bibitem [{\citenamefont {Pal}\ \emph {et~al.}(2019{\natexlab{c}})\citenamefont
  {Pal}, \citenamefont {Hua}, \citenamefont {Xia},\ and\ \citenamefont
  {Wolverton}}]{pal2019unraveling}%
  \BibitemOpen
  \bibfield  {author} {\bibinfo {author} {\bibfnamefont {K.}~\bibnamefont
  {Pal}}, \bibinfo {author} {\bibfnamefont {X.}~\bibnamefont {Hua}}, \bibinfo
  {author} {\bibfnamefont {Y.}~\bibnamefont {Xia}},\ and\ \bibinfo {author}
  {\bibfnamefont {C.}~\bibnamefont {Wolverton}},\ }\bibfield  {title} {\bibinfo
  {title} {Unraveling the structure-valence-property relationships in
  {AMM'}{Q}$_3$ chalcogenides with promising thermoelectric performance},\
  }\href {https://doi.org/10.1021/acsaem.9b02139} {\bibfield  {journal}
  {\bibinfo  {journal} {ACS Applied Energy Materials}\ }\textbf {\bibinfo
  {volume} {3}},\ \bibinfo {pages} {2110} (\bibinfo {year}
  {2019}{\natexlab{c}})}\BibitemShut {NoStop}%
\bibitem [{\citenamefont {Yu}\ \emph {et~al.}(2024)\citenamefont {Yu},
  \citenamefont {Wang}, \citenamefont {Cai}, \citenamefont {Guo}, \citenamefont
  {Lin}, \citenamefont {Li}, \citenamefont {Xing}, \citenamefont {Zhang},
  \citenamefont {Yang},\ and\ \citenamefont {Zhao}}]{yu2024substitution}%
  \BibitemOpen
  \bibfield  {author} {\bibinfo {author} {\bibfnamefont {X.}~\bibnamefont
  {Yu}}, \bibinfo {author} {\bibfnamefont {Z.}~\bibnamefont {Wang}}, \bibinfo
  {author} {\bibfnamefont {P.}~\bibnamefont {Cai}}, \bibinfo {author}
  {\bibfnamefont {K.}~\bibnamefont {Guo}}, \bibinfo {author} {\bibfnamefont
  {J.}~\bibnamefont {Lin}}, \bibinfo {author} {\bibfnamefont {S.}~\bibnamefont
  {Li}}, \bibinfo {author} {\bibfnamefont {J.}~\bibnamefont {Xing}}, \bibinfo
  {author} {\bibfnamefont {J.}~\bibnamefont {Zhang}}, \bibinfo {author}
  {\bibfnamefont {X.}~\bibnamefont {Yang}},\ and\ \bibinfo {author}
  {\bibfnamefont {J.-T.}\ \bibnamefont {Zhao}},\ }\bibfield  {title} {\bibinfo
  {title} {The substitution of rare-earth {G}d in {B}a{S}c{C}u{T}e$_3$
  realizing the band degeneracy and the point-defect scattering toward enhanced
  thermoelectric performance},\ }\href
  {https://pubs.acs.org/doi/full/10.1021/acs.inorgchem.3c03955} {\bibfield
  {journal} {\bibinfo  {journal} {Inorganic Chemistry}\ }\textbf {\bibinfo
  {volume} {63}},\ \bibinfo {pages} {20093} (\bibinfo {year}
  {2024})}\BibitemShut {NoStop}%
\bibitem [{sup()}]{supmat}%
  \BibitemOpen
  \href@noop {} {\bibinfo {title} {See {S}upplemental {M}aterial at [url will
  be inserted by publisher] for the calculated properties of {AMM'Q}$_3$
  compounds and experimentally known photovoltaic materials. {I}t includes
  {T}ables {SI}--{SVII}, {F}igures {S}1--{S}11, and convergence
  tests.}}\BibitemShut {Stop}%
\bibitem [{\citenamefont {Fabini}\ \emph {et~al.}(2019)\citenamefont {Fabini},
  \citenamefont {Koerner},\ and\ \citenamefont
  {Seshadri}}]{fabini2019candidate}%
  \BibitemOpen
  \bibfield  {author} {\bibinfo {author} {\bibfnamefont {D.~H.}\ \bibnamefont
  {Fabini}}, \bibinfo {author} {\bibfnamefont {M.}~\bibnamefont {Koerner}},\
  and\ \bibinfo {author} {\bibfnamefont {R.}~\bibnamefont {Seshadri}},\
  }\bibfield  {title} {\bibinfo {title} {Candidate inorganic photovoltaic
  materials from electronic structure-based optical absorption and charge
  transport proxies},\ }\href {https://doi.org/10.1021/acs.chemmater.8b04542}
  {\bibfield  {journal} {\bibinfo  {journal} {Chemistry of Materials}\ }\textbf
  {\bibinfo {volume} {31}},\ \bibinfo {pages} {1561} (\bibinfo {year}
  {2019})}\BibitemShut {NoStop}%
\bibitem [{\citenamefont {Kresse}\ and\ \citenamefont
  {Furthm{\"u}ller}(1996)}]{kresse1996efficiency}%
  \BibitemOpen
  \bibfield  {author} {\bibinfo {author} {\bibfnamefont {G.}~\bibnamefont
  {Kresse}}\ and\ \bibinfo {author} {\bibfnamefont {J.}~\bibnamefont
  {Furthm{\"u}ller}},\ }\bibfield  {title} {\bibinfo {title} {Efficiency of
  ab-initio total energy calculations for metals and semiconductors using a
  plane-wave basis set},\ }\href
  {https://www.sciencedirect.com/science/article/pii/0927025696000080}
  {\bibfield  {journal} {\bibinfo  {journal} {Comput. Mater. Sci.}\ }\textbf
  {\bibinfo {volume} {6}},\ \bibinfo {pages} {15} (\bibinfo {year}
  {1996})}\BibitemShut {NoStop}%
\bibitem [{\citenamefont {Kresse}\ and\ \citenamefont
  {Joubert}(1999)}]{kresse1999ultrasoft}%
  \BibitemOpen
  \bibfield  {author} {\bibinfo {author} {\bibfnamefont {G.}~\bibnamefont
  {Kresse}}\ and\ \bibinfo {author} {\bibfnamefont {D.}~\bibnamefont
  {Joubert}},\ }\bibfield  {title} {\bibinfo {title} {From ultrasoft
  pseudopotentials to the projector augmented-wave method},\ }\href
  {https://journals.aps.org/prb/abstract/10.1103/PhysRevB.59.1758} {\bibfield
  {journal} {\bibinfo  {journal} {Physical Review B}\ }\textbf {\bibinfo
  {volume} {59}},\ \bibinfo {pages} {1758} (\bibinfo {year}
  {1999})}\BibitemShut {NoStop}%
\bibitem [{\citenamefont {Perdew}\ \emph {et~al.}(1996)\citenamefont {Perdew},
  \citenamefont {Burke},\ and\ \citenamefont
  {Ernzerhof}}]{perdew1996generalized}%
  \BibitemOpen
  \bibfield  {author} {\bibinfo {author} {\bibfnamefont {J.~P.}\ \bibnamefont
  {Perdew}}, \bibinfo {author} {\bibfnamefont {K.}~\bibnamefont {Burke}},\ and\
  \bibinfo {author} {\bibfnamefont {M.}~\bibnamefont {Ernzerhof}},\ }\bibfield
  {title} {\bibinfo {title} {Generalized gradient approximation made simple},\
  }\href {https://journals.aps.org/prl/abstract/10.1103/PhysRevLett.77.3865}
  {\bibfield  {journal} {\bibinfo  {journal} {Phys. Rev. Lett.}\ }\textbf
  {\bibinfo {volume} {77}},\ \bibinfo {pages} {3865} (\bibinfo {year}
  {1996})}\BibitemShut {NoStop}%
\bibitem [{\citenamefont {Setyawan}\ and\ \citenamefont
  {Curtarolo}(2010)}]{setyawan2010high}%
  \BibitemOpen
  \bibfield  {author} {\bibinfo {author} {\bibfnamefont {W.}~\bibnamefont
  {Setyawan}}\ and\ \bibinfo {author} {\bibfnamefont {S.}~\bibnamefont
  {Curtarolo}},\ }\bibfield  {title} {\bibinfo {title} {High-throughput
  electronic band structure calculations: Challenges and tools},\ }\href@noop
  {} {\bibfield  {journal} {\bibinfo  {journal} {Computational materials
  science}\ }\textbf {\bibinfo {volume} {49}},\ \bibinfo {pages} {299}
  (\bibinfo {year} {2010})}\BibitemShut {NoStop}%
\bibitem [{\citenamefont {Heyd}\ \emph {et~al.}(2003)\citenamefont {Heyd},
  \citenamefont {Scuseria},\ and\ \citenamefont {Ernzerhof}}]{heyd2003hybrid}%
  \BibitemOpen
  \bibfield  {author} {\bibinfo {author} {\bibfnamefont {J.}~\bibnamefont
  {Heyd}}, \bibinfo {author} {\bibfnamefont {G.~E.}\ \bibnamefont {Scuseria}},\
  and\ \bibinfo {author} {\bibfnamefont {M.}~\bibnamefont {Ernzerhof}},\
  }\bibfield  {title} {\bibinfo {title} {Hybrid functionals based on a screened
  coulomb potential},\ }\href
  {https://pubs.aip.org/aip/jcp/article/118/18/8207/460359/Hybrid-functionals-based-on-a-screened-Coulomb}
  {\bibfield  {journal} {\bibinfo  {journal} {J. Chem. Phys.}\ }\textbf
  {\bibinfo {volume} {118}},\ \bibinfo {pages} {8207} (\bibinfo {year}
  {2003})}\BibitemShut {NoStop}%
\bibitem [{\citenamefont {Ganose}\ \emph {et~al.}(2018)\citenamefont {Ganose},
  \citenamefont {Jackson},\ and\ \citenamefont {Scanlon}}]{ganose2018sumo}%
  \BibitemOpen
  \bibfield  {author} {\bibinfo {author} {\bibfnamefont {A.~M.}\ \bibnamefont
  {Ganose}}, \bibinfo {author} {\bibfnamefont {A.~J.}\ \bibnamefont
  {Jackson}},\ and\ \bibinfo {author} {\bibfnamefont {D.~O.}\ \bibnamefont
  {Scanlon}},\ }\bibfield  {title} {\bibinfo {title} {sumo: Command-line tools
  for plotting and analysis of periodic ab initio calculations},\ }\href
  {https://doi.org/10.21105/joss.00717} {\bibfield  {journal} {\bibinfo
  {journal} {Journal of Open Source Software}\ }\textbf {\bibinfo {volume}
  {3}},\ \bibinfo {pages} {717} (\bibinfo {year} {2018})}\BibitemShut {NoStop}%
\bibitem [{\citenamefont {Fox}(2010)}]{fox2010optical}%
  \BibitemOpen
  \bibfield  {author} {\bibinfo {author} {\bibfnamefont {M.}~\bibnamefont
  {Fox}},\ }\href
  {https://books.google.co.in/books?hl=en&lr=&id=5WkVDAAAQBAJ&oi=fnd&pg=PP1&dq=M.+Fox,+Optical+Properties+of+Solids,+2nd+ed.,+Oxford+Master+Series+in+Physics+(Oxford+University+Press,+London,+2010).&ots=v54ItgdvRB&sig=wxxBTDby72lA3s2nB86k5kr4VHQ&redir_esc=y#v=onepage&q&f=false}
  {\emph {\bibinfo {title} {Optical properties of solids}}},\ Vol.~\bibinfo
  {volume} {3}\ (\bibinfo  {publisher} {Oxford university press},\ \bibinfo
  {year} {2010})\BibitemShut {NoStop}%
\bibitem [{\citenamefont {Grunert}\ \emph {et~al.}(2024)\citenamefont
  {Grunert}, \citenamefont {Gro{\ss}mann},\ and\ \citenamefont
  {Runge}}]{grunert2024predicting}%
  \BibitemOpen
  \bibfield  {author} {\bibinfo {author} {\bibfnamefont {M.}~\bibnamefont
  {Grunert}}, \bibinfo {author} {\bibfnamefont {M.}~\bibnamefont
  {Gro{\ss}mann}},\ and\ \bibinfo {author} {\bibfnamefont {E.}~\bibnamefont
  {Runge}},\ }\bibfield  {title} {\bibinfo {title} {Predicting exciton binding
  energies from ground-state properties},\ }\href
  {https://journals.aps.org/prb/abstract/10.1103/PhysRevB.110.075204}
  {\bibfield  {journal} {\bibinfo  {journal} {Physical Review B}\ }\textbf
  {\bibinfo {volume} {110}},\ \bibinfo {pages} {075204} (\bibinfo {year}
  {2024})}\BibitemShut {NoStop}%
\bibitem [{\citenamefont {Wu}\ \emph {et~al.}(2005)\citenamefont {Wu},
  \citenamefont {Vanderbilt},\ and\ \citenamefont {Hamann}}]{wu2005systematic}%
  \BibitemOpen
  \bibfield  {author} {\bibinfo {author} {\bibfnamefont {X.}~\bibnamefont
  {Wu}}, \bibinfo {author} {\bibfnamefont {D.}~\bibnamefont {Vanderbilt}},\
  and\ \bibinfo {author} {\bibfnamefont {D.}~\bibnamefont {Hamann}},\
  }\bibfield  {title} {\bibinfo {title} {Systematic treatment of displacements,
  strains, and electric fields in density-functional perturbation theory},\
  }\href {https://doi.org/10.1103/PhysRevB.72.035105} {\bibfield  {journal}
  {\bibinfo  {journal} {Physical Review B—Condensed Matter and Materials
  Physics}\ }\textbf {\bibinfo {volume} {72}},\ \bibinfo {pages} {035105}
  (\bibinfo {year} {2005})}\BibitemShut {NoStop}%
\bibitem [{\citenamefont {Gajdo{\v{s}}}\ \emph {et~al.}(2006)\citenamefont
  {Gajdo{\v{s}}}, \citenamefont {Hummer}, \citenamefont {Kresse}, \citenamefont
  {Furthm{\"u}ller},\ and\ \citenamefont {Bechstedt}}]{gajdovs2006linear}%
  \BibitemOpen
  \bibfield  {author} {\bibinfo {author} {\bibfnamefont {M.}~\bibnamefont
  {Gajdo{\v{s}}}}, \bibinfo {author} {\bibfnamefont {K.}~\bibnamefont
  {Hummer}}, \bibinfo {author} {\bibfnamefont {G.}~\bibnamefont {Kresse}},
  \bibinfo {author} {\bibfnamefont {J.}~\bibnamefont {Furthm{\"u}ller}},\ and\
  \bibinfo {author} {\bibfnamefont {F.}~\bibnamefont {Bechstedt}},\ }\bibfield
  {title} {\bibinfo {title} {Linear optical properties in the
  projector-augmented wave methodology},\ }\href
  {https://doi.org/10.1103/PhysRevB.73.045112} {\bibfield  {journal} {\bibinfo
  {journal} {Physical Review B—Condensed Matter and Materials Physics}\
  }\textbf {\bibinfo {volume} {73}},\ \bibinfo {pages} {045112} (\bibinfo
  {year} {2006})}\BibitemShut {NoStop}%
\bibitem [{\citenamefont {Togo}\ and\ \citenamefont {Tanaka}(2015)}]{phonopy}%
  \BibitemOpen
  \bibfield  {author} {\bibinfo {author} {\bibfnamefont {A.}~\bibnamefont
  {Togo}}\ and\ \bibinfo {author} {\bibfnamefont {I.}~\bibnamefont {Tanaka}},\
  }\bibfield  {title} {\bibinfo {title} {First principles phonon calculations
  in materials science},\ }\href
  {https://www.sciencedirect.com/science/article/pii/S1359646215003127}
  {\bibfield  {journal} {\bibinfo  {journal} {Scr. Mater.}\ }\textbf {\bibinfo
  {volume} {108}},\ \bibinfo {pages} {1} (\bibinfo {year} {2015})}\BibitemShut
  {NoStop}%
\bibitem [{\citenamefont {Wang}\ \emph {et~al.}(2019)\citenamefont {Wang},
  \citenamefont {Chen}, \citenamefont {Wei},\ and\ \citenamefont
  {Yin}}]{wang2019materials}%
  \BibitemOpen
  \bibfield  {author} {\bibinfo {author} {\bibfnamefont {J.}~\bibnamefont
  {Wang}}, \bibinfo {author} {\bibfnamefont {H.}~\bibnamefont {Chen}}, \bibinfo
  {author} {\bibfnamefont {S.-H.}\ \bibnamefont {Wei}},\ and\ \bibinfo {author}
  {\bibfnamefont {W.-J.}\ \bibnamefont {Yin}},\ }\bibfield  {title} {\bibinfo
  {title} {Materials design of solar cell absorbers beyond perovskites and
  conventional semiconductors via combining tetrahedral and octahedral
  coordination},\ }\href
  {https://advanced.onlinelibrary.wiley.com/doi/full/10.1002/adma.201806593}
  {\bibfield  {journal} {\bibinfo  {journal} {Advanced Materials}\ ,\ \bibinfo
  {pages} {1806593}} (\bibinfo {year} {2019})}\BibitemShut {NoStop}%
\bibitem [{\citenamefont {Yu}\ and\ \citenamefont
  {Zunger}(2012)}]{yu2012identification}%
  \BibitemOpen
  \bibfield  {author} {\bibinfo {author} {\bibfnamefont {L.}~\bibnamefont
  {Yu}}\ and\ \bibinfo {author} {\bibfnamefont {A.}~\bibnamefont {Zunger}},\
  }\bibfield  {title} {\bibinfo {title} {Identification of potential
  photovoltaic absorbers based on first-principles spectroscopic screening of
  materials},\ }\href {https://doi.org/10.1103/PhysRevLett.108.068701}
  {\bibfield  {journal} {\bibinfo  {journal} {Physical review letters}\
  }\textbf {\bibinfo {volume} {108}},\ \bibinfo {pages} {068701} (\bibinfo
  {year} {2012})}\BibitemShut {NoStop}%
\bibitem [{\citenamefont {Wang}\ \emph {et~al.}(2021)\citenamefont {Wang},
  \citenamefont {Xu}, \citenamefont {Liu}, \citenamefont {Tang},\ and\
  \citenamefont {Geng}}]{VASPKIT}%
  \BibitemOpen
  \bibfield  {author} {\bibinfo {author} {\bibfnamefont {V.}~\bibnamefont
  {Wang}}, \bibinfo {author} {\bibfnamefont {N.}~\bibnamefont {Xu}}, \bibinfo
  {author} {\bibfnamefont {J.-C.}\ \bibnamefont {Liu}}, \bibinfo {author}
  {\bibfnamefont {G.}~\bibnamefont {Tang}},\ and\ \bibinfo {author}
  {\bibfnamefont {W.-T.}\ \bibnamefont {Geng}},\ }\bibfield  {title} {\bibinfo
  {title} {Vaspkit: A user-friendly interface facilitating high-throughput
  computing and analysis using vasp code},\ }\href
  {https://doi.org/https://doi.org/10.1016/j.cpc.2021.108033} {\bibfield
  {journal} {\bibinfo  {journal} {Computer Physics Communications}\ }\textbf
  {\bibinfo {volume} {267}},\ \bibinfo {pages} {108033} (\bibinfo {year}
  {2021})}\BibitemShut {NoStop}%
\bibitem [{\citenamefont {Zhang}\ \emph {et~al.}(1998)\citenamefont {Zhang},
  \citenamefont {Wei}, \citenamefont {Zunger},\ and\ \citenamefont
  {Katayama-Yoshida}}]{zhang1998defect}%
  \BibitemOpen
  \bibfield  {author} {\bibinfo {author} {\bibfnamefont {S.}~\bibnamefont
  {Zhang}}, \bibinfo {author} {\bibfnamefont {S.-H.}\ \bibnamefont {Wei}},
  \bibinfo {author} {\bibfnamefont {A.}~\bibnamefont {Zunger}},\ and\ \bibinfo
  {author} {\bibfnamefont {H.}~\bibnamefont {Katayama-Yoshida}},\ }\bibfield
  {title} {\bibinfo {title} {Defect physics of the {C}u{I}n{S}e$_2$
  chalcopyrite semiconductor},\ }\href
  {https://doi.org/10.1103/PhysRevB.57.9642} {\bibfield  {journal} {\bibinfo
  {journal} {Physical Review B}\ }\textbf {\bibinfo {volume} {57}},\ \bibinfo
  {pages} {9642} (\bibinfo {year} {1998})}\BibitemShut {NoStop}%
\bibitem [{\citenamefont {Freysoldt}\ \emph {et~al.}(2014)\citenamefont
  {Freysoldt}, \citenamefont {Grabowski}, \citenamefont {Hickel}, \citenamefont
  {Neugebauer}, \citenamefont {Kresse}, \citenamefont {Janotti},\ and\
  \citenamefont {Van~de Walle}}]{freysoldt2014first}%
  \BibitemOpen
  \bibfield  {author} {\bibinfo {author} {\bibfnamefont {C.}~\bibnamefont
  {Freysoldt}}, \bibinfo {author} {\bibfnamefont {B.}~\bibnamefont
  {Grabowski}}, \bibinfo {author} {\bibfnamefont {T.}~\bibnamefont {Hickel}},
  \bibinfo {author} {\bibfnamefont {J.}~\bibnamefont {Neugebauer}}, \bibinfo
  {author} {\bibfnamefont {G.}~\bibnamefont {Kresse}}, \bibinfo {author}
  {\bibfnamefont {A.}~\bibnamefont {Janotti}},\ and\ \bibinfo {author}
  {\bibfnamefont {C.~G.}\ \bibnamefont {Van~de Walle}},\ }\bibfield  {title}
  {\bibinfo {title} {First-principles calculations for point defects in
  solids},\ }\href {https://doi.org/10.1103/RevModPhys.86.253} {\bibfield
  {journal} {\bibinfo  {journal} {Reviews of modern physics}\ }\textbf
  {\bibinfo {volume} {86}},\ \bibinfo {pages} {253} (\bibinfo {year}
  {2014})}\BibitemShut {NoStop}%
\bibitem [{\citenamefont {Akimov}\ and\ \citenamefont
  {Prezhdo}(2013)}]{Akimov2013}%
  \BibitemOpen
  \bibfield  {author} {\bibinfo {author} {\bibfnamefont {A.~V.}\ \bibnamefont
  {Akimov}}\ and\ \bibinfo {author} {\bibfnamefont {O.~V.}\ \bibnamefont
  {Prezhdo}},\ }\bibfield  {title} {\bibinfo {title} {The pyxaid program for
  non-adiabatic molecular dynamics in condensed matter systems},\ }\href
  {https://doi.org/10.1021/ct400641n} {\bibfield  {journal} {\bibinfo
  {journal} {Journal of Chemical Theory and Computation}\ }\textbf {\bibinfo
  {volume} {9}},\ \bibinfo {pages} {4959–4972} (\bibinfo {year}
  {2013})}\BibitemShut {NoStop}%
\bibitem [{\citenamefont {Akimov}(2016)}]{Akimov2016}%
  \BibitemOpen
  \bibfield  {author} {\bibinfo {author} {\bibfnamefont {A.~V.}\ \bibnamefont
  {Akimov}},\ }\bibfield  {title} {\bibinfo {title} {Libra: An open-source
  “methodology discovery” library for quantum and classical dynamics
  simulations},\ }\href {https://doi.org/10.1002/JCC.24367} {\bibfield
  {journal} {\bibinfo  {journal} {Journal of Computational Chemistry}\ }\textbf
  {\bibinfo {volume} {37}},\ \bibinfo {pages} {1626–1649} (\bibinfo {year}
  {2016})}\BibitemShut {NoStop}%
\bibitem [{\citenamefont {Hamm}(2005)}]{hamm2005principles}%
  \BibitemOpen
  \bibfield  {author} {\bibinfo {author} {\bibfnamefont {P.}~\bibnamefont
  {Hamm}},\ }\bibfield  {title} {\bibinfo {title} {Principles of nonlinear
  optical spectroscopy: A practical approach or: Mukamel for dummies},\ }\href
  {https://www.chem.uci.edu/~dmitryf/manuals/Fundamentals/Mukamel%20for%20dummies.pdf}
  {\bibfield  {journal} {\bibinfo  {journal} {University of Zurich}\ }\textbf
  {\bibinfo {volume} {41}},\ \bibinfo {pages} {77} (\bibinfo {year}
  {2005})}\BibitemShut {NoStop}%
\bibitem [{\citenamefont {Bouarissa}\ and\ \citenamefont
  {Aourag}(1999)}]{bouarissa1999effective}%
  \BibitemOpen
  \bibfield  {author} {\bibinfo {author} {\bibfnamefont {N.}~\bibnamefont
  {Bouarissa}}\ and\ \bibinfo {author} {\bibfnamefont {H.}~\bibnamefont
  {Aourag}},\ }\bibfield  {title} {\bibinfo {title} {Effective masses of
  electrons and heavy holes in {I}n{A}s, {I}n{S}b, {G}a{S}b, {G}a{A}s and some
  of their ternary compounds},\ }\href
  {https://doi.org/10.1016/S1350-4495(99)00020-1} {\bibfield  {journal}
  {\bibinfo  {journal} {Infrared physics \& technology}\ }\textbf {\bibinfo
  {volume} {40}},\ \bibinfo {pages} {343} (\bibinfo {year} {1999})}\BibitemShut
  {NoStop}%
\bibitem [{\citenamefont {Yang}\ \emph {et~al.}(2016)\citenamefont {Yang},
  \citenamefont {Peng}, \citenamefont {Sun},\ and\ \citenamefont
  {Perdew}}]{yang2016more}%
  \BibitemOpen
  \bibfield  {author} {\bibinfo {author} {\bibfnamefont {Z.-h.}\ \bibnamefont
  {Yang}}, \bibinfo {author} {\bibfnamefont {H.}~\bibnamefont {Peng}}, \bibinfo
  {author} {\bibfnamefont {J.}~\bibnamefont {Sun}},\ and\ \bibinfo {author}
  {\bibfnamefont {J.~P.}\ \bibnamefont {Perdew}},\ }\bibfield  {title}
  {\bibinfo {title} {More realistic band gaps from meta-generalized gradient
  approximations: Only in a generalized kohn-sham scheme},\ }\href
  {https://doi.org/10.1103/PhysRevB.93.205205} {\bibfield  {journal} {\bibinfo
  {journal} {Physical review B}\ }\textbf {\bibinfo {volume} {93}},\ \bibinfo
  {pages} {205205} (\bibinfo {year} {2016})}\BibitemShut {NoStop}%
\bibitem [{\citenamefont {Shabaev}\ \emph {et~al.}(2015)\citenamefont
  {Shabaev}, \citenamefont {Mehl},\ and\ \citenamefont
  {Efros}}]{shabaev2015energy}%
  \BibitemOpen
  \bibfield  {author} {\bibinfo {author} {\bibfnamefont {A.}~\bibnamefont
  {Shabaev}}, \bibinfo {author} {\bibfnamefont {M.}~\bibnamefont {Mehl}},\ and\
  \bibinfo {author} {\bibfnamefont {A.~L.}\ \bibnamefont {Efros}},\ }\bibfield
  {title} {\bibinfo {title} {Energy band structure of {C}u{I}n{S}$_2$ and
  optical spectra of cuins 2 nanocrystals},\ }\href
  {https://doi.org/10.1103/PhysRevB.92.035431} {\bibfield  {journal} {\bibinfo
  {journal} {Physical Review B}\ }\textbf {\bibinfo {volume} {92}},\ \bibinfo
  {pages} {035431} (\bibinfo {year} {2015})}\BibitemShut {NoStop}%
\bibitem [{\citenamefont {Sa}\ and\ \citenamefont
  {Liu}(2022)}]{sa2022unveiling}%
  \BibitemOpen
  \bibfield  {author} {\bibinfo {author} {\bibfnamefont {R.}~\bibnamefont
  {Sa}}\ and\ \bibinfo {author} {\bibfnamefont {D.}~\bibnamefont {Liu}},\
  }\bibfield  {title} {\bibinfo {title} {Unveiling the fundamental physical
  properties of {C}u$_{2-x}${N}a$_x${Z}n{S}n{X}$_4$ ({X} = {S}, {S}e) alloys
  for solar cell applications: a theoretical investigation},\ }\href
  {https://doi.org/10.1016/j.jmrt.2022.08.070} {\bibfield  {journal} {\bibinfo
  {journal} {journal of materials research and technology}\ }\textbf {\bibinfo
  {volume} {20}},\ \bibinfo {pages} {2680} (\bibinfo {year}
  {2022})}\BibitemShut {NoStop}%
\bibitem [{\citenamefont {Pandech}\ \emph {et~al.}(2020)\citenamefont
  {Pandech}, \citenamefont {Kongnok}, \citenamefont {Palakawong}, \citenamefont
  {Limpijumnong}, \citenamefont {Lambrecht},\ and\ \citenamefont
  {Jungthawan}}]{pandech2020effects}%
  \BibitemOpen
  \bibfield  {author} {\bibinfo {author} {\bibfnamefont {N.}~\bibnamefont
  {Pandech}}, \bibinfo {author} {\bibfnamefont {T.}~\bibnamefont {Kongnok}},
  \bibinfo {author} {\bibfnamefont {N.}~\bibnamefont {Palakawong}}, \bibinfo
  {author} {\bibfnamefont {S.}~\bibnamefont {Limpijumnong}}, \bibinfo {author}
  {\bibfnamefont {W.~R.}\ \bibnamefont {Lambrecht}},\ and\ \bibinfo {author}
  {\bibfnamefont {S.}~\bibnamefont {Jungthawan}},\ }\bibfield  {title}
  {\bibinfo {title} {Effects of the van der waals interactions on structural
  and electronic properties of {CH}$_3${NH}$_3$ ({P}b, {S}n)({I}, {B}r,
  {C}l)$_3$ halide perovskites},\ }\href
  {https://doi.org/10.1021/acsomega.0c03016} {\bibfield  {journal} {\bibinfo
  {journal} {ACS omega}\ }\textbf {\bibinfo {volume} {5}},\ \bibinfo {pages}
  {25723} (\bibinfo {year} {2020})}\BibitemShut {NoStop}%
\bibitem [{\citenamefont {Ghaithan}\ \emph {et~al.}(2020)\citenamefont
  {Ghaithan}, \citenamefont {Alahmed}, \citenamefont {Qaid}, \citenamefont
  {Hezam},\ and\ \citenamefont {Aldwayyan}}]{ghaithan2020density}%
  \BibitemOpen
  \bibfield  {author} {\bibinfo {author} {\bibfnamefont {H.~M.}\ \bibnamefont
  {Ghaithan}}, \bibinfo {author} {\bibfnamefont {Z.~A.}\ \bibnamefont
  {Alahmed}}, \bibinfo {author} {\bibfnamefont {S.~M.}\ \bibnamefont {Qaid}},
  \bibinfo {author} {\bibfnamefont {M.}~\bibnamefont {Hezam}},\ and\ \bibinfo
  {author} {\bibfnamefont {A.~S.}\ \bibnamefont {Aldwayyan}},\ }\bibfield
  {title} {\bibinfo {title} {Density functional study of cubic, tetragonal, and
  orthorhombic {C}s{P}b{B}r$_3$ perovskite},\ }\href
  {https://doi.org/10.1021/acsomega.0c00197} {\bibfield  {journal} {\bibinfo
  {journal} {ACS omega}\ }\textbf {\bibinfo {volume} {5}},\ \bibinfo {pages}
  {7468} (\bibinfo {year} {2020})}\BibitemShut {NoStop}%
\bibitem [{\citenamefont {Wang}\ \emph {et~al.}(2020)\citenamefont {Wang},
  \citenamefont {Xiao},\ and\ \citenamefont {Wang}}]{wang2020structural}%
  \BibitemOpen
  \bibfield  {author} {\bibinfo {author} {\bibfnamefont {S.}~\bibnamefont
  {Wang}}, \bibinfo {author} {\bibfnamefont {W.-b.}\ \bibnamefont {Xiao}},\
  and\ \bibinfo {author} {\bibfnamefont {F.}~\bibnamefont {Wang}},\ }\bibfield
  {title} {\bibinfo {title} {Structural, electronic, and optical properties of
  cubic formamidinium lead iodide perovskite: a first-principles
  investigation},\ }\href {https://doi.org/10.1039/D0RA06028C} {\bibfield
  {journal} {\bibinfo  {journal} {RSC advances}\ }\textbf {\bibinfo {volume}
  {10}},\ \bibinfo {pages} {32364} (\bibinfo {year} {2020})}\BibitemShut
  {NoStop}%
\bibitem [{\citenamefont {Wallace}\ \emph {et~al.}(2017)\citenamefont
  {Wallace}, \citenamefont {Svane}, \citenamefont {Huhn}, \citenamefont {Zhu},
  \citenamefont {Mitzi}, \citenamefont {Blum},\ and\ \citenamefont
  {Walsh}}]{wallace2017candidate}%
  \BibitemOpen
  \bibfield  {author} {\bibinfo {author} {\bibfnamefont {S.~K.}\ \bibnamefont
  {Wallace}}, \bibinfo {author} {\bibfnamefont {K.~L.}\ \bibnamefont {Svane}},
  \bibinfo {author} {\bibfnamefont {W.~P.}\ \bibnamefont {Huhn}}, \bibinfo
  {author} {\bibfnamefont {T.}~\bibnamefont {Zhu}}, \bibinfo {author}
  {\bibfnamefont {D.~B.}\ \bibnamefont {Mitzi}}, \bibinfo {author}
  {\bibfnamefont {V.}~\bibnamefont {Blum}},\ and\ \bibinfo {author}
  {\bibfnamefont {A.}~\bibnamefont {Walsh}},\ }\bibfield  {title} {\bibinfo
  {title} {Candidate photoferroic absorber materials for thin-film solar cells
  from naturally occurring minerals: enargite, stephanite, and bournonite},\
  }\href {https://pubs.rsc.org/en/content/articlehtml/2017/se/c7se00277g}
  {\bibfield  {journal} {\bibinfo  {journal} {Sustainable Energy \& Fuels}\
  }\textbf {\bibinfo {volume} {1}},\ \bibinfo {pages} {1339} (\bibinfo {year}
  {2017})}\BibitemShut {NoStop}%
\bibitem [{\citenamefont {Frost}\ \emph {et~al.}(2014)\citenamefont {Frost},
  \citenamefont {Butler}, \citenamefont {Brivio}, \citenamefont {Hendon},
  \citenamefont {Van~Schilfgaarde},\ and\ \citenamefont
  {Walsh}}]{frost2014atomistic}%
  \BibitemOpen
  \bibfield  {author} {\bibinfo {author} {\bibfnamefont {J.~M.}\ \bibnamefont
  {Frost}}, \bibinfo {author} {\bibfnamefont {K.~T.}\ \bibnamefont {Butler}},
  \bibinfo {author} {\bibfnamefont {F.}~\bibnamefont {Brivio}}, \bibinfo
  {author} {\bibfnamefont {C.~H.}\ \bibnamefont {Hendon}}, \bibinfo {author}
  {\bibfnamefont {M.}~\bibnamefont {Van~Schilfgaarde}},\ and\ \bibinfo {author}
  {\bibfnamefont {A.}~\bibnamefont {Walsh}},\ }\bibfield  {title} {\bibinfo
  {title} {Atomistic origins of high-performance in hybrid halide perovskite
  solar cells},\ }\href {https://pubs.acs.org/doi/10.1021/nl500390f} {\bibfield
   {journal} {\bibinfo  {journal} {Nano letters}\ }\textbf {\bibinfo {volume}
  {14}},\ \bibinfo {pages} {2584} (\bibinfo {year} {2014})}\BibitemShut
  {NoStop}%
\bibitem [{\citenamefont {Chen}\ \emph {et~al.}(2009)\citenamefont {Chen},
  \citenamefont {Gong}, \citenamefont {Walsh},\ and\ \citenamefont
  {Wei}}]{chen2009crystal}%
  \BibitemOpen
  \bibfield  {author} {\bibinfo {author} {\bibfnamefont {S.}~\bibnamefont
  {Chen}}, \bibinfo {author} {\bibfnamefont {X.}~\bibnamefont {Gong}}, \bibinfo
  {author} {\bibfnamefont {A.}~\bibnamefont {Walsh}},\ and\ \bibinfo {author}
  {\bibfnamefont {S.-H.}\ \bibnamefont {Wei}},\ }\bibfield  {title} {\bibinfo
  {title} {Crystal and electronic band structure of {C}u$_2${Z}n{S}n{X}$_4$({X}
  = {S} and {S}e) photovoltaic absorbers: First-principles insights},\ }\href
  {https://pubs.aip.org/aip/apl/article/94/4/041903/337567/Crystal-and-electronic-band-structure-of-Cu2ZnSnX4}
  {\bibfield  {journal} {\bibinfo  {journal} {Applied Physics Letters}\
  }\textbf {\bibinfo {volume} {94}},\ \bibinfo {pages} {041903} (\bibinfo
  {year} {2009})}\BibitemShut {NoStop}%
\bibitem [{\citenamefont {Eames}\ \emph {et~al.}(2015)\citenamefont {Eames},
  \citenamefont {Frost}, \citenamefont {Barnes}, \citenamefont {O’regan},
  \citenamefont {Walsh},\ and\ \citenamefont {Islam}}]{eames2015ionic}%
  \BibitemOpen
  \bibfield  {author} {\bibinfo {author} {\bibfnamefont {C.}~\bibnamefont
  {Eames}}, \bibinfo {author} {\bibfnamefont {J.~M.}\ \bibnamefont {Frost}},
  \bibinfo {author} {\bibfnamefont {P.~R.}\ \bibnamefont {Barnes}}, \bibinfo
  {author} {\bibfnamefont {B.~C.}\ \bibnamefont {O’regan}}, \bibinfo {author}
  {\bibfnamefont {A.}~\bibnamefont {Walsh}},\ and\ \bibinfo {author}
  {\bibfnamefont {M.~S.}\ \bibnamefont {Islam}},\ }\bibfield  {title} {\bibinfo
  {title} {Ionic transport in hybrid lead iodide perovskite solar cells},\
  }\href {https://www.nature.com/articles/ncomms8497} {\bibfield  {journal}
  {\bibinfo  {journal} {Nature communications}\ }\textbf {\bibinfo {volume}
  {6}},\ \bibinfo {pages} {7497} (\bibinfo {year} {2015})}\BibitemShut
  {NoStop}%
\bibitem [{\citenamefont {Xiao}\ \emph {et~al.}(2017)\citenamefont {Xiao},
  \citenamefont {Meng}, \citenamefont {Wang}, \citenamefont {Mitzi},\ and\
  \citenamefont {Yan}}]{xiao2017searching}%
  \BibitemOpen
  \bibfield  {author} {\bibinfo {author} {\bibfnamefont {Z.}~\bibnamefont
  {Xiao}}, \bibinfo {author} {\bibfnamefont {W.}~\bibnamefont {Meng}}, \bibinfo
  {author} {\bibfnamefont {J.}~\bibnamefont {Wang}}, \bibinfo {author}
  {\bibfnamefont {D.~B.}\ \bibnamefont {Mitzi}},\ and\ \bibinfo {author}
  {\bibfnamefont {Y.}~\bibnamefont {Yan}},\ }\bibfield  {title} {\bibinfo
  {title} {Searching for promising new perovskite-based photovoltaic absorbers:
  the importance of electronic dimensionality},\ }\href
  {https://pubs.rsc.org/en/content/articlehtml/2017/mh/c6mh00519e} {\bibfield
  {journal} {\bibinfo  {journal} {Materials Horizons}\ }\textbf {\bibinfo
  {volume} {4}},\ \bibinfo {pages} {206} (\bibinfo {year} {2017})}\BibitemShut
  {NoStop}%
\bibitem [{\citenamefont {Zhu}\ \emph {et~al.}(2017)\citenamefont {Zhu},
  \citenamefont {Huhn}, \citenamefont {Wessler}, \citenamefont {Shin},
  \citenamefont {Saparov}, \citenamefont {Mitzi},\ and\ \citenamefont
  {Blum}}]{zhu2017i2}%
  \BibitemOpen
  \bibfield  {author} {\bibinfo {author} {\bibfnamefont {T.}~\bibnamefont
  {Zhu}}, \bibinfo {author} {\bibfnamefont {W.~P.}\ \bibnamefont {Huhn}},
  \bibinfo {author} {\bibfnamefont {G.~C.}\ \bibnamefont {Wessler}}, \bibinfo
  {author} {\bibfnamefont {D.}~\bibnamefont {Shin}}, \bibinfo {author}
  {\bibfnamefont {B.}~\bibnamefont {Saparov}}, \bibinfo {author} {\bibfnamefont
  {D.~B.}\ \bibnamefont {Mitzi}},\ and\ \bibinfo {author} {\bibfnamefont
  {V.}~\bibnamefont {Blum}},\ }\bibfield  {title} {\bibinfo {title}
  {{I}$_2$--{II}--{IV}--{VI}$_4$ ({I} = {C}u, {A}g; {II} = {S}r, {B}a; {IV} =
  {G}e, {S}n; {VI} = {S}, {S}e): Chalcogenides for thin-film photovoltaics},\
  }\href {https://pubs.acs.org/doi/10.1021/acs.chemmater.7b02638} {\bibfield
  {journal} {\bibinfo  {journal} {Chemistry of Materials}\ }\textbf {\bibinfo
  {volume} {29}},\ \bibinfo {pages} {7868} (\bibinfo {year}
  {2017})}\BibitemShut {NoStop}%
\bibitem [{\citenamefont {Brandt}\ \emph {et~al.}(2015)\citenamefont {Brandt},
  \citenamefont {Stevanovi{\'c}}, \citenamefont {Ginley},\ and\ \citenamefont
  {Buonassisi}}]{brandt2015identifying}%
  \BibitemOpen
  \bibfield  {author} {\bibinfo {author} {\bibfnamefont {R.~E.}\ \bibnamefont
  {Brandt}}, \bibinfo {author} {\bibfnamefont {V.}~\bibnamefont
  {Stevanovi{\'c}}}, \bibinfo {author} {\bibfnamefont {D.~S.}\ \bibnamefont
  {Ginley}},\ and\ \bibinfo {author} {\bibfnamefont {T.}~\bibnamefont
  {Buonassisi}},\ }\bibfield  {title} {\bibinfo {title} {Identifying
  defect-tolerant semiconductors with high minority-carrier lifetimes: beyond
  hybrid lead halide perovskites},\ }\href
  {https://link.springer.com/article/10.1557/mrc.2015.26} {\bibfield  {journal}
  {\bibinfo  {journal} {Mrs Communications}\ }\textbf {\bibinfo {volume} {5}},\
  \bibinfo {pages} {265} (\bibinfo {year} {2015})}\BibitemShut {NoStop}%
\bibitem [{\citenamefont {Brivio}\ \emph {et~al.}(2013)\citenamefont {Brivio},
  \citenamefont {Walker},\ and\ \citenamefont {Walsh}}]{brivio2013structural}%
  \BibitemOpen
  \bibfield  {author} {\bibinfo {author} {\bibfnamefont {F.}~\bibnamefont
  {Brivio}}, \bibinfo {author} {\bibfnamefont {A.~B.}\ \bibnamefont {Walker}},\
  and\ \bibinfo {author} {\bibfnamefont {A.}~\bibnamefont {Walsh}},\ }\bibfield
   {title} {\bibinfo {title} {Structural and electronic properties of hybrid
  perovskites for high-efficiency thin-film photovoltaics from
  first-principles},\ }\href
  {https://pubs.aip.org/aip/apm/article/1/4/042111/120006/Structural-and-electronic-properties-of-hybrid}
  {\bibfield  {journal} {\bibinfo  {journal} {Apl Materials}\ }\textbf
  {\bibinfo {volume} {1}},\ \bibinfo {pages} {042111} (\bibinfo {year}
  {2013})}\BibitemShut {NoStop}%
\bibitem [{\citenamefont {De~Wolf}\ \emph {et~al.}(2014)\citenamefont
  {De~Wolf}, \citenamefont {Holovsky}, \citenamefont {Moon}, \citenamefont
  {Loper}, \citenamefont {Niesen}, \citenamefont {Ledinsky}, \citenamefont
  {Haug}, \citenamefont {Yum},\ and\ \citenamefont
  {Ballif}}]{de2014organometallic}%
  \BibitemOpen
  \bibfield  {author} {\bibinfo {author} {\bibfnamefont {S.}~\bibnamefont
  {De~Wolf}}, \bibinfo {author} {\bibfnamefont {J.}~\bibnamefont {Holovsky}},
  \bibinfo {author} {\bibfnamefont {S.-J.}\ \bibnamefont {Moon}}, \bibinfo
  {author} {\bibfnamefont {P.}~\bibnamefont {Loper}}, \bibinfo {author}
  {\bibfnamefont {B.}~\bibnamefont {Niesen}}, \bibinfo {author} {\bibfnamefont
  {M.}~\bibnamefont {Ledinsky}}, \bibinfo {author} {\bibfnamefont {F.-J.}\
  \bibnamefont {Haug}}, \bibinfo {author} {\bibfnamefont {J.-H.}\ \bibnamefont
  {Yum}},\ and\ \bibinfo {author} {\bibfnamefont {C.}~\bibnamefont {Ballif}},\
  }\bibfield  {title} {\bibinfo {title} {Organometallic halide perovskites:
  sharp optical absorption edge and its relation to photovoltaic performance},\
  }\href {https://doi.org/10.1021/jz500279b} {\bibfield  {journal} {\bibinfo
  {journal} {The journal of physical chemistry letters}\ }\textbf {\bibinfo
  {volume} {5}},\ \bibinfo {pages} {1035} (\bibinfo {year} {2014})}\BibitemShut
  {NoStop}%
\bibitem [{\citenamefont {D’innocenzo}\ \emph {et~al.}(2014)\citenamefont
  {D’innocenzo}, \citenamefont {Grancini}, \citenamefont {Alcocer},
  \citenamefont {Kandada}, \citenamefont {Stranks}, \citenamefont {Lee},
  \citenamefont {Lanzani}, \citenamefont {Snaith},\ and\ \citenamefont
  {Petrozza}}]{d2014excitons}%
  \BibitemOpen
  \bibfield  {author} {\bibinfo {author} {\bibfnamefont {V.}~\bibnamefont
  {D’innocenzo}}, \bibinfo {author} {\bibfnamefont {G.}~\bibnamefont
  {Grancini}}, \bibinfo {author} {\bibfnamefont {M.~J.}\ \bibnamefont
  {Alcocer}}, \bibinfo {author} {\bibfnamefont {A.~R.~S.}\ \bibnamefont
  {Kandada}}, \bibinfo {author} {\bibfnamefont {S.~D.}\ \bibnamefont
  {Stranks}}, \bibinfo {author} {\bibfnamefont {M.~M.}\ \bibnamefont {Lee}},
  \bibinfo {author} {\bibfnamefont {G.}~\bibnamefont {Lanzani}}, \bibinfo
  {author} {\bibfnamefont {H.~J.}\ \bibnamefont {Snaith}},\ and\ \bibinfo
  {author} {\bibfnamefont {A.}~\bibnamefont {Petrozza}},\ }\bibfield  {title}
  {\bibinfo {title} {Excitons versus free charges in organo-lead tri-halide
  perovskites},\ }\href {https://doi.org/10.1038/ncomms4586} {\bibfield
  {journal} {\bibinfo  {journal} {Nature communications}\ }\textbf {\bibinfo
  {volume} {5}},\ \bibinfo {pages} {3586} (\bibinfo {year} {2014})}\BibitemShut
  {NoStop}%
\bibitem [{\citenamefont {Kaur}\ and\ \citenamefont
  {Chakraborty}(2022)}]{kaur2022tuning}%
  \BibitemOpen
  \bibfield  {author} {\bibinfo {author} {\bibfnamefont {J.}~\bibnamefont
  {Kaur}}\ and\ \bibinfo {author} {\bibfnamefont {S.}~\bibnamefont
  {Chakraborty}},\ }\bibfield  {title} {\bibinfo {title} {Tuning spin texture
  and spectroscopic limited maximum efficiency through chemical composition
  space in double halide perovskites},\ }\href
  {https://doi.org/10.1021/acsaem.1c03824} {\bibfield  {journal} {\bibinfo
  {journal} {ACS Applied Energy Materials}\ }\textbf {\bibinfo {volume} {5}},\
  \bibinfo {pages} {5579} (\bibinfo {year} {2022})}\BibitemShut {NoStop}%
\bibitem [{\citenamefont {D.~Ghosh}\ and\ \citenamefont
  {Neukirch}(2022)}]{Ghosh2022}%
  \BibitemOpen
  \bibfield  {author} {\bibinfo {author} {\bibfnamefont {O.~P. W. N. S.~T.}\
  \bibnamefont {D.~Ghosh}, \bibfnamefont {C.~Mora~Perez}}\ and\ \bibinfo
  {author} {\bibfnamefont {A.~J.}\ \bibnamefont {Neukirch}},\ }\bibfield
  {title} {\bibinfo {title} {Impact of composition engineering on charge
  carrier cooling in hybrid perovskites: computational insights},\ }\href
  {https://doi.org/10.1039/D2TC01413K} {\bibfield  {journal} {\bibinfo
  {journal} {Journal of Materials Chemistry C}\ }\textbf {\bibinfo {volume}
  {10}},\ \bibinfo {pages} {9563} (\bibinfo {year} {2022})}\BibitemShut
  {NoStop}%
\bibitem [{\citenamefont {Nayak}\ and\ \citenamefont
  {Ghosh}(2025)}]{Nayak2025}%
  \BibitemOpen
  \bibfield  {author} {\bibinfo {author} {\bibfnamefont {P.~K.}\ \bibnamefont
  {Nayak}}\ and\ \bibinfo {author} {\bibfnamefont {D.}~\bibnamefont {Ghosh}},\
  }\bibfield  {title} {\bibinfo {title} {Optimizing excited charge dynamics in
  layered halide perovskites through compositional engineering},\ }\href
  {https://doi.org/10.1021/acs.nanolett.5c01223} {\bibfield  {journal}
  {\bibinfo  {journal} {Nano Letters}\ }\textbf {\bibinfo {volume} {25}},\
  \bibinfo {pages} {5520} (\bibinfo {year} {2025})}\BibitemShut {NoStop}%
\bibitem [{\citenamefont {Nayak}\ \emph {et~al.}(2024)\citenamefont {Nayak},
  \citenamefont {Mora~Perez}, \citenamefont {Liu}, \citenamefont {Prezhdo},\
  and\ \citenamefont {Ghosh}}]{Nayak2024}%
  \BibitemOpen
  \bibfield  {author} {\bibinfo {author} {\bibfnamefont {P.~K.}\ \bibnamefont
  {Nayak}}, \bibinfo {author} {\bibfnamefont {C.}~\bibnamefont {Mora~Perez}},
  \bibinfo {author} {\bibfnamefont {D.}~\bibnamefont {Liu}}, \bibinfo {author}
  {\bibfnamefont {O.~V.}\ \bibnamefont {Prezhdo}},\ and\ \bibinfo {author}
  {\bibfnamefont {D.}~\bibnamefont {Ghosh}},\ }\bibfield  {title} {\bibinfo
  {title} {A-cation-dependent excited state charge carrier dynamics in
  vacancy-ordered halide perovskites: Insights from computational and machine
  learning models},\ }\href {https://doi.org/10.1021/acs.chemmater.4c00290}
  {\bibfield  {journal} {\bibinfo  {journal} {Chemistry of Materials}\ }\textbf
  {\bibinfo {volume} {36}},\ \bibinfo {pages} {3875} (\bibinfo {year}
  {2024})}\BibitemShut {NoStop}%
\bibitem [{\citenamefont {Zhang}\ \emph {et~al.}(2018)\citenamefont {Zhang},
  \citenamefont {Fang}, \citenamefont {Tokina}, \citenamefont {Long},\ and\
  \citenamefont {Prezhdo}}]{Zhang2018}%
  \BibitemOpen
  \bibfield  {author} {\bibinfo {author} {\bibfnamefont {Z.}~\bibnamefont
  {Zhang}}, \bibinfo {author} {\bibfnamefont {W.-H.}\ \bibnamefont {Fang}},
  \bibinfo {author} {\bibfnamefont {M.~V.}\ \bibnamefont {Tokina}}, \bibinfo
  {author} {\bibfnamefont {R.}~\bibnamefont {Long}},\ and\ \bibinfo {author}
  {\bibfnamefont {O.~V.}\ \bibnamefont {Prezhdo}},\ }\bibfield  {title}
  {\bibinfo {title} {Rapid decoherence suppresses charge recombination in
  multi-layer {2D} halide perovskites: Time-domain ab initio analysis},\ }\href
  {https://doi.org/10.1021/acs.nanolett.8b00035} {\bibfield  {journal}
  {\bibinfo  {journal} {Nano Letters}\ }\textbf {\bibinfo {volume} {18}},\
  \bibinfo {pages} {2459} (\bibinfo {year} {2018})}\BibitemShut {NoStop}%
\bibitem [{\citenamefont {Ghosh}\ \emph {et~al.}(2020)\citenamefont {Ghosh},
  \citenamefont {Acharya}, \citenamefont {Pedesseau}, \citenamefont {Katan},
  \citenamefont {Even}, \citenamefont {Tretiak},\ and\ \citenamefont
  {Neukirch}}]{Ghosh2020}%
  \BibitemOpen
  \bibfield  {author} {\bibinfo {author} {\bibfnamefont {D.}~\bibnamefont
  {Ghosh}}, \bibinfo {author} {\bibfnamefont {D.}~\bibnamefont {Acharya}},
  \bibinfo {author} {\bibfnamefont {L.}~\bibnamefont {Pedesseau}}, \bibinfo
  {author} {\bibfnamefont {C.}~\bibnamefont {Katan}}, \bibinfo {author}
  {\bibfnamefont {J.}~\bibnamefont {Even}}, \bibinfo {author} {\bibfnamefont
  {S.}~\bibnamefont {Tretiak}},\ and\ \bibinfo {author} {\bibfnamefont {A.~J.}\
  \bibnamefont {Neukirch}},\ }\bibfield  {title} {\bibinfo {title} {Charge
  carrier dynamics in two-dimensional hybrid perovskites: Dion–jacobson vs.
  ruddlesden–popper phases},\ }\href {https://doi.org/10.1039/D0TA07205B}
  {\bibfield  {journal} {\bibinfo  {journal} {Journal of Materials Chemistry
  A}\ }\textbf {\bibinfo {volume} {8}},\ \bibinfo {pages} {22009} (\bibinfo
  {year} {2020})}\BibitemShut {NoStop}%
\bibitem [{\citenamefont {Li}\ \emph {et~al.}(2017)\citenamefont {Li},
  \citenamefont {Zhang}, \citenamefont {Zhang},\ and\ \citenamefont
  {Yin}}]{li2017pbcl2}%
  \BibitemOpen
  \bibfield  {author} {\bibinfo {author} {\bibfnamefont {B.}~\bibnamefont
  {Li}}, \bibinfo {author} {\bibfnamefont {Y.}~\bibnamefont {Zhang}}, \bibinfo
  {author} {\bibfnamefont {L.}~\bibnamefont {Zhang}},\ and\ \bibinfo {author}
  {\bibfnamefont {L.}~\bibnamefont {Yin}},\ }\bibfield  {title} {\bibinfo
  {title} {Pbcl2-tuned inorganic cubic {C}s{P}b{B}r$_3$ ({C}l) perovskite solar
  cells with enhanced electron lifetime, diffusion length and photovoltaic
  performance},\ }\href {https://doi.org/10.1016/j.jpowsour.2017.05.050}
  {\bibfield  {journal} {\bibinfo  {journal} {Journal of Power Sources}\
  }\textbf {\bibinfo {volume} {360}},\ \bibinfo {pages} {11} (\bibinfo {year}
  {2017})}\BibitemShut {NoStop}%
\bibitem [{\citenamefont {He}\ \emph {et~al.}(2018)\citenamefont {He},
  \citenamefont {Vasenko}, \citenamefont {Long},\ and\ \citenamefont
  {Prezhdo}}]{he2018halide}%
  \BibitemOpen
  \bibfield  {author} {\bibinfo {author} {\bibfnamefont {J.}~\bibnamefont
  {He}}, \bibinfo {author} {\bibfnamefont {A.~S.}\ \bibnamefont {Vasenko}},
  \bibinfo {author} {\bibfnamefont {R.}~\bibnamefont {Long}},\ and\ \bibinfo
  {author} {\bibfnamefont {O.~V.}\ \bibnamefont {Prezhdo}},\ }\bibfield
  {title} {\bibinfo {title} {Halide composition controls electron--hole
  recombination in cesium--lead halide perovskite quantum dots: a time domain
  ab initio study},\ }\href {https://doi.org/10.1021/acs.jpclett.8b00446}
  {\bibfield  {journal} {\bibinfo  {journal} {The journal of physical chemistry
  letters}\ }\textbf {\bibinfo {volume} {9}},\ \bibinfo {pages} {1872}
  (\bibinfo {year} {2018})}\BibitemShut {NoStop}%
\bibitem [{\citenamefont {Chu}\ \emph {et~al.}(2020)\citenamefont {Chu},
  \citenamefont {Zheng}, \citenamefont {Prezhdo}, \citenamefont {Zhao},\ and\
  \citenamefont {Saidi}}]{chu2020low}%
  \BibitemOpen
  \bibfield  {author} {\bibinfo {author} {\bibfnamefont {W.}~\bibnamefont
  {Chu}}, \bibinfo {author} {\bibfnamefont {Q.}~\bibnamefont {Zheng}}, \bibinfo
  {author} {\bibfnamefont {O.~V.}\ \bibnamefont {Prezhdo}}, \bibinfo {author}
  {\bibfnamefont {J.}~\bibnamefont {Zhao}},\ and\ \bibinfo {author}
  {\bibfnamefont {W.~A.}\ \bibnamefont {Saidi}},\ }\bibfield  {title} {\bibinfo
  {title} {Low-frequency lattice phonons in halide perovskites explain high
  defect tolerance toward electron-hole recombination},\ }\href
  {https://www.science.org/doi/10.1126/sciadv.aaw7453} {\bibfield  {journal}
  {\bibinfo  {journal} {Science advances}\ }\textbf {\bibinfo {volume} {6}},\
  \bibinfo {pages} {eaaw7453} (\bibinfo {year} {2020})}\BibitemShut {NoStop}%
\bibitem [{\citenamefont {Shan}\ and\ \citenamefont
  {Saidi}(2017)}]{shan2017segregation}%
  \BibitemOpen
  \bibfield  {author} {\bibinfo {author} {\bibfnamefont {W.}~\bibnamefont
  {Shan}}\ and\ \bibinfo {author} {\bibfnamefont {W.~A.}\ \bibnamefont
  {Saidi}},\ }\bibfield  {title} {\bibinfo {title} {Segregation of native
  defects to the grain boundaries in methylammonium lead iodide perovskite},\
  }\href {https://pubs.acs.org/doi/10.1021/acs.jpclett.7b02727} {\bibfield
  {journal} {\bibinfo  {journal} {The journal of physical chemistry letters}\
  }\textbf {\bibinfo {volume} {8}},\ \bibinfo {pages} {5935} (\bibinfo {year}
  {2017})}\BibitemShut {NoStop}%
\bibitem [{\citenamefont {Zakutayev}\ \emph {et~al.}(2014)\citenamefont
  {Zakutayev}, \citenamefont {Caskey}, \citenamefont {Fioretti}, \citenamefont
  {Ginley}, \citenamefont {Vidal}, \citenamefont {Stevanovic}, \citenamefont
  {Tea},\ and\ \citenamefont {Lany}}]{zakutayev2014defect}%
  \BibitemOpen
  \bibfield  {author} {\bibinfo {author} {\bibfnamefont {A.}~\bibnamefont
  {Zakutayev}}, \bibinfo {author} {\bibfnamefont {C.~M.}\ \bibnamefont
  {Caskey}}, \bibinfo {author} {\bibfnamefont {A.~N.}\ \bibnamefont
  {Fioretti}}, \bibinfo {author} {\bibfnamefont {D.~S.}\ \bibnamefont
  {Ginley}}, \bibinfo {author} {\bibfnamefont {J.}~\bibnamefont {Vidal}},
  \bibinfo {author} {\bibfnamefont {V.}~\bibnamefont {Stevanovic}}, \bibinfo
  {author} {\bibfnamefont {E.}~\bibnamefont {Tea}},\ and\ \bibinfo {author}
  {\bibfnamefont {S.}~\bibnamefont {Lany}},\ }\bibfield  {title} {\bibinfo
  {title} {Defect tolerant semiconductors for solar energy conversion},\ }\href
  {https://pubs.acs.org/doi/10.1021/jz5001787} {\bibfield  {journal} {\bibinfo
  {journal} {The journal of physical chemistry letters}\ }\textbf {\bibinfo
  {volume} {5}},\ \bibinfo {pages} {1117} (\bibinfo {year} {2014})}\BibitemShut
  {NoStop}%
\end{thebibliography}%

\end{document}